\documentclass[submission, Phys]{SciPost}

\usepackage[export]{adjustbox}
\usepackage{amsmath,amssymb}
\usepackage{bm}
\usepackage{tikz}

\def\plaquettev{\tikz[baseline=.1ex]{
\fill (0,0) circle (1pt) coordinate (A);
\fill (1.5ex,0) circle (1pt) coordinate (B);
\fill (2.25ex,1.3ex) circle (1pt) coordinate (C);
\fill (0.75ex,1.3ex) circle (1pt) coordinate (D);
\draw (A)--(B);
\draw [ultra thick] (B)--(C);
\draw (C)--(D);
\draw [ultra thick] (D)--(A);
\draw (B)--(D);}
}

\def\plaquetteh{\tikz[baseline=.1ex]{
\fill (0,0) circle (1pt) coordinate (A);
\fill (1.5ex,0) circle (1pt) coordinate (B);
\fill (2.25ex,1.3ex) circle (1pt) coordinate (C);
\fill (0.75ex,1.3ex) circle (1pt) coordinate (D);
\draw [ultra thick] (A)--(B);
\draw (B)--(C);
\draw [ultra thick] (C)--(D);
\draw (D)--(A);
\draw (B)--(D);}
}

\usepackage{subcaption}
\def\plaquettev{\tikz[baseline=.1ex]{
\fill (0,0) circle (1pt) coordinate (A);
\fill (1.5ex,0) circle (1pt) coordinate (B);
\fill (2.25ex,1.3ex) circle (1pt) coordinate (C);
\fill (0.75ex,1.3ex) circle (1pt) coordinate (D);
\draw (A)--(B);
\draw [ultra thick] (B)--(C);
\draw (C)--(D);
\draw [ultra thick] (D)--(A);
\draw (B)--(D);}
}

\def\plaquetteh{\tikz[baseline=.1ex]{
\fill (0,0) circle (1pt) coordinate (A);
\fill (1.5ex,0) circle (1pt) coordinate (B);
\fill (2.25ex,1.3ex) circle (1pt) coordinate (C);
\fill (0.75ex,1.3ex) circle (1pt) coordinate (D);
\draw [ultra thick] (A)--(B);
\draw (B)--(C);
\draw [ultra thick] (C)--(D);
\draw (D)--(A);
\draw (B)--(D);}
}

\def\dimersIbra{ \tikz[baseline=-0.5ex]{ \fill (-1.125ex,-0.649519ex)
circle (1pt); \fill (-0.375ex,0.649519ex) circle (1pt); \fill
(0.375ex,-0.649519ex) circle (1pt); \fill (1.125ex,0.649519ex) circle
(1pt); \draw[ultra thick]
(-1.125ex,-0.649519ex)--(0.375ex,-0.649519ex); \draw[ultra thick]
(-0.375ex,0.649519ex)--(1.125ex,0.649519ex); \draw[thin]
(-1.125ex,-0.649519ex)--(-0.375ex,0.649519ex); \draw[thin]
(0.375ex,-0.649519ex)--(1.125ex,0.649519ex); } } \def\dimersIIbra{
\tikz[baseline=-0.5ex]{ \fill (-1.125ex,-0.649519ex) circle (1pt);
\fill (0.375ex,-0.649519ex) circle (1pt); \fill (-0.375ex,0.649519ex)
circle (1pt); \fill (1.125ex,0.649519ex) circle (1pt); \draw[ultra
thick] (-1.125ex,-0.649519ex)--(-0.375ex,0.649519ex); \draw[ultra
thick] (0.375ex,-0.649519ex)--(1.125ex,0.649519ex); \draw[thin]
(-1.125ex,-0.649519ex)--(0.375ex,-0.649519ex); \draw[thin]
(-0.375ex,0.649519ex)--(1.125ex,0.649519ex); } } \def\dimersIIIbra{
\tikz[baseline=-0.5ex]{ \fill (-1.875ex,-0.649519ex) circle (1pt);
\fill (-1.125ex,0.649519ex) circle (1pt); \fill (-0.375ex,-0.649519ex)
circle (1pt); \fill (1.125ex,-0.649519ex) circle (1pt); \fill
(0.375ex,0.649519ex) circle (1pt); \fill (1.875ex,0.649519ex) circle
(1pt); \draw[ultra thick]
(-1.875ex,-0.649519ex)--(-0.375ex,-0.649519ex); \draw[ultra thick]
(-1.125ex,0.649519ex)--(0.375ex,0.649519ex); \draw[ultra thick]
(1.125ex,-0.649519ex)--(1.875ex,0.649519ex); \draw[thin]
(-1.875ex,-0.649519ex)--(-1.125ex,0.649519ex); \draw[thin]
(-0.375ex,-0.649519ex)--(1.125ex,-0.649519ex); \draw[thin]
(0.375ex,0.649519ex)--(1.875ex,0.649519ex); } } \def\dimersIVbra{
\tikz[baseline=-0.5ex]{ \fill (-1.875ex,-0.649519ex) circle (1pt);
\fill (-0.375ex,-0.649519ex) circle (1pt); \fill (-1.125ex,0.649519ex)
circle (1pt); \fill (0.375ex,0.649519ex) circle (1pt); \fill
(1.125ex,-0.649519ex) circle (1pt); \fill (1.875ex,0.649519ex) circle
(1pt); \draw[ultra thick]
(-1.875ex,-0.649519ex)--(-1.125ex,0.649519ex); \draw[ultra thick]
(-0.375ex,-0.649519ex)--(1.125ex,-0.649519ex); \draw[ultra thick]
(0.375ex,0.649519ex)--(1.875ex,0.649519ex); \draw[thin]
(-1.875ex,-0.649519ex)--(-0.375ex,-0.649519ex); \draw[thin]
(-1.125ex,0.649519ex)--(0.375ex,0.649519ex); \draw[thin]
(1.125ex,-0.649519ex)--(1.875ex,0.649519ex); } } \def\dimersVbra{
\tikz[baseline=-0.5ex]{ \fill (-1.ex,-1.29904ex) circle (1pt); \fill
(-0.25ex,0.ex) circle (1pt); \fill (-1.ex,1.29904ex) circle (1pt);
\fill (0.5ex,1.29904ex) circle (1pt); \fill (0.5ex,-1.29904ex) circle
(1pt); \fill (1.25ex,0.ex) circle (1pt); \draw[ultra thick]
(-1.ex,-1.29904ex)--(0.5ex,-1.29904ex); \draw[ultra thick]
(-1.ex,1.29904ex)--(-0.25ex,0.ex); \draw[ultra thick]
(0.5ex,1.29904ex)--(1.25ex,0.ex); \draw[thin]
(-1.ex,-1.29904ex)--(-0.25ex,0.ex); \draw[thin]
(-1.ex,1.29904ex)--(0.5ex,1.29904ex); \draw[thin]
(0.5ex,-1.29904ex)--(1.25ex,0.ex); } } \def\dimersVIbra{
\tikz[baseline=-0.5ex]{ \fill (-1.ex,-1.29904ex) circle (1pt); \fill
(0.5ex,-1.29904ex) circle (1pt); \fill (-1.ex,1.29904ex) circle (1pt);
\fill (-0.25ex,0.ex) circle (1pt); \fill (0.5ex,1.29904ex) circle
(1pt); \fill (1.25ex,0.ex) circle (1pt); \draw[ultra thick]
(-1.ex,-1.29904ex)--(-0.25ex,0.ex); \draw[ultra thick]
(-1.ex,1.29904ex)--(0.5ex,1.29904ex); \draw[ultra thick]
(0.5ex,-1.29904ex)--(1.25ex,0.ex); \draw[thin]
(-1.ex,-1.29904ex)--(0.5ex,-1.29904ex); \draw[thin]
(-1.ex,1.29904ex)--(-0.25ex,0.ex); \draw[thin]
(0.5ex,1.29904ex)--(1.25ex,0.ex); } } \def\dimersVIIbra{
\tikz[baseline=-0.5ex]{ \fill (-1.125ex,-0.649519ex) circle (1pt);
\fill (-0.375ex,0.649519ex) circle (1pt); \fill (0.375ex,-0.649519ex)
circle (1pt); \fill (1.125ex,0.649519ex) circle (1pt); \draw[ultra
thick] (-1.125ex,-0.649519ex)--(-0.375ex,0.649519ex); \draw[ultra
thick] (0.375ex,-0.649519ex)--(1.125ex,0.649519ex); \draw[thin]
(-1.125ex,-0.649519ex)--(-0.375ex,0.649519ex); \draw[thin]
(0.375ex,-0.649519ex)--(1.125ex,0.649519ex); } } \def\dimersVIIIbra{
\tikz[baseline=-0.5ex]{ \fill (-1.125ex,-0.649519ex) circle (1pt);
\fill (0.375ex,-0.649519ex) circle (1pt); \fill (-0.375ex,0.649519ex)
circle (1pt); \fill (1.125ex,0.649519ex) circle (1pt); \draw[ultra
thick] (-1.125ex,-0.649519ex)--(0.375ex,-0.649519ex); \draw[ultra
thick] (-0.375ex,0.649519ex)--(1.125ex,0.649519ex); \draw[thin]
(-1.125ex,-0.649519ex)--(0.375ex,-0.649519ex); \draw[thin]
(-0.375ex,0.649519ex)--(1.125ex,0.649519ex); } }   \def\dimersXIbra{
\tikz[baseline=-0.5ex]{ \fill (-1.96875ex,-1.13666ex) circle (1pt);
\fill (-1.21875ex,0.16238ex) circle (1pt); \fill
(-0.46875ex,-1.13666ex) circle (1pt); \fill (1.03125ex,-1.13666ex)
circle (1pt); \fill (-0.46875ex,1.46142ex) circle (1pt); \fill
(0.28125ex,0.16238ex) circle (1pt); \fill (1.03125ex,1.46142ex) circle
(1pt); \fill (1.78125ex,0.16238ex) circle (1pt); \draw[ultra thick]
(-1.96875ex,-1.13666ex)--(-0.46875ex,-1.13666ex); \draw[ultra thick]
(-1.21875ex,0.16238ex)--(-0.46875ex,1.46142ex); \draw[ultra thick]
(0.28125ex,0.16238ex)--(1.03125ex,1.46142ex); \draw[ultra thick]
(1.03125ex,-1.13666ex)--(1.78125ex,0.16238ex); \draw[thin]
(-1.96875ex,-1.13666ex)--(-1.21875ex,0.16238ex); \draw[thin]
(-0.46875ex,-1.13666ex)--(1.03125ex,-1.13666ex); \draw[thin]
(-0.46875ex,1.46142ex)--(0.28125ex,0.16238ex); \draw[thin]
(1.03125ex,1.46142ex)--(1.78125ex,0.16238ex); } } \def\dimersXIIbra{
\tikz[baseline=-0.5ex]{ \fill (-1.96875ex,-1.13666ex) circle (1pt);
\fill (-0.46875ex,-1.13666ex) circle (1pt); \fill
(-1.21875ex,0.16238ex) circle (1pt); \fill (-0.46875ex,1.46142ex)
circle (1pt); \fill (0.28125ex,0.16238ex) circle (1pt); \fill
(1.03125ex,1.46142ex) circle (1pt); \fill (1.03125ex,-1.13666ex)
circle (1pt); \fill (1.78125ex,0.16238ex) circle (1pt); \draw[ultra
thick] (-1.96875ex,-1.13666ex)--(-1.21875ex,0.16238ex); \draw[ultra
thick] (-0.46875ex,-1.13666ex)--(1.03125ex,-1.13666ex); \draw[ultra
thick] (-0.46875ex,1.46142ex)--(0.28125ex,0.16238ex); \draw[ultra
thick] (1.03125ex,1.46142ex)--(1.78125ex,0.16238ex); \draw[thin]
(-1.96875ex,-1.13666ex)--(-0.46875ex,-1.13666ex); \draw[thin]
(-1.21875ex,0.16238ex)--(-0.46875ex,1.46142ex); \draw[thin]
(0.28125ex,0.16238ex)--(1.03125ex,1.46142ex); \draw[thin]
(1.03125ex,-1.13666ex)--(1.78125ex,0.16238ex); } } \def\dimersXIIIbra{
\tikz[baseline=-0.5ex]{ \fill (-1.5ex,1.29904ex) circle (1pt); \fill
(-0.75ex,0.ex) circle (1pt); \fill (-1.5ex,-1.29904ex) circle (1pt);
\fill (0.ex,-1.29904ex) circle (1pt); \fill (0.ex,1.29904ex) circle
(1pt); \fill (1.5ex,1.29904ex) circle (1pt); \fill (0.75ex,0.ex)
circle (1pt); \fill (1.5ex,-1.29904ex) circle (1pt); \draw[ultra
thick] (-1.5ex,1.29904ex)--(0.ex,1.29904ex); \draw[ultra thick]
(-1.5ex,-1.29904ex)--(-0.75ex,0.ex); \draw[ultra thick]
(0.ex,-1.29904ex)--(1.5ex,-1.29904ex); \draw[ultra thick]
(0.75ex,0.ex)--(1.5ex,1.29904ex); \draw[thin]
(-1.5ex,1.29904ex)--(-0.75ex,0.ex); \draw[thin]
(-1.5ex,-1.29904ex)--(0.ex,-1.29904ex); \draw[thin]
(0.ex,1.29904ex)--(1.5ex,1.29904ex); \draw[thin]
(0.75ex,0.ex)--(1.5ex,-1.29904ex); } } \def\dimersXIVbra{
\tikz[baseline=-0.5ex]{ \fill (-1.5ex,1.29904ex) circle (1pt); \fill
(0.ex,1.29904ex) circle (1pt); \fill (-1.5ex,-1.29904ex) circle (1pt);
\fill (-0.75ex,0.ex) circle (1pt); \fill (0.ex,-1.29904ex) circle
(1pt); \fill (1.5ex,-1.29904ex) circle (1pt); \fill (0.75ex,0.ex)
circle (1pt); \fill (1.5ex,1.29904ex) circle (1pt); \draw[ultra thick]
(-1.5ex,1.29904ex)--(-0.75ex,0.ex); \draw[ultra thick]
(-1.5ex,-1.29904ex)--(0.ex,-1.29904ex); \draw[ultra thick]
(0.ex,1.29904ex)--(1.5ex,1.29904ex); \draw[ultra thick]
(0.75ex,0.ex)--(1.5ex,-1.29904ex); \draw[thin]
(-1.5ex,1.29904ex)--(0.ex,1.29904ex); \draw[thin]
(-1.5ex,-1.29904ex)--(-0.75ex,0.ex); \draw[thin]
(0.ex,-1.29904ex)--(1.5ex,-1.29904ex); \draw[thin]
(0.75ex,0.ex)--(1.5ex,1.29904ex); } } \def\dimersXVbra{
\tikz[baseline=-0.5ex]{ \fill (-2.34375ex,0.811899ex) circle (1pt);
\fill (-1.59375ex,-0.487139ex) circle (1pt); \fill
(-0.84375ex,0.811899ex) circle (1pt); \fill (0.65625ex,0.811899ex)
circle (1pt); \fill (-0.09375ex,-0.487139ex) circle (1pt); \fill
(0.65625ex,-1.78618ex) circle (1pt); \fill (1.40625ex,-0.487139ex)
circle (1pt); \fill (2.15625ex,0.811899ex) circle (1pt); \draw[ultra
thick] (-2.34375ex,0.811899ex)--(-0.84375ex,0.811899ex); \draw[ultra
thick] (-1.59375ex,-0.487139ex)--(-0.09375ex,-0.487139ex); \draw[ultra
thick] (0.65625ex,0.811899ex)--(2.15625ex,0.811899ex); \draw[ultra
thick] (0.65625ex,-1.78618ex)--(1.40625ex,-0.487139ex); \draw[thin]
(-2.34375ex,0.811899ex)--(-1.59375ex,-0.487139ex); \draw[thin]
(-0.84375ex,0.811899ex)--(0.65625ex,0.811899ex); \draw[thin]
(-0.09375ex,-0.487139ex)--(0.65625ex,-1.78618ex); \draw[thin]
(1.40625ex,-0.487139ex)--(2.15625ex,0.811899ex); } }
\def\dimersXVIbra{ \tikz[baseline=-0.5ex]{ \fill
(-2.34375ex,0.811899ex) circle (1pt); \fill (-0.84375ex,0.811899ex)
circle (1pt); \fill (-1.59375ex,-0.487139ex) circle (1pt); \fill
(-0.09375ex,-0.487139ex) circle (1pt); \fill (0.65625ex,0.811899ex)
circle (1pt); \fill (2.15625ex,0.811899ex) circle (1pt); \fill
(0.65625ex,-1.78618ex) circle (1pt); \fill (1.40625ex,-0.487139ex)
circle (1pt); \draw[ultra thick]
(-2.34375ex,0.811899ex)--(-1.59375ex,-0.487139ex); \draw[ultra thick]
(-0.84375ex,0.811899ex)--(0.65625ex,0.811899ex); \draw[ultra thick]
(-0.09375ex,-0.487139ex)--(0.65625ex,-1.78618ex); \draw[ultra thick]
(1.40625ex,-0.487139ex)--(2.15625ex,0.811899ex); \draw[thin]
(-2.34375ex,0.811899ex)--(-0.84375ex,0.811899ex); \draw[thin]
(-1.59375ex,-0.487139ex)--(-0.09375ex,-0.487139ex); \draw[thin]
(0.65625ex,0.811899ex)--(2.15625ex,0.811899ex); \draw[thin]
(0.65625ex,-1.78618ex)--(1.40625ex,-0.487139ex); } }
        \def\dimersXXVbra{
\tikz[baseline=-0.5ex]{ \fill (-2.34375ex,-1.13666ex) circle (1pt);
\fill (-1.59375ex,0.16238ex) circle (1pt); \fill
(-0.84375ex,-1.13666ex) circle (1pt); \fill (0.65625ex,-1.13666ex)
circle (1pt); \fill (-0.09375ex,0.16238ex) circle (1pt); \fill
(0.65625ex,1.46142ex) circle (1pt); \fill (1.40625ex,0.16238ex) circle
(1pt); \fill (2.15625ex,1.46142ex) circle (1pt); \draw[ultra thick]
(-2.34375ex,-1.13666ex)--(-0.84375ex,-1.13666ex); \draw[ultra thick]
(-1.59375ex,0.16238ex)--(-0.09375ex,0.16238ex); \draw[ultra thick]
(0.65625ex,-1.13666ex)--(1.40625ex,0.16238ex); \draw[ultra thick]
(0.65625ex,1.46142ex)--(2.15625ex,1.46142ex); \draw[thin]
(-2.34375ex,-1.13666ex)--(-1.59375ex,0.16238ex); \draw[thin]
(-0.84375ex,-1.13666ex)--(0.65625ex,-1.13666ex); \draw[thin]
(-0.09375ex,0.16238ex)--(0.65625ex,1.46142ex); \draw[thin]
(1.40625ex,0.16238ex)--(2.15625ex,1.46142ex); } } \def\dimersXXVIbra{
\tikz[baseline=-0.5ex]{ \fill (-2.34375ex,-1.13666ex) circle (1pt);
\fill (-0.84375ex,-1.13666ex) circle (1pt); \fill
(-1.59375ex,0.16238ex) circle (1pt); \fill (-0.09375ex,0.16238ex)
circle (1pt); \fill (0.65625ex,-1.13666ex) circle (1pt); \fill
(1.40625ex,0.16238ex) circle (1pt); \fill (0.65625ex,1.46142ex) circle
(1pt); \fill (2.15625ex,1.46142ex) circle (1pt); \draw[ultra thick]
(-2.34375ex,-1.13666ex)--(-1.59375ex,0.16238ex); \draw[ultra thick]
(-0.84375ex,-1.13666ex)--(0.65625ex,-1.13666ex); \draw[ultra thick]
(-0.09375ex,0.16238ex)--(0.65625ex,1.46142ex); \draw[ultra thick]
(1.40625ex,0.16238ex)--(2.15625ex,1.46142ex); \draw[thin]
(-2.34375ex,-1.13666ex)--(-0.84375ex,-1.13666ex); \draw[thin]
(-1.59375ex,0.16238ex)--(-0.09375ex,0.16238ex); \draw[thin]
(0.65625ex,-1.13666ex)--(1.40625ex,0.16238ex); \draw[thin]
(0.65625ex,1.46142ex)--(2.15625ex,1.46142ex); } } \def\dimersXXVIIbra{
\tikz[baseline=-0.5ex]{ \fill (-2.25ex,0.32476ex) circle (1pt); \fill
(-1.5ex,-0.974279ex) circle (1pt); \fill (-0.75ex,0.32476ex) circle
(1pt); \fill (0.ex,1.6238ex) circle (1pt); \fill (0.ex,-0.974279ex)
circle (1pt); \fill (1.5ex,-0.974279ex) circle (1pt); \fill
(0.75ex,0.32476ex) circle (1pt); \fill (2.25ex,0.32476ex) circle
(1pt); \draw[ultra thick] (-2.25ex,0.32476ex)--(-0.75ex,0.32476ex);
\draw[ultra thick] (-1.5ex,-0.974279ex)--(0.ex,-0.974279ex);
\draw[ultra thick] (0.ex,1.6238ex)--(0.75ex,0.32476ex); \draw[ultra
thick] (1.5ex,-0.974279ex)--(2.25ex,0.32476ex); \draw[thin]
(-2.25ex,0.32476ex)--(-1.5ex,-0.974279ex); \draw[thin]
(-0.75ex,0.32476ex)--(0.ex,1.6238ex); \draw[thin]
(0.ex,-0.974279ex)--(1.5ex,-0.974279ex); \draw[thin]
(0.75ex,0.32476ex)--(2.25ex,0.32476ex); } } \def\dimersXXVIIIbra{
\tikz[baseline=-0.5ex]{ \fill (-2.25ex,0.32476ex) circle (1pt); \fill
(-0.75ex,0.32476ex) circle (1pt); \fill (-1.5ex,-0.974279ex) circle
(1pt); \fill (0.ex,-0.974279ex) circle (1pt); \fill (0.ex,1.6238ex)
circle (1pt); \fill (0.75ex,0.32476ex) circle (1pt); \fill
(1.5ex,-0.974279ex) circle (1pt); \fill (2.25ex,0.32476ex) circle
(1pt); \draw[ultra thick] (-2.25ex,0.32476ex)--(-1.5ex,-0.974279ex);
\draw[ultra thick] (-0.75ex,0.32476ex)--(0.ex,1.6238ex); \draw[ultra
thick] (0.ex,-0.974279ex)--(1.5ex,-0.974279ex); \draw[ultra thick]
(0.75ex,0.32476ex)--(2.25ex,0.32476ex); \draw[thin]
(-2.25ex,0.32476ex)--(-0.75ex,0.32476ex); \draw[thin]
(-1.5ex,-0.974279ex)--(0.ex,-0.974279ex); \draw[thin]
(0.ex,1.6238ex)--(0.75ex,0.32476ex); \draw[thin]
(1.5ex,-0.974279ex)--(2.25ex,0.32476ex); } } \def\dimersXXIXbra{
\tikz[baseline=-0.5ex]{ \fill (-1.59375ex,-1.13666ex) circle (1pt);
\fill (-0.84375ex,0.16238ex) circle (1pt); \fill
(-1.59375ex,1.46142ex) circle (1pt); \fill (-0.09375ex,1.46142ex)
circle (1pt); \fill (-0.09375ex,-1.13666ex) circle (1pt); \fill
(1.40625ex,-1.13666ex) circle (1pt); \fill (0.65625ex,0.16238ex)
circle (1pt); \fill (2.15625ex,0.16238ex) circle (1pt); \draw[ultra
thick] (-1.59375ex,-1.13666ex)--(-0.09375ex,-1.13666ex); \draw[ultra
thick] (-1.59375ex,1.46142ex)--(-0.84375ex,0.16238ex); \draw[ultra
thick] (-0.09375ex,1.46142ex)--(0.65625ex,0.16238ex); \draw[ultra
thick] (1.40625ex,-1.13666ex)--(2.15625ex,0.16238ex); \draw[thin]
(-1.59375ex,-1.13666ex)--(-0.84375ex,0.16238ex); \draw[thin]
(-1.59375ex,1.46142ex)--(-0.09375ex,1.46142ex); \draw[thin]
(-0.09375ex,-1.13666ex)--(1.40625ex,-1.13666ex); \draw[thin]
(0.65625ex,0.16238ex)--(2.15625ex,0.16238ex); } } \def\dimersXXXbra{
\tikz[baseline=-0.5ex]{ \fill (-1.59375ex,-1.13666ex) circle (1pt);
\fill (-0.09375ex,-1.13666ex) circle (1pt); \fill
(-1.59375ex,1.46142ex) circle (1pt); \fill (-0.84375ex,0.16238ex)
circle (1pt); \fill (-0.09375ex,1.46142ex) circle (1pt); \fill
(0.65625ex,0.16238ex) circle (1pt); \fill (1.40625ex,-1.13666ex)
circle (1pt); \fill (2.15625ex,0.16238ex) circle (1pt); \draw[ultra
thick] (-1.59375ex,-1.13666ex)--(-0.84375ex,0.16238ex); \draw[ultra
thick] (-1.59375ex,1.46142ex)--(-0.09375ex,1.46142ex); \draw[ultra
thick] (-0.09375ex,-1.13666ex)--(1.40625ex,-1.13666ex); \draw[ultra
thick] (0.65625ex,0.16238ex)--(2.15625ex,0.16238ex); \draw[thin]
(-1.59375ex,-1.13666ex)--(-0.09375ex,-1.13666ex); \draw[thin]
(-1.59375ex,1.46142ex)--(-0.84375ex,0.16238ex); \draw[thin]
(-0.09375ex,1.46142ex)--(0.65625ex,0.16238ex); \draw[thin]
(1.40625ex,-1.13666ex)--(2.15625ex,0.16238ex); } } \def\dimersXXXIbra{
\tikz[baseline=-0.5ex]{ \fill (-2.625ex,-0.649519ex) circle (1pt);
\fill (-1.875ex,0.649519ex) circle (1pt); \fill (-1.125ex,-0.649519ex)
circle (1pt); \fill (0.375ex,-0.649519ex) circle (1pt); \fill
(-0.375ex,0.649519ex) circle (1pt); \fill (1.125ex,0.649519ex) circle
(1pt); \fill (1.875ex,-0.649519ex) circle (1pt); \fill
(2.625ex,0.649519ex) circle (1pt); \draw[ultra thick]
(-2.625ex,-0.649519ex)--(-1.125ex,-0.649519ex); \draw[ultra thick]
(-1.875ex,0.649519ex)--(-0.375ex,0.649519ex); \draw[ultra thick]
(0.375ex,-0.649519ex)--(1.875ex,-0.649519ex); \draw[ultra thick]
(1.125ex,0.649519ex)--(2.625ex,0.649519ex); \draw[thin]
(-2.625ex,-0.649519ex)--(-1.875ex,0.649519ex); \draw[thin]
(-1.125ex,-0.649519ex)--(0.375ex,-0.649519ex); \draw[thin]
(-0.375ex,0.649519ex)--(1.125ex,0.649519ex); \draw[thin]
(1.875ex,-0.649519ex)--(2.625ex,0.649519ex); } } \def\dimersXXXIIbra{
\tikz[baseline=-0.5ex]{ \fill (-2.625ex,-0.649519ex) circle (1pt);
\fill (-1.125ex,-0.649519ex) circle (1pt); \fill (-1.875ex,0.649519ex)
circle (1pt); \fill (-0.375ex,0.649519ex) circle (1pt); \fill
(0.375ex,-0.649519ex) circle (1pt); \fill (1.875ex,-0.649519ex) circle
(1pt); \fill (1.125ex,0.649519ex) circle (1pt); \fill
(2.625ex,0.649519ex) circle (1pt); \draw[ultra thick]
(-2.625ex,-0.649519ex)--(-1.875ex,0.649519ex); \draw[ultra thick]
(-1.125ex,-0.649519ex)--(0.375ex,-0.649519ex); \draw[ultra thick]
(-0.375ex,0.649519ex)--(1.125ex,0.649519ex); \draw[ultra thick]
(1.875ex,-0.649519ex)--(2.625ex,0.649519ex); \draw[thin]
(-2.625ex,-0.649519ex)--(-1.125ex,-0.649519ex); \draw[thin]
(-1.875ex,0.649519ex)--(-0.375ex,0.649519ex); \draw[thin]
(0.375ex,-0.649519ex)--(1.875ex,-0.649519ex); \draw[thin]
(1.125ex,0.649519ex)--(2.625ex,0.649519ex); } } \def\dimersXXXIIIbra{
\tikz[baseline=-0.5ex]{ \fill (-2.4375ex,0.32476ex) circle (1pt);
\fill (-1.6875ex,-0.974279ex) circle (1pt); \fill
(-0.9375ex,0.32476ex) circle (1pt); \fill (0.5625ex,0.32476ex) circle
(1pt); \fill (-0.1875ex,-0.974279ex) circle (1pt); \fill
(1.3125ex,-0.974279ex) circle (1pt); \fill (1.3125ex,1.6238ex) circle
(1pt); \fill (2.0625ex,0.32476ex) circle (1pt); \draw[ultra thick]
(-2.4375ex,0.32476ex)--(-0.9375ex,0.32476ex); \draw[ultra thick]
(-1.6875ex,-0.974279ex)--(-0.1875ex,-0.974279ex); \draw[ultra thick]
(0.5625ex,0.32476ex)--(1.3125ex,1.6238ex); \draw[ultra thick]
(1.3125ex,-0.974279ex)--(2.0625ex,0.32476ex); \draw[thin]
(-2.4375ex,0.32476ex)--(-1.6875ex,-0.974279ex); \draw[thin]
(-0.9375ex,0.32476ex)--(0.5625ex,0.32476ex); \draw[thin]
(-0.1875ex,-0.974279ex)--(1.3125ex,-0.974279ex); \draw[thin]
(1.3125ex,1.6238ex)--(2.0625ex,0.32476ex); } } \def\dimersXXXIVbra{
\tikz[baseline=-0.5ex]{ \fill (-2.4375ex,0.32476ex) circle (1pt);
\fill (-0.9375ex,0.32476ex) circle (1pt); \fill
(-1.6875ex,-0.974279ex) circle (1pt); \fill (-0.1875ex,-0.974279ex)
circle (1pt); \fill (0.5625ex,0.32476ex) circle (1pt); \fill
(1.3125ex,1.6238ex) circle (1pt); \fill (1.3125ex,-0.974279ex) circle
(1pt); \fill (2.0625ex,0.32476ex) circle (1pt); \draw[ultra thick]
(-2.4375ex,0.32476ex)--(-1.6875ex,-0.974279ex); \draw[ultra thick]
(-0.9375ex,0.32476ex)--(0.5625ex,0.32476ex); \draw[ultra thick]
(-0.1875ex,-0.974279ex)--(1.3125ex,-0.974279ex); \draw[ultra thick]
(1.3125ex,1.6238ex)--(2.0625ex,0.32476ex); \draw[thin]
(-2.4375ex,0.32476ex)--(-0.9375ex,0.32476ex); \draw[thin]
(-1.6875ex,-0.974279ex)--(-0.1875ex,-0.974279ex); \draw[thin]
(0.5625ex,0.32476ex)--(1.3125ex,1.6238ex); \draw[thin]
(1.3125ex,-0.974279ex)--(2.0625ex,0.32476ex); } } \def\dimersXXXVbra{
\tikz[baseline=-0.5ex]{ \fill (-1.125ex,-1.94856ex) circle (1pt);
\fill (-0.375ex,-0.649519ex) circle (1pt); \fill (-1.125ex,0.649519ex)
circle (1pt); \fill (-0.375ex,1.94856ex) circle (1pt); \fill
(0.375ex,-1.94856ex) circle (1pt); \fill (1.125ex,-0.649519ex) circle
(1pt); \fill (0.375ex,0.649519ex) circle (1pt); \fill
(1.125ex,1.94856ex) circle (1pt); \draw[ultra thick]
(-1.125ex,-1.94856ex)--(0.375ex,-1.94856ex); \draw[ultra thick]
(-1.125ex,0.649519ex)--(-0.375ex,-0.649519ex); \draw[ultra thick]
(-0.375ex,1.94856ex)--(1.125ex,1.94856ex); \draw[ultra thick]
(0.375ex,0.649519ex)--(1.125ex,-0.649519ex); \draw[thin]
(-1.125ex,-1.94856ex)--(-0.375ex,-0.649519ex); \draw[thin]
(-1.125ex,0.649519ex)--(-0.375ex,1.94856ex); \draw[thin]
(0.375ex,-1.94856ex)--(1.125ex,-0.649519ex); \draw[thin]
(0.375ex,0.649519ex)--(1.125ex,1.94856ex); } } \def\dimersXXXVIbra{
\tikz[baseline=-0.5ex]{ \fill (-1.125ex,-1.94856ex) circle (1pt);
\fill (0.375ex,-1.94856ex) circle (1pt); \fill (-1.125ex,0.649519ex)
circle (1pt); \fill (-0.375ex,-0.649519ex) circle (1pt); \fill
(-0.375ex,1.94856ex) circle (1pt); \fill (1.125ex,1.94856ex) circle
(1pt); \fill (0.375ex,0.649519ex) circle (1pt); \fill
(1.125ex,-0.649519ex) circle (1pt); \draw[ultra thick]
(-1.125ex,-1.94856ex)--(-0.375ex,-0.649519ex); \draw[ultra thick]
(-1.125ex,0.649519ex)--(-0.375ex,1.94856ex); \draw[ultra thick]
(0.375ex,-1.94856ex)--(1.125ex,-0.649519ex); \draw[ultra thick]
(0.375ex,0.649519ex)--(1.125ex,1.94856ex); \draw[thin]
(-1.125ex,-1.94856ex)--(0.375ex,-1.94856ex); \draw[thin]
(-1.125ex,0.649519ex)--(-0.375ex,-0.649519ex); \draw[thin]
(-0.375ex,1.94856ex)--(1.125ex,1.94856ex); \draw[thin]
(0.375ex,0.649519ex)--(1.125ex,-0.649519ex); } } \def\dimersXXXVIIbra{
\tikz[baseline=-0.5ex]{ \fill (-1.875ex,-0.649519ex) circle (1pt);
\fill (-1.125ex,0.649519ex) circle (1pt); \fill (-0.375ex,-0.649519ex)
circle (1pt); \fill (0.375ex,0.649519ex) circle (1pt); \fill
(1.125ex,-0.649519ex) circle (1pt); \fill (1.875ex,0.649519ex) circle
(1pt); \draw[ultra thick]
(-1.875ex,-0.649519ex)--(-1.125ex,0.649519ex); \draw[ultra thick]
(-0.375ex,-0.649519ex)--(1.125ex,-0.649519ex); \draw[ultra thick]
(0.375ex,0.649519ex)--(1.875ex,0.649519ex); \draw[thin]
(-1.875ex,-0.649519ex)--(-1.125ex,0.649519ex); \draw[thin]
(-0.375ex,-0.649519ex)--(0.375ex,0.649519ex); \draw[thin]
(1.125ex,-0.649519ex)--(1.875ex,0.649519ex); } }
\def\dimersXXXVIIIbra{ \tikz[baseline=-0.5ex]{ \fill
(-1.875ex,-0.649519ex) circle (1pt); \fill (-1.125ex,0.649519ex)
circle (1pt); \fill (-0.375ex,-0.649519ex) circle (1pt); \fill
(0.375ex,0.649519ex) circle (1pt); \fill (1.125ex,-0.649519ex) circle
(1pt); \fill (1.875ex,0.649519ex) circle (1pt); \draw[ultra thick]
(-1.875ex,-0.649519ex)--(-0.375ex,-0.649519ex); \draw[ultra thick]
(-1.125ex,0.649519ex)--(0.375ex,0.649519ex); \draw[ultra thick]
(1.125ex,-0.649519ex)--(1.875ex,0.649519ex); \draw[thin]
(-1.875ex,-0.649519ex)--(-1.125ex,0.649519ex); \draw[thin]
(-0.375ex,-0.649519ex)--(0.375ex,0.649519ex); \draw[thin]
(1.125ex,-0.649519ex)--(1.875ex,0.649519ex); } } \def\dimersXXXIXbra{
\tikz[baseline=-0.5ex]{ \fill (-1.875ex,-0.649519ex) circle (1pt);
\fill (-1.125ex,0.649519ex) circle (1pt); \fill (-0.375ex,-0.649519ex)
circle (1pt); \fill (1.125ex,-0.649519ex) circle (1pt); \fill
(0.375ex,0.649519ex) circle (1pt); \fill (1.875ex,0.649519ex) circle
(1pt); \draw[ultra thick]
(-1.875ex,-0.649519ex)--(-1.125ex,0.649519ex); \draw[ultra thick]
(-0.375ex,-0.649519ex)--(0.375ex,0.649519ex); \draw[ultra thick]
(1.125ex,-0.649519ex)--(1.875ex,0.649519ex); \draw[thin]
(-1.875ex,-0.649519ex)--(-1.125ex,0.649519ex); \draw[thin]
(-0.375ex,-0.649519ex)--(1.125ex,-0.649519ex); \draw[thin]
(0.375ex,0.649519ex)--(1.875ex,0.649519ex); } } \def\dimersXLbra{
\tikz[baseline=-0.5ex]{ \fill (-1.875ex,-0.649519ex) circle (1pt);
\fill (-0.375ex,-0.649519ex) circle (1pt); \fill (-1.125ex,0.649519ex)
circle (1pt); \fill (0.375ex,0.649519ex) circle (1pt); \fill
(1.125ex,-0.649519ex) circle (1pt); \fill (1.875ex,0.649519ex) circle
(1pt); \draw[ultra thick]
(-1.875ex,-0.649519ex)--(-1.125ex,0.649519ex); \draw[ultra thick]
(-0.375ex,-0.649519ex)--(0.375ex,0.649519ex); \draw[ultra thick]
(1.125ex,-0.649519ex)--(1.875ex,0.649519ex); \draw[thin]
(-1.875ex,-0.649519ex)--(-0.375ex,-0.649519ex); \draw[thin]
(-1.125ex,0.649519ex)--(0.375ex,0.649519ex); \draw[thin]
(1.125ex,-0.649519ex)--(1.875ex,0.649519ex); } } \def\dimersXLIbra{
\tikz[baseline=-0.5ex]{ \fill (-1.ex,-1.29904ex) circle (1pt); \fill
(-0.25ex,0.ex) circle (1pt); \fill (-1.ex,1.29904ex) circle (1pt);
\fill (0.5ex,1.29904ex) circle (1pt); \fill (0.5ex,-1.29904ex) circle
(1pt); \fill (1.25ex,0.ex) circle (1pt); \draw[ultra thick]
(-1.ex,-1.29904ex)--(0.5ex,-1.29904ex); \draw[ultra thick]
(-1.ex,1.29904ex)--(0.5ex,1.29904ex); \draw[ultra thick]
(-0.25ex,0.ex)--(1.25ex,0.ex); \draw[thin]
(-1.ex,-1.29904ex)--(-0.25ex,0.ex); \draw[thin]
(-1.ex,1.29904ex)--(0.5ex,1.29904ex); \draw[thin]
(0.5ex,-1.29904ex)--(1.25ex,0.ex); } } \def\dimersXLIIbra{
\tikz[baseline=-0.5ex]{ \fill (-1.ex,-1.29904ex) circle (1pt); \fill
(0.5ex,-1.29904ex) circle (1pt); \fill (-1.ex,1.29904ex) circle (1pt);
\fill (-0.25ex,0.ex) circle (1pt); \fill (0.5ex,1.29904ex) circle
(1pt); \fill (1.25ex,0.ex) circle (1pt); \draw[ultra thick]
(-1.ex,-1.29904ex)--(0.5ex,-1.29904ex); \draw[ultra thick]
(-1.ex,1.29904ex)--(0.5ex,1.29904ex); \draw[ultra thick]
(-0.25ex,0.ex)--(1.25ex,0.ex); \draw[thin]
(-1.ex,-1.29904ex)--(0.5ex,-1.29904ex); \draw[thin]
(-1.ex,1.29904ex)--(-0.25ex,0.ex); \draw[thin]
(0.5ex,1.29904ex)--(1.25ex,0.ex); } } \def\dimersXLIIIbra{
\tikz[baseline=-0.5ex]{ \fill (-1.ex,-1.29904ex) circle (1pt); \fill
(0.5ex,-1.29904ex) circle (1pt); \fill (-1.ex,1.29904ex) circle (1pt);
\fill (0.5ex,1.29904ex) circle (1pt); \fill (-0.25ex,0.ex) circle
(1pt); \fill (1.25ex,0.ex) circle (1pt); \draw[ultra thick]
(-1.ex,-1.29904ex)--(-0.25ex,0.ex); \draw[ultra thick]
(-1.ex,1.29904ex)--(0.5ex,1.29904ex); \draw[ultra thick]
(0.5ex,-1.29904ex)--(1.25ex,0.ex); \draw[thin]
(-1.ex,-1.29904ex)--(0.5ex,-1.29904ex); \draw[thin]
(-1.ex,1.29904ex)--(0.5ex,1.29904ex); \draw[thin]
(-0.25ex,0.ex)--(1.25ex,0.ex); } } \def\dimersXLIVbra{
\tikz[baseline=-0.5ex]{ \fill (-1.ex,-1.29904ex) circle (1pt); \fill
(0.5ex,-1.29904ex) circle (1pt); \fill (-1.ex,1.29904ex) circle (1pt);
\fill (0.5ex,1.29904ex) circle (1pt); \fill (-0.25ex,0.ex) circle
(1pt); \fill (1.25ex,0.ex) circle (1pt); \draw[ultra thick]
(-1.ex,-1.29904ex)--(0.5ex,-1.29904ex); \draw[ultra thick]
(-1.ex,1.29904ex)--(-0.25ex,0.ex); \draw[ultra thick]
(0.5ex,1.29904ex)--(1.25ex,0.ex); \draw[thin]
(-1.ex,-1.29904ex)--(0.5ex,-1.29904ex); \draw[thin]
(-1.ex,1.29904ex)--(0.5ex,1.29904ex); \draw[thin]
(-0.25ex,0.ex)--(1.25ex,0.ex); } } \def\dimersXLVbra{
\tikz[baseline=-0.5ex]{ \fill (-1.5ex,0.866025ex) circle (1pt); \fill
(-0.75ex,-0.433013ex) circle (1pt); \fill (0.ex,0.866025ex) circle
(1pt); \fill (1.5ex,0.866025ex) circle (1pt); \fill (0.ex,-1.73205ex)
circle (1pt); \fill (0.75ex,-0.433013ex) circle (1pt); \draw[ultra
thick] (-1.5ex,0.866025ex)--(0.ex,0.866025ex); \draw[ultra thick]
(-0.75ex,-0.433013ex)--(0.ex,-1.73205ex); \draw[ultra thick]
(0.75ex,-0.433013ex)--(1.5ex,0.866025ex); \draw[thin]
(-1.5ex,0.866025ex)--(-0.75ex,-0.433013ex); \draw[thin]
(0.ex,0.866025ex)--(1.5ex,0.866025ex); \draw[thin]
(0.ex,-1.73205ex)--(0.75ex,-0.433013ex); } } \def\dimersXLVIbra{
\tikz[baseline=-0.5ex]{ \fill (-1.5ex,0.866025ex) circle (1pt); \fill
(0.ex,0.866025ex) circle (1pt); \fill (-0.75ex,-0.433013ex) circle
(1pt); \fill (0.ex,-1.73205ex) circle (1pt); \fill
(0.75ex,-0.433013ex) circle (1pt); \fill (1.5ex,0.866025ex) circle
(1pt); \draw[ultra thick] (-1.5ex,0.866025ex)--(-0.75ex,-0.433013ex);
\draw[ultra thick] (0.ex,0.866025ex)--(1.5ex,0.866025ex); \draw[ultra
thick] (0.ex,-1.73205ex)--(0.75ex,-0.433013ex); \draw[thin]
(-1.5ex,0.866025ex)--(0.ex,0.866025ex); \draw[thin]
(-0.75ex,-0.433013ex)--(0.ex,-1.73205ex); \draw[thin]
(0.75ex,-0.433013ex)--(1.5ex,0.866025ex); } } \def\dimersXLVIIbra{
\tikz[baseline=-0.5ex]{ \fill (-1.96875ex,-1.13666ex) circle (1pt);
\fill (-1.21875ex,0.16238ex) circle (1pt); \fill
(-0.46875ex,-1.13666ex) circle (1pt); \fill (1.03125ex,-1.13666ex)
circle (1pt); \fill (-0.46875ex,1.46142ex) circle (1pt); \fill
(0.28125ex,0.16238ex) circle (1pt); \fill (1.03125ex,1.46142ex) circle
(1pt); \fill (1.78125ex,0.16238ex) circle (1pt); \draw[ultra thick]
(-1.96875ex,-1.13666ex)--(-0.46875ex,-1.13666ex); \draw[ultra thick]
(-1.21875ex,0.16238ex)--(0.28125ex,0.16238ex); \draw[ultra thick]
(-0.46875ex,1.46142ex)--(1.03125ex,1.46142ex); \draw[ultra thick]
(1.03125ex,-1.13666ex)--(1.78125ex,0.16238ex); \draw[thin]
(-1.96875ex,-1.13666ex)--(-1.21875ex,0.16238ex); \draw[thin]
(-0.46875ex,-1.13666ex)--(1.03125ex,-1.13666ex); \draw[thin]
(-0.46875ex,1.46142ex)--(0.28125ex,0.16238ex); \draw[thin]
(1.03125ex,1.46142ex)--(1.78125ex,0.16238ex); } }
\def\dimersXLVIIIbra{ \tikz[baseline=-0.5ex]{ \fill
(-1.96875ex,-1.13666ex) circle (1pt); \fill (-1.21875ex,0.16238ex)
circle (1pt); \fill (-0.46875ex,-1.13666ex) circle (1pt); \fill
(1.03125ex,-1.13666ex) circle (1pt); \fill (-0.46875ex,1.46142ex)
circle (1pt); \fill (1.03125ex,1.46142ex) circle (1pt); \fill
(0.28125ex,0.16238ex) circle (1pt); \fill (1.78125ex,0.16238ex) circle
(1pt); \draw[ultra thick]
(-1.96875ex,-1.13666ex)--(-0.46875ex,-1.13666ex); \draw[ultra thick]
(-1.21875ex,0.16238ex)--(-0.46875ex,1.46142ex); \draw[ultra thick]
(0.28125ex,0.16238ex)--(1.03125ex,1.46142ex); \draw[ultra thick]
(1.03125ex,-1.13666ex)--(1.78125ex,0.16238ex); \draw[thin]
(-1.96875ex,-1.13666ex)--(-1.21875ex,0.16238ex); \draw[thin]
(-0.46875ex,-1.13666ex)--(1.03125ex,-1.13666ex); \draw[thin]
(-0.46875ex,1.46142ex)--(1.03125ex,1.46142ex); \draw[thin]
(0.28125ex,0.16238ex)--(1.78125ex,0.16238ex); } } \def\dimersXLIXbra{
\tikz[baseline=-0.5ex]{ \fill (-1.96875ex,-1.13666ex) circle (1pt);
\fill (-0.46875ex,-1.13666ex) circle (1pt); \fill
(-1.21875ex,0.16238ex) circle (1pt); \fill (-0.46875ex,1.46142ex)
circle (1pt); \fill (0.28125ex,0.16238ex) circle (1pt); \fill
(1.03125ex,1.46142ex) circle (1pt); \fill (1.03125ex,-1.13666ex)
circle (1pt); \fill (1.78125ex,0.16238ex) circle (1pt); \draw[ultra
thick] (-1.96875ex,-1.13666ex)--(-1.21875ex,0.16238ex); \draw[ultra
thick] (-0.46875ex,-1.13666ex)--(1.03125ex,-1.13666ex); \draw[ultra
thick] (-0.46875ex,1.46142ex)--(1.03125ex,1.46142ex); \draw[ultra
thick] (0.28125ex,0.16238ex)--(1.78125ex,0.16238ex); \draw[thin]
(-1.96875ex,-1.13666ex)--(-0.46875ex,-1.13666ex); \draw[thin]
(-1.21875ex,0.16238ex)--(-0.46875ex,1.46142ex); \draw[thin]
(0.28125ex,0.16238ex)--(1.03125ex,1.46142ex); \draw[thin]
(1.03125ex,-1.13666ex)--(1.78125ex,0.16238ex); } } \def\dimersLbra{
\tikz[baseline=-0.5ex]{ \fill (-1.96875ex,-1.13666ex) circle (1pt);
\fill (-0.46875ex,-1.13666ex) circle (1pt); \fill
(-1.21875ex,0.16238ex) circle (1pt); \fill (0.28125ex,0.16238ex)
circle (1pt); \fill (-0.46875ex,1.46142ex) circle (1pt); \fill
(1.03125ex,1.46142ex) circle (1pt); \fill (1.03125ex,-1.13666ex)
circle (1pt); \fill (1.78125ex,0.16238ex) circle (1pt); \draw[ultra
thick] (-1.96875ex,-1.13666ex)--(-1.21875ex,0.16238ex); \draw[ultra
thick] (-0.46875ex,-1.13666ex)--(1.03125ex,-1.13666ex); \draw[ultra
thick] (-0.46875ex,1.46142ex)--(0.28125ex,0.16238ex); \draw[ultra
thick] (1.03125ex,1.46142ex)--(1.78125ex,0.16238ex); \draw[thin]
(-1.96875ex,-1.13666ex)--(-0.46875ex,-1.13666ex); \draw[thin]
(-1.21875ex,0.16238ex)--(0.28125ex,0.16238ex); \draw[thin]
(-0.46875ex,1.46142ex)--(1.03125ex,1.46142ex); \draw[thin]
(1.03125ex,-1.13666ex)--(1.78125ex,0.16238ex); } } \def\dimersLIbra{
\tikz[baseline=-0.5ex]{ \fill (-2.25ex,0.ex) circle (1pt); \fill
(-1.5ex,-1.29904ex) circle (1pt); \fill (-1.5ex,1.29904ex) circle
(1pt); \fill (0.ex,1.29904ex) circle (1pt); \fill (0.ex,-1.29904ex)
circle (1pt); \fill (1.5ex,-1.29904ex) circle (1pt); \fill
(1.5ex,1.29904ex) circle (1pt); \fill (2.25ex,0.ex) circle (1pt);
\draw[ultra thick] (-2.25ex,0.ex)--(-1.5ex,1.29904ex); \draw[ultra
thick] (-1.5ex,-1.29904ex)--(0.ex,-1.29904ex); \draw[ultra thick]
(0.ex,1.29904ex)--(1.5ex,1.29904ex); \draw[ultra thick]
(1.5ex,-1.29904ex)--(2.25ex,0.ex); \draw[thin]
(-2.25ex,0.ex)--(-1.5ex,-1.29904ex); \draw[thin]
(-1.5ex,1.29904ex)--(0.ex,1.29904ex); \draw[thin]
(0.ex,-1.29904ex)--(1.5ex,-1.29904ex); \draw[thin]
(1.5ex,1.29904ex)--(2.25ex,0.ex); } } \def\dimersLIIbra{
\tikz[baseline=-0.5ex]{ \fill (-2.25ex,0.ex) circle (1pt); \fill
(-1.5ex,1.29904ex) circle (1pt); \fill (-1.5ex,-1.29904ex) circle
(1pt); \fill (0.ex,-1.29904ex) circle (1pt); \fill (0.ex,1.29904ex)
circle (1pt); \fill (1.5ex,1.29904ex) circle (1pt); \fill
(1.5ex,-1.29904ex) circle (1pt); \fill (2.25ex,0.ex) circle (1pt);
\draw[ultra thick] (-2.25ex,0.ex)--(-1.5ex,-1.29904ex); \draw[ultra
thick] (-1.5ex,1.29904ex)--(0.ex,1.29904ex); \draw[ultra thick]
(0.ex,-1.29904ex)--(1.5ex,-1.29904ex); \draw[ultra thick]
(1.5ex,1.29904ex)--(2.25ex,0.ex); \draw[thin]
(-2.25ex,0.ex)--(-1.5ex,1.29904ex); \draw[thin]
(-1.5ex,-1.29904ex)--(0.ex,-1.29904ex); \draw[thin]
(0.ex,1.29904ex)--(1.5ex,1.29904ex); \draw[thin]
(1.5ex,-1.29904ex)--(2.25ex,0.ex); } } \def\dimersIket{
\tikz[baseline=-0.5ex]{ \fill (-1.125ex,-0.649519ex) circle (1pt);
\fill (0.375ex,-0.649519ex) circle (1pt); \fill (-0.375ex,0.649519ex)
circle (1pt); \fill (1.125ex,0.649519ex) circle (1pt); \draw[ultra
thick] (-1.125ex,-0.649519ex)--(-0.375ex,0.649519ex); \draw[ultra
thick] (0.375ex,-0.649519ex)--(1.125ex,0.649519ex); \draw[thin]
(-1.125ex,-0.649519ex)--(0.375ex,-0.649519ex); \draw[thin]
(-0.375ex,0.649519ex)--(1.125ex,0.649519ex); } } \def\dimersIIket{
\tikz[baseline=-0.5ex]{ \fill (-1.125ex,-0.649519ex) circle (1pt);
\fill (-0.375ex,0.649519ex) circle (1pt); \fill (0.375ex,-0.649519ex)
circle (1pt); \fill (1.125ex,0.649519ex) circle (1pt); \draw[ultra
thick] (-1.125ex,-0.649519ex)--(0.375ex,-0.649519ex); \draw[ultra
thick] (-0.375ex,0.649519ex)--(1.125ex,0.649519ex); \draw[thin]
(-1.125ex,-0.649519ex)--(-0.375ex,0.649519ex); \draw[thin]
(0.375ex,-0.649519ex)--(1.125ex,0.649519ex); } } \def\dimersIIIket{
\tikz[baseline=-0.5ex]{ \fill (-1.875ex,-0.649519ex) circle (1pt);
\fill (-0.375ex,-0.649519ex) circle (1pt); \fill (-1.125ex,0.649519ex)
circle (1pt); \fill (0.375ex,0.649519ex) circle (1pt); \fill
(1.125ex,-0.649519ex) circle (1pt); \fill (1.875ex,0.649519ex) circle
(1pt); \draw[ultra thick]
(-1.875ex,-0.649519ex)--(-1.125ex,0.649519ex); \draw[ultra thick]
(-0.375ex,-0.649519ex)--(1.125ex,-0.649519ex); \draw[ultra thick]
(0.375ex,0.649519ex)--(1.875ex,0.649519ex); \draw[thin]
(-1.875ex,-0.649519ex)--(-0.375ex,-0.649519ex); \draw[thin]
(-1.125ex,0.649519ex)--(0.375ex,0.649519ex); \draw[thin]
(1.125ex,-0.649519ex)--(1.875ex,0.649519ex); } } \def\dimersIVket{
\tikz[baseline=-0.5ex]{ \fill (-1.875ex,-0.649519ex) circle (1pt);
\fill (-1.125ex,0.649519ex) circle (1pt); \fill (-0.375ex,-0.649519ex)
circle (1pt); \fill (1.125ex,-0.649519ex) circle (1pt); \fill
(0.375ex,0.649519ex) circle (1pt); \fill (1.875ex,0.649519ex) circle
(1pt); \draw[ultra thick]
(-1.875ex,-0.649519ex)--(-0.375ex,-0.649519ex); \draw[ultra thick]
(-1.125ex,0.649519ex)--(0.375ex,0.649519ex); \draw[ultra thick]
(1.125ex,-0.649519ex)--(1.875ex,0.649519ex); \draw[thin]
(-1.875ex,-0.649519ex)--(-1.125ex,0.649519ex); \draw[thin]
(-0.375ex,-0.649519ex)--(1.125ex,-0.649519ex); \draw[thin]
(0.375ex,0.649519ex)--(1.875ex,0.649519ex); } } \def\dimersVket{
\tikz[baseline=-0.5ex]{ \fill (-1.ex,-1.29904ex) circle (1pt); \fill
(0.5ex,-1.29904ex) circle (1pt); \fill (-1.ex,1.29904ex) circle (1pt);
\fill (-0.25ex,0.ex) circle (1pt); \fill (0.5ex,1.29904ex) circle
(1pt); \fill (1.25ex,0.ex) circle (1pt); \draw[ultra thick]
(-1.ex,-1.29904ex)--(-0.25ex,0.ex); \draw[ultra thick]
(-1.ex,1.29904ex)--(0.5ex,1.29904ex); \draw[ultra thick]
(0.5ex,-1.29904ex)--(1.25ex,0.ex); \draw[thin]
(-1.ex,-1.29904ex)--(0.5ex,-1.29904ex); \draw[thin]
(-1.ex,1.29904ex)--(-0.25ex,0.ex); \draw[thin]
(0.5ex,1.29904ex)--(1.25ex,0.ex); } } \def\dimersVIket{
\tikz[baseline=-0.5ex]{ \fill (-1.ex,-1.29904ex) circle (1pt); \fill
(-0.25ex,0.ex) circle (1pt); \fill (-1.ex,1.29904ex) circle (1pt);
\fill (0.5ex,1.29904ex) circle (1pt); \fill (0.5ex,-1.29904ex) circle
(1pt); \fill (1.25ex,0.ex) circle (1pt); \draw[ultra thick]
(-1.ex,-1.29904ex)--(0.5ex,-1.29904ex); \draw[ultra thick]
(-1.ex,1.29904ex)--(-0.25ex,0.ex); \draw[ultra thick]
(0.5ex,1.29904ex)--(1.25ex,0.ex); \draw[thin]
(-1.ex,-1.29904ex)--(-0.25ex,0.ex); \draw[thin]
(-1.ex,1.29904ex)--(0.5ex,1.29904ex); \draw[thin]
(0.5ex,-1.29904ex)--(1.25ex,0.ex); } } \def\dimersVIIket{
\tikz[baseline=-0.5ex]{ \fill (-1.125ex,-0.649519ex) circle (1pt);
\fill (-0.375ex,0.649519ex) circle (1pt); \fill (0.375ex,-0.649519ex)
circle (1pt); \fill (1.125ex,0.649519ex) circle (1pt); \draw[ultra
thick] (-1.125ex,-0.649519ex)--(-0.375ex,0.649519ex); \draw[ultra
thick] (0.375ex,-0.649519ex)--(1.125ex,0.649519ex); \draw[thin]
(-1.125ex,-0.649519ex)--(-0.375ex,0.649519ex); \draw[thin]
(0.375ex,-0.649519ex)--(1.125ex,0.649519ex); } } \def\dimersVIIIket{
\tikz[baseline=-0.5ex]{ \fill (-1.125ex,-0.649519ex) circle (1pt);
\fill (0.375ex,-0.649519ex) circle (1pt); \fill (-0.375ex,0.649519ex)
circle (1pt); \fill (1.125ex,0.649519ex) circle (1pt); \draw[ultra
thick] (-1.125ex,-0.649519ex)--(0.375ex,-0.649519ex); \draw[ultra
thick] (-0.375ex,0.649519ex)--(1.125ex,0.649519ex); \draw[thin]
(-1.125ex,-0.649519ex)--(0.375ex,-0.649519ex); \draw[thin]
(-0.375ex,0.649519ex)--(1.125ex,0.649519ex); } }   \def\dimersXIket{
\tikz[baseline=-0.5ex]{ \fill (-1.96875ex,-1.13666ex) circle (1pt);
\fill (-0.46875ex,-1.13666ex) circle (1pt); \fill
(-1.21875ex,0.16238ex) circle (1pt); \fill (-0.46875ex,1.46142ex)
circle (1pt); \fill (0.28125ex,0.16238ex) circle (1pt); \fill
(1.03125ex,1.46142ex) circle (1pt); \fill (1.03125ex,-1.13666ex)
circle (1pt); \fill (1.78125ex,0.16238ex) circle (1pt); \draw[ultra
thick] (-1.96875ex,-1.13666ex)--(-1.21875ex,0.16238ex); \draw[ultra
thick] (-0.46875ex,-1.13666ex)--(1.03125ex,-1.13666ex); \draw[ultra
thick] (-0.46875ex,1.46142ex)--(0.28125ex,0.16238ex); \draw[ultra
thick] (1.03125ex,1.46142ex)--(1.78125ex,0.16238ex); \draw[thin]
(-1.96875ex,-1.13666ex)--(-0.46875ex,-1.13666ex); \draw[thin]
(-1.21875ex,0.16238ex)--(-0.46875ex,1.46142ex); \draw[thin]
(0.28125ex,0.16238ex)--(1.03125ex,1.46142ex); \draw[thin]
(1.03125ex,-1.13666ex)--(1.78125ex,0.16238ex); } } \def\dimersXIIket{
\tikz[baseline=-0.5ex]{ \fill (-1.96875ex,-1.13666ex) circle (1pt);
\fill (-1.21875ex,0.16238ex) circle (1pt); \fill
(-0.46875ex,-1.13666ex) circle (1pt); \fill (1.03125ex,-1.13666ex)
circle (1pt); \fill (-0.46875ex,1.46142ex) circle (1pt); \fill
(0.28125ex,0.16238ex) circle (1pt); \fill (1.03125ex,1.46142ex) circle
(1pt); \fill (1.78125ex,0.16238ex) circle (1pt); \draw[ultra thick]
(-1.96875ex,-1.13666ex)--(-0.46875ex,-1.13666ex); \draw[ultra thick]
(-1.21875ex,0.16238ex)--(-0.46875ex,1.46142ex); \draw[ultra thick]
(0.28125ex,0.16238ex)--(1.03125ex,1.46142ex); \draw[ultra thick]
(1.03125ex,-1.13666ex)--(1.78125ex,0.16238ex); \draw[thin]
(-1.96875ex,-1.13666ex)--(-1.21875ex,0.16238ex); \draw[thin]
(-0.46875ex,-1.13666ex)--(1.03125ex,-1.13666ex); \draw[thin]
(-0.46875ex,1.46142ex)--(0.28125ex,0.16238ex); \draw[thin]
(1.03125ex,1.46142ex)--(1.78125ex,0.16238ex); } } \def\dimersXIIIket{
\tikz[baseline=-0.5ex]{ \fill (-1.5ex,1.29904ex) circle (1pt); \fill
(0.ex,1.29904ex) circle (1pt); \fill (-1.5ex,-1.29904ex) circle (1pt);
\fill (-0.75ex,0.ex) circle (1pt); \fill (0.ex,-1.29904ex) circle
(1pt); \fill (1.5ex,-1.29904ex) circle (1pt); \fill (0.75ex,0.ex)
circle (1pt); \fill (1.5ex,1.29904ex) circle (1pt); \draw[ultra thick]
(-1.5ex,1.29904ex)--(-0.75ex,0.ex); \draw[ultra thick]
(-1.5ex,-1.29904ex)--(0.ex,-1.29904ex); \draw[ultra thick]
(0.ex,1.29904ex)--(1.5ex,1.29904ex); \draw[ultra thick]
(0.75ex,0.ex)--(1.5ex,-1.29904ex); \draw[thin]
(-1.5ex,1.29904ex)--(0.ex,1.29904ex); \draw[thin]
(-1.5ex,-1.29904ex)--(-0.75ex,0.ex); \draw[thin]
(0.ex,-1.29904ex)--(1.5ex,-1.29904ex); \draw[thin]
(0.75ex,0.ex)--(1.5ex,1.29904ex); } } \def\dimersXIVket{
\tikz[baseline=-0.5ex]{ \fill (-1.5ex,1.29904ex) circle (1pt); \fill
(-0.75ex,0.ex) circle (1pt); \fill (-1.5ex,-1.29904ex) circle (1pt);
\fill (0.ex,-1.29904ex) circle (1pt); \fill (0.ex,1.29904ex) circle
(1pt); \fill (1.5ex,1.29904ex) circle (1pt); \fill (0.75ex,0.ex)
circle (1pt); \fill (1.5ex,-1.29904ex) circle (1pt); \draw[ultra
thick] (-1.5ex,1.29904ex)--(0.ex,1.29904ex); \draw[ultra thick]
(-1.5ex,-1.29904ex)--(-0.75ex,0.ex); \draw[ultra thick]
(0.ex,-1.29904ex)--(1.5ex,-1.29904ex); \draw[ultra thick]
(0.75ex,0.ex)--(1.5ex,1.29904ex); \draw[thin]
(-1.5ex,1.29904ex)--(-0.75ex,0.ex); \draw[thin]
(-1.5ex,-1.29904ex)--(0.ex,-1.29904ex); \draw[thin]
(0.ex,1.29904ex)--(1.5ex,1.29904ex); \draw[thin]
(0.75ex,0.ex)--(1.5ex,-1.29904ex); } } \def\dimersXVket{
\tikz[baseline=-0.5ex]{ \fill (-2.34375ex,0.811899ex) circle (1pt);
\fill (-0.84375ex,0.811899ex) circle (1pt); \fill
(-1.59375ex,-0.487139ex) circle (1pt); \fill (-0.09375ex,-0.487139ex)
circle (1pt); \fill (0.65625ex,0.811899ex) circle (1pt); \fill
(2.15625ex,0.811899ex) circle (1pt); \fill (0.65625ex,-1.78618ex)
circle (1pt); \fill (1.40625ex,-0.487139ex) circle (1pt); \draw[ultra
thick] (-2.34375ex,0.811899ex)--(-1.59375ex,-0.487139ex); \draw[ultra
thick] (-0.84375ex,0.811899ex)--(0.65625ex,0.811899ex); \draw[ultra
thick] (-0.09375ex,-0.487139ex)--(0.65625ex,-1.78618ex); \draw[ultra
thick] (1.40625ex,-0.487139ex)--(2.15625ex,0.811899ex); \draw[thin]
(-2.34375ex,0.811899ex)--(-0.84375ex,0.811899ex); \draw[thin]
(-1.59375ex,-0.487139ex)--(-0.09375ex,-0.487139ex); \draw[thin]
(0.65625ex,0.811899ex)--(2.15625ex,0.811899ex); \draw[thin]
(0.65625ex,-1.78618ex)--(1.40625ex,-0.487139ex); } }
\def\dimersXVIket{ \tikz[baseline=-0.5ex]{ \fill
(-2.34375ex,0.811899ex) circle (1pt); \fill (-1.59375ex,-0.487139ex)
circle (1pt); \fill (-0.84375ex,0.811899ex) circle (1pt); \fill
(0.65625ex,0.811899ex) circle (1pt); \fill (-0.09375ex,-0.487139ex)
circle (1pt); \fill (0.65625ex,-1.78618ex) circle (1pt); \fill
(1.40625ex,-0.487139ex) circle (1pt); \fill (2.15625ex,0.811899ex)
circle (1pt); \draw[ultra thick]
(-2.34375ex,0.811899ex)--(-0.84375ex,0.811899ex); \draw[ultra thick]
(-1.59375ex,-0.487139ex)--(-0.09375ex,-0.487139ex); \draw[ultra thick]
(0.65625ex,0.811899ex)--(2.15625ex,0.811899ex); \draw[ultra thick]
(0.65625ex,-1.78618ex)--(1.40625ex,-0.487139ex); \draw[thin]
(-2.34375ex,0.811899ex)--(-1.59375ex,-0.487139ex); \draw[thin]
(-0.84375ex,0.811899ex)--(0.65625ex,0.811899ex); \draw[thin]
(-0.09375ex,-0.487139ex)--(0.65625ex,-1.78618ex); \draw[thin]
(1.40625ex,-0.487139ex)--(2.15625ex,0.811899ex); } }

       \def\dimersXXVket{
\tikz[baseline=-0.5ex]{ \fill (-2.34375ex,-1.13666ex) circle (1pt);
\fill (-0.84375ex,-1.13666ex) circle (1pt); \fill
(-1.59375ex,0.16238ex) circle (1pt); \fill (-0.09375ex,0.16238ex)
circle (1pt); \fill (0.65625ex,-1.13666ex) circle (1pt); \fill
(1.40625ex,0.16238ex) circle (1pt); \fill (0.65625ex,1.46142ex) circle
(1pt); \fill (2.15625ex,1.46142ex) circle (1pt); \draw[ultra thick]
(-2.34375ex,-1.13666ex)--(-1.59375ex,0.16238ex); \draw[ultra thick]
(-0.84375ex,-1.13666ex)--(0.65625ex,-1.13666ex); \draw[ultra thick]
(-0.09375ex,0.16238ex)--(0.65625ex,1.46142ex); \draw[ultra thick]
(1.40625ex,0.16238ex)--(2.15625ex,1.46142ex); \draw[thin]
(-2.34375ex,-1.13666ex)--(-0.84375ex,-1.13666ex); \draw[thin]
(-1.59375ex,0.16238ex)--(-0.09375ex,0.16238ex); \draw[thin]
(0.65625ex,-1.13666ex)--(1.40625ex,0.16238ex); \draw[thin]
(0.65625ex,1.46142ex)--(2.15625ex,1.46142ex); } } \def\dimersXXVIket{
\tikz[baseline=-0.5ex]{ \fill (-2.34375ex,-1.13666ex) circle (1pt);
\fill (-1.59375ex,0.16238ex) circle (1pt); \fill
(-0.84375ex,-1.13666ex) circle (1pt); \fill (0.65625ex,-1.13666ex)
circle (1pt); \fill (-0.09375ex,0.16238ex) circle (1pt); \fill
(0.65625ex,1.46142ex) circle (1pt); \fill (1.40625ex,0.16238ex) circle
(1pt); \fill (2.15625ex,1.46142ex) circle (1pt); \draw[ultra thick]
(-2.34375ex,-1.13666ex)--(-0.84375ex,-1.13666ex); \draw[ultra thick]
(-1.59375ex,0.16238ex)--(-0.09375ex,0.16238ex); \draw[ultra thick]
(0.65625ex,-1.13666ex)--(1.40625ex,0.16238ex); \draw[ultra thick]
(0.65625ex,1.46142ex)--(2.15625ex,1.46142ex); \draw[thin]
(-2.34375ex,-1.13666ex)--(-1.59375ex,0.16238ex); \draw[thin]
(-0.84375ex,-1.13666ex)--(0.65625ex,-1.13666ex); \draw[thin]
(-0.09375ex,0.16238ex)--(0.65625ex,1.46142ex); \draw[thin]
(1.40625ex,0.16238ex)--(2.15625ex,1.46142ex); } } \def\dimersXXVIIket{
\tikz[baseline=-0.5ex]{ \fill (-2.25ex,0.32476ex) circle (1pt); \fill
(-0.75ex,0.32476ex) circle (1pt); \fill (-1.5ex,-0.974279ex) circle
(1pt); \fill (0.ex,-0.974279ex) circle (1pt); \fill (0.ex,1.6238ex)
circle (1pt); \fill (0.75ex,0.32476ex) circle (1pt); \fill
(1.5ex,-0.974279ex) circle (1pt); \fill (2.25ex,0.32476ex) circle
(1pt); \draw[ultra thick] (-2.25ex,0.32476ex)--(-1.5ex,-0.974279ex);
\draw[ultra thick] (-0.75ex,0.32476ex)--(0.ex,1.6238ex); \draw[ultra
thick] (0.ex,-0.974279ex)--(1.5ex,-0.974279ex); \draw[ultra thick]
(0.75ex,0.32476ex)--(2.25ex,0.32476ex); \draw[thin]
(-2.25ex,0.32476ex)--(-0.75ex,0.32476ex); \draw[thin]
(-1.5ex,-0.974279ex)--(0.ex,-0.974279ex); \draw[thin]
(0.ex,1.6238ex)--(0.75ex,0.32476ex); \draw[thin]
(1.5ex,-0.974279ex)--(2.25ex,0.32476ex); } } \def\dimersXXVIIIket{
\tikz[baseline=-0.5ex]{ \fill (-2.25ex,0.32476ex) circle (1pt); \fill
(-1.5ex,-0.974279ex) circle (1pt); \fill (-0.75ex,0.32476ex) circle
(1pt); \fill (0.ex,1.6238ex) circle (1pt); \fill (0.ex,-0.974279ex)
circle (1pt); \fill (1.5ex,-0.974279ex) circle (1pt); \fill
(0.75ex,0.32476ex) circle (1pt); \fill (2.25ex,0.32476ex) circle
(1pt); \draw[ultra thick] (-2.25ex,0.32476ex)--(-0.75ex,0.32476ex);
\draw[ultra thick] (-1.5ex,-0.974279ex)--(0.ex,-0.974279ex);
\draw[ultra thick] (0.ex,1.6238ex)--(0.75ex,0.32476ex); \draw[ultra
thick] (1.5ex,-0.974279ex)--(2.25ex,0.32476ex); \draw[thin]
(-2.25ex,0.32476ex)--(-1.5ex,-0.974279ex); \draw[thin]
(-0.75ex,0.32476ex)--(0.ex,1.6238ex); \draw[thin]
(0.ex,-0.974279ex)--(1.5ex,-0.974279ex); \draw[thin]
(0.75ex,0.32476ex)--(2.25ex,0.32476ex); } } \def\dimersXXIXket{
\tikz[baseline=-0.5ex]{ \fill (-1.59375ex,-1.13666ex) circle (1pt);
\fill (-0.09375ex,-1.13666ex) circle (1pt); \fill
(-1.59375ex,1.46142ex) circle (1pt); \fill (-0.84375ex,0.16238ex)
circle (1pt); \fill (-0.09375ex,1.46142ex) circle (1pt); \fill
(0.65625ex,0.16238ex) circle (1pt); \fill (1.40625ex,-1.13666ex)
circle (1pt); \fill (2.15625ex,0.16238ex) circle (1pt); \draw[ultra
thick] (-1.59375ex,-1.13666ex)--(-0.84375ex,0.16238ex); \draw[ultra
thick] (-1.59375ex,1.46142ex)--(-0.09375ex,1.46142ex); \draw[ultra
thick] (-0.09375ex,-1.13666ex)--(1.40625ex,-1.13666ex); \draw[ultra
thick] (0.65625ex,0.16238ex)--(2.15625ex,0.16238ex); \draw[thin]
(-1.59375ex,-1.13666ex)--(-0.09375ex,-1.13666ex); \draw[thin]
(-1.59375ex,1.46142ex)--(-0.84375ex,0.16238ex); \draw[thin]
(-0.09375ex,1.46142ex)--(0.65625ex,0.16238ex); \draw[thin]
(1.40625ex,-1.13666ex)--(2.15625ex,0.16238ex); } } \def\dimersXXXket{
\tikz[baseline=-0.5ex]{ \fill (-1.59375ex,-1.13666ex) circle (1pt);
\fill (-0.84375ex,0.16238ex) circle (1pt); \fill
(-1.59375ex,1.46142ex) circle (1pt); \fill (-0.09375ex,1.46142ex)
circle (1pt); \fill (-0.09375ex,-1.13666ex) circle (1pt); \fill
(1.40625ex,-1.13666ex) circle (1pt); \fill (0.65625ex,0.16238ex)
circle (1pt); \fill (2.15625ex,0.16238ex) circle (1pt); \draw[ultra
thick] (-1.59375ex,-1.13666ex)--(-0.09375ex,-1.13666ex); \draw[ultra
thick] (-1.59375ex,1.46142ex)--(-0.84375ex,0.16238ex); \draw[ultra
thick] (-0.09375ex,1.46142ex)--(0.65625ex,0.16238ex); \draw[ultra
thick] (1.40625ex,-1.13666ex)--(2.15625ex,0.16238ex); \draw[thin]
(-1.59375ex,-1.13666ex)--(-0.84375ex,0.16238ex); \draw[thin]
(-1.59375ex,1.46142ex)--(-0.09375ex,1.46142ex); \draw[thin]
(-0.09375ex,-1.13666ex)--(1.40625ex,-1.13666ex); \draw[thin]
(0.65625ex,0.16238ex)--(2.15625ex,0.16238ex); } } \def\dimersXXXIket{
\tikz[baseline=-0.5ex]{ \fill (-2.625ex,-0.649519ex) circle (1pt);
\fill (-1.125ex,-0.649519ex) circle (1pt); \fill (-1.875ex,0.649519ex)
circle (1pt); \fill (-0.375ex,0.649519ex) circle (1pt); \fill
(0.375ex,-0.649519ex) circle (1pt); \fill (1.875ex,-0.649519ex) circle
(1pt); \fill (1.125ex,0.649519ex) circle (1pt); \fill
(2.625ex,0.649519ex) circle (1pt); \draw[ultra thick]
(-2.625ex,-0.649519ex)--(-1.875ex,0.649519ex); \draw[ultra thick]
(-1.125ex,-0.649519ex)--(0.375ex,-0.649519ex); \draw[ultra thick]
(-0.375ex,0.649519ex)--(1.125ex,0.649519ex); \draw[ultra thick]
(1.875ex,-0.649519ex)--(2.625ex,0.649519ex); \draw[thin]
(-2.625ex,-0.649519ex)--(-1.125ex,-0.649519ex); \draw[thin]
(-1.875ex,0.649519ex)--(-0.375ex,0.649519ex); \draw[thin]
(0.375ex,-0.649519ex)--(1.875ex,-0.649519ex); \draw[thin]
(1.125ex,0.649519ex)--(2.625ex,0.649519ex); } } \def\dimersXXXIIket{
\tikz[baseline=-0.5ex]{ \fill (-2.625ex,-0.649519ex) circle (1pt);
\fill (-1.875ex,0.649519ex) circle (1pt); \fill (-1.125ex,-0.649519ex)
circle (1pt); \fill (0.375ex,-0.649519ex) circle (1pt); \fill
(-0.375ex,0.649519ex) circle (1pt); \fill (1.125ex,0.649519ex) circle
(1pt); \fill (1.875ex,-0.649519ex) circle (1pt); \fill
(2.625ex,0.649519ex) circle (1pt); \draw[ultra thick]
(-2.625ex,-0.649519ex)--(-1.125ex,-0.649519ex); \draw[ultra thick]
(-1.875ex,0.649519ex)--(-0.375ex,0.649519ex); \draw[ultra thick]
(0.375ex,-0.649519ex)--(1.875ex,-0.649519ex); \draw[ultra thick]
(1.125ex,0.649519ex)--(2.625ex,0.649519ex); \draw[thin]
(-2.625ex,-0.649519ex)--(-1.875ex,0.649519ex); \draw[thin]
(-1.125ex,-0.649519ex)--(0.375ex,-0.649519ex); \draw[thin]
(-0.375ex,0.649519ex)--(1.125ex,0.649519ex); \draw[thin]
(1.875ex,-0.649519ex)--(2.625ex,0.649519ex); } } \def\dimersXXXIIIket{
\tikz[baseline=-0.5ex]{ \fill (-2.4375ex,0.32476ex) circle (1pt);
\fill (-0.9375ex,0.32476ex) circle (1pt); \fill
(-1.6875ex,-0.974279ex) circle (1pt); \fill (-0.1875ex,-0.974279ex)
circle (1pt); \fill (0.5625ex,0.32476ex) circle (1pt); \fill
(1.3125ex,1.6238ex) circle (1pt); \fill (1.3125ex,-0.974279ex) circle
(1pt); \fill (2.0625ex,0.32476ex) circle (1pt); \draw[ultra thick]
(-2.4375ex,0.32476ex)--(-1.6875ex,-0.974279ex); \draw[ultra thick]
(-0.9375ex,0.32476ex)--(0.5625ex,0.32476ex); \draw[ultra thick]
(-0.1875ex,-0.974279ex)--(1.3125ex,-0.974279ex); \draw[ultra thick]
(1.3125ex,1.6238ex)--(2.0625ex,0.32476ex); \draw[thin]
(-2.4375ex,0.32476ex)--(-0.9375ex,0.32476ex); \draw[thin]
(-1.6875ex,-0.974279ex)--(-0.1875ex,-0.974279ex); \draw[thin]
(0.5625ex,0.32476ex)--(1.3125ex,1.6238ex); \draw[thin]
(1.3125ex,-0.974279ex)--(2.0625ex,0.32476ex); } } \def\dimersXXXIVket{
\tikz[baseline=-0.5ex]{ \fill (-2.4375ex,0.32476ex) circle (1pt);
\fill (-1.6875ex,-0.974279ex) circle (1pt); \fill
(-0.9375ex,0.32476ex) circle (1pt); \fill (0.5625ex,0.32476ex) circle
(1pt); \fill (-0.1875ex,-0.974279ex) circle (1pt); \fill
(1.3125ex,-0.974279ex) circle (1pt); \fill (1.3125ex,1.6238ex) circle
(1pt); \fill (2.0625ex,0.32476ex) circle (1pt); \draw[ultra thick]
(-2.4375ex,0.32476ex)--(-0.9375ex,0.32476ex); \draw[ultra thick]
(-1.6875ex,-0.974279ex)--(-0.1875ex,-0.974279ex); \draw[ultra thick]
(0.5625ex,0.32476ex)--(1.3125ex,1.6238ex); \draw[ultra thick]
(1.3125ex,-0.974279ex)--(2.0625ex,0.32476ex); \draw[thin]
(-2.4375ex,0.32476ex)--(-1.6875ex,-0.974279ex); \draw[thin]
(-0.9375ex,0.32476ex)--(0.5625ex,0.32476ex); \draw[thin]
(-0.1875ex,-0.974279ex)--(1.3125ex,-0.974279ex); \draw[thin]
(1.3125ex,1.6238ex)--(2.0625ex,0.32476ex); } } \def\dimersXXXVket{
\tikz[baseline=-0.5ex]{ \fill (-1.125ex,-1.94856ex) circle (1pt);
\fill (0.375ex,-1.94856ex) circle (1pt); \fill (-1.125ex,0.649519ex)
circle (1pt); \fill (-0.375ex,-0.649519ex) circle (1pt); \fill
(-0.375ex,1.94856ex) circle (1pt); \fill (1.125ex,1.94856ex) circle
(1pt); \fill (0.375ex,0.649519ex) circle (1pt); \fill
(1.125ex,-0.649519ex) circle (1pt); \draw[ultra thick]
(-1.125ex,-1.94856ex)--(-0.375ex,-0.649519ex); \draw[ultra thick]
(-1.125ex,0.649519ex)--(-0.375ex,1.94856ex); \draw[ultra thick]
(0.375ex,-1.94856ex)--(1.125ex,-0.649519ex); \draw[ultra thick]
(0.375ex,0.649519ex)--(1.125ex,1.94856ex); \draw[thin]
(-1.125ex,-1.94856ex)--(0.375ex,-1.94856ex); \draw[thin]
(-1.125ex,0.649519ex)--(-0.375ex,-0.649519ex); \draw[thin]
(-0.375ex,1.94856ex)--(1.125ex,1.94856ex); \draw[thin]
(0.375ex,0.649519ex)--(1.125ex,-0.649519ex); } } \def\dimersXXXVIket{
\tikz[baseline=-0.5ex]{ \fill (-1.125ex,-1.94856ex) circle (1pt);
\fill (-0.375ex,-0.649519ex) circle (1pt); \fill (-1.125ex,0.649519ex)
circle (1pt); \fill (-0.375ex,1.94856ex) circle (1pt); \fill
(0.375ex,-1.94856ex) circle (1pt); \fill (1.125ex,-0.649519ex) circle
(1pt); \fill (0.375ex,0.649519ex) circle (1pt); \fill
(1.125ex,1.94856ex) circle (1pt); \draw[ultra thick]
(-1.125ex,-1.94856ex)--(0.375ex,-1.94856ex); \draw[ultra thick]
(-1.125ex,0.649519ex)--(-0.375ex,-0.649519ex); \draw[ultra thick]
(-0.375ex,1.94856ex)--(1.125ex,1.94856ex); \draw[ultra thick]
(0.375ex,0.649519ex)--(1.125ex,-0.649519ex); \draw[thin]
(-1.125ex,-1.94856ex)--(-0.375ex,-0.649519ex); \draw[thin]
(-1.125ex,0.649519ex)--(-0.375ex,1.94856ex); \draw[thin]
(0.375ex,-1.94856ex)--(1.125ex,-0.649519ex); \draw[thin]
(0.375ex,0.649519ex)--(1.125ex,1.94856ex); } } \def\dimersXXXVIIket{
\tikz[baseline=-0.5ex]{ \fill (-1.875ex,-0.649519ex) circle (1pt);
\fill (-1.125ex,0.649519ex) circle (1pt); \fill (-0.375ex,-0.649519ex)
circle (1pt); \fill (1.125ex,-0.649519ex) circle (1pt); \fill
(0.375ex,0.649519ex) circle (1pt); \fill (1.875ex,0.649519ex) circle
(1pt); \draw[ultra thick]
(-1.875ex,-0.649519ex)--(-1.125ex,0.649519ex); \draw[ultra thick]
(-0.375ex,-0.649519ex)--(0.375ex,0.649519ex); \draw[ultra thick]
(1.125ex,-0.649519ex)--(1.875ex,0.649519ex); \draw[thin]
(-1.875ex,-0.649519ex)--(-1.125ex,0.649519ex); \draw[thin]
(-0.375ex,-0.649519ex)--(1.125ex,-0.649519ex); \draw[thin]
(0.375ex,0.649519ex)--(1.875ex,0.649519ex); } } \def\dimersXXXVIIIket{
\tikz[baseline=-0.5ex]{ \fill (-1.875ex,-0.649519ex) circle (1pt);
\fill (-0.375ex,-0.649519ex) circle (1pt); \fill (-1.125ex,0.649519ex)
circle (1pt); \fill (0.375ex,0.649519ex) circle (1pt); \fill
(1.125ex,-0.649519ex) circle (1pt); \fill (1.875ex,0.649519ex) circle
(1pt); \draw[ultra thick]
(-1.875ex,-0.649519ex)--(-1.125ex,0.649519ex); \draw[ultra thick]
(-0.375ex,-0.649519ex)--(0.375ex,0.649519ex); \draw[ultra thick]
(1.125ex,-0.649519ex)--(1.875ex,0.649519ex); \draw[thin]
(-1.875ex,-0.649519ex)--(-0.375ex,-0.649519ex); \draw[thin]
(-1.125ex,0.649519ex)--(0.375ex,0.649519ex); \draw[thin]
(1.125ex,-0.649519ex)--(1.875ex,0.649519ex); } } \def\dimersXXXIXket{
\tikz[baseline=-0.5ex]{ \fill (-1.875ex,-0.649519ex) circle (1pt);
\fill (-1.125ex,0.649519ex) circle (1pt); \fill (-0.375ex,-0.649519ex)
circle (1pt); \fill (0.375ex,0.649519ex) circle (1pt); \fill
(1.125ex,-0.649519ex) circle (1pt); \fill (1.875ex,0.649519ex) circle
(1pt); \draw[ultra thick]
(-1.875ex,-0.649519ex)--(-1.125ex,0.649519ex); \draw[ultra thick]
(-0.375ex,-0.649519ex)--(1.125ex,-0.649519ex); \draw[ultra thick]
(0.375ex,0.649519ex)--(1.875ex,0.649519ex); \draw[thin]
(-1.875ex,-0.649519ex)--(-1.125ex,0.649519ex); \draw[thin]
(-0.375ex,-0.649519ex)--(0.375ex,0.649519ex); \draw[thin]
(1.125ex,-0.649519ex)--(1.875ex,0.649519ex); } } \def\dimersXLket{
\tikz[baseline=-0.5ex]{ \fill (-1.875ex,-0.649519ex) circle (1pt);
\fill (-1.125ex,0.649519ex) circle (1pt); \fill (-0.375ex,-0.649519ex)
circle (1pt); \fill (0.375ex,0.649519ex) circle (1pt); \fill
(1.125ex,-0.649519ex) circle (1pt); \fill (1.875ex,0.649519ex) circle
(1pt); \draw[ultra thick]
(-1.875ex,-0.649519ex)--(-0.375ex,-0.649519ex); \draw[ultra thick]
(-1.125ex,0.649519ex)--(0.375ex,0.649519ex); \draw[ultra thick]
(1.125ex,-0.649519ex)--(1.875ex,0.649519ex); \draw[thin]
(-1.875ex,-0.649519ex)--(-1.125ex,0.649519ex); \draw[thin]
(-0.375ex,-0.649519ex)--(0.375ex,0.649519ex); \draw[thin]
(1.125ex,-0.649519ex)--(1.875ex,0.649519ex); } } \def\dimersXLIket{
\tikz[baseline=-0.5ex]{ \fill (-1.ex,-1.29904ex) circle (1pt); \fill
(0.5ex,-1.29904ex) circle (1pt); \fill (-1.ex,1.29904ex) circle (1pt);
\fill (0.5ex,1.29904ex) circle (1pt); \fill (-0.25ex,0.ex) circle
(1pt); \fill (1.25ex,0.ex) circle (1pt); \draw[ultra thick]
(-1.ex,-1.29904ex)--(-0.25ex,0.ex); \draw[ultra thick]
(-1.ex,1.29904ex)--(0.5ex,1.29904ex); \draw[ultra thick]
(0.5ex,-1.29904ex)--(1.25ex,0.ex); \draw[thin]
(-1.ex,-1.29904ex)--(0.5ex,-1.29904ex); \draw[thin]
(-1.ex,1.29904ex)--(0.5ex,1.29904ex); \draw[thin]
(-0.25ex,0.ex)--(1.25ex,0.ex); } } \def\dimersXLIIket{
\tikz[baseline=-0.5ex]{ \fill (-1.ex,-1.29904ex) circle (1pt); \fill
(0.5ex,-1.29904ex) circle (1pt); \fill (-1.ex,1.29904ex) circle (1pt);
\fill (0.5ex,1.29904ex) circle (1pt); \fill (-0.25ex,0.ex) circle
(1pt); \fill (1.25ex,0.ex) circle (1pt); \draw[ultra thick]
(-1.ex,-1.29904ex)--(0.5ex,-1.29904ex); \draw[ultra thick]
(-1.ex,1.29904ex)--(-0.25ex,0.ex); \draw[ultra thick]
(0.5ex,1.29904ex)--(1.25ex,0.ex); \draw[thin]
(-1.ex,-1.29904ex)--(0.5ex,-1.29904ex); \draw[thin]
(-1.ex,1.29904ex)--(0.5ex,1.29904ex); \draw[thin]
(-0.25ex,0.ex)--(1.25ex,0.ex); } } \def\dimersXLIIIket{
\tikz[baseline=-0.5ex]{ \fill (-1.ex,-1.29904ex) circle (1pt); \fill
(-0.25ex,0.ex) circle (1pt); \fill (-1.ex,1.29904ex) circle (1pt);
\fill (0.5ex,1.29904ex) circle (1pt); \fill (0.5ex,-1.29904ex) circle
(1pt); \fill (1.25ex,0.ex) circle (1pt); \draw[ultra thick]
(-1.ex,-1.29904ex)--(0.5ex,-1.29904ex); \draw[ultra thick]
(-1.ex,1.29904ex)--(0.5ex,1.29904ex); \draw[ultra thick]
(-0.25ex,0.ex)--(1.25ex,0.ex); \draw[thin]
(-1.ex,-1.29904ex)--(-0.25ex,0.ex); \draw[thin]
(-1.ex,1.29904ex)--(0.5ex,1.29904ex); \draw[thin]
(0.5ex,-1.29904ex)--(1.25ex,0.ex); } } \def\dimersXLIVket{
\tikz[baseline=-0.5ex]{ \fill (-1.ex,-1.29904ex) circle (1pt); \fill
(0.5ex,-1.29904ex) circle (1pt); \fill (-1.ex,1.29904ex) circle (1pt);
\fill (-0.25ex,0.ex) circle (1pt); \fill (0.5ex,1.29904ex) circle
(1pt); \fill (1.25ex,0.ex) circle (1pt); \draw[ultra thick]
(-1.ex,-1.29904ex)--(0.5ex,-1.29904ex); \draw[ultra thick]
(-1.ex,1.29904ex)--(0.5ex,1.29904ex); \draw[ultra thick]
(-0.25ex,0.ex)--(1.25ex,0.ex); \draw[thin]
(-1.ex,-1.29904ex)--(0.5ex,-1.29904ex); \draw[thin]
(-1.ex,1.29904ex)--(-0.25ex,0.ex); \draw[thin]
(0.5ex,1.29904ex)--(1.25ex,0.ex); } } \def\dimersXLVket{
\tikz[baseline=-0.5ex]{ \fill (-1.5ex,0.866025ex) circle (1pt); \fill
(0.ex,0.866025ex) circle (1pt); \fill (-0.75ex,-0.433013ex) circle
(1pt); \fill (0.ex,-1.73205ex) circle (1pt); \fill
(0.75ex,-0.433013ex) circle (1pt); \fill (1.5ex,0.866025ex) circle
(1pt); \draw[ultra thick] (-1.5ex,0.866025ex)--(-0.75ex,-0.433013ex);
\draw[ultra thick] (0.ex,0.866025ex)--(1.5ex,0.866025ex); \draw[ultra
thick] (0.ex,-1.73205ex)--(0.75ex,-0.433013ex); \draw[thin]
(-1.5ex,0.866025ex)--(0.ex,0.866025ex); \draw[thin]
(-0.75ex,-0.433013ex)--(0.ex,-1.73205ex); \draw[thin]
(0.75ex,-0.433013ex)--(1.5ex,0.866025ex); } } \def\dimersXLVIket{
\tikz[baseline=-0.5ex]{ \fill (-1.5ex,0.866025ex) circle (1pt); \fill
(-0.75ex,-0.433013ex) circle (1pt); \fill (0.ex,0.866025ex) circle
(1pt); \fill (1.5ex,0.866025ex) circle (1pt); \fill (0.ex,-1.73205ex)
circle (1pt); \fill (0.75ex,-0.433013ex) circle (1pt); \draw[ultra
thick] (-1.5ex,0.866025ex)--(0.ex,0.866025ex); \draw[ultra thick]
(-0.75ex,-0.433013ex)--(0.ex,-1.73205ex); \draw[ultra thick]
(0.75ex,-0.433013ex)--(1.5ex,0.866025ex); \draw[thin]
(-1.5ex,0.866025ex)--(-0.75ex,-0.433013ex); \draw[thin]
(0.ex,0.866025ex)--(1.5ex,0.866025ex); \draw[thin]
(0.ex,-1.73205ex)--(0.75ex,-0.433013ex); } } \def\dimersXLVIIket{
\tikz[baseline=-0.5ex]{ \fill (-1.96875ex,-1.13666ex) circle (1pt);
\fill (-0.46875ex,-1.13666ex) circle (1pt); \fill
(-1.21875ex,0.16238ex) circle (1pt); \fill (0.28125ex,0.16238ex)
circle (1pt); \fill (-0.46875ex,1.46142ex) circle (1pt); \fill
(1.03125ex,1.46142ex) circle (1pt); \fill (1.03125ex,-1.13666ex)
circle (1pt); \fill (1.78125ex,0.16238ex) circle (1pt); \draw[ultra
thick] (-1.96875ex,-1.13666ex)--(-1.21875ex,0.16238ex); \draw[ultra
thick] (-0.46875ex,-1.13666ex)--(1.03125ex,-1.13666ex); \draw[ultra
thick] (-0.46875ex,1.46142ex)--(0.28125ex,0.16238ex); \draw[ultra
thick] (1.03125ex,1.46142ex)--(1.78125ex,0.16238ex); \draw[thin]
(-1.96875ex,-1.13666ex)--(-0.46875ex,-1.13666ex); \draw[thin]
(-1.21875ex,0.16238ex)--(0.28125ex,0.16238ex); \draw[thin]
(-0.46875ex,1.46142ex)--(1.03125ex,1.46142ex); \draw[thin]
(1.03125ex,-1.13666ex)--(1.78125ex,0.16238ex); } }
\def\dimersXLVIIIket{ \tikz[baseline=-0.5ex]{ \fill
(-1.96875ex,-1.13666ex) circle (1pt); \fill (-0.46875ex,-1.13666ex)
circle (1pt); \fill (-1.21875ex,0.16238ex) circle (1pt); \fill
(-0.46875ex,1.46142ex) circle (1pt); \fill (0.28125ex,0.16238ex)
circle (1pt); \fill (1.03125ex,1.46142ex) circle (1pt); \fill
(1.03125ex,-1.13666ex) circle (1pt); \fill (1.78125ex,0.16238ex)
circle (1pt); \draw[ultra thick]
(-1.96875ex,-1.13666ex)--(-1.21875ex,0.16238ex); \draw[ultra thick]
(-0.46875ex,-1.13666ex)--(1.03125ex,-1.13666ex); \draw[ultra thick]
(-0.46875ex,1.46142ex)--(1.03125ex,1.46142ex); \draw[ultra thick]
(0.28125ex,0.16238ex)--(1.78125ex,0.16238ex); \draw[thin]
(-1.96875ex,-1.13666ex)--(-0.46875ex,-1.13666ex); \draw[thin]
(-1.21875ex,0.16238ex)--(-0.46875ex,1.46142ex); \draw[thin]
(0.28125ex,0.16238ex)--(1.03125ex,1.46142ex); \draw[thin]
(1.03125ex,-1.13666ex)--(1.78125ex,0.16238ex); } } \def\dimersXLIXket{
\tikz[baseline=-0.5ex]{ \fill (-1.96875ex,-1.13666ex) circle (1pt);
\fill (-1.21875ex,0.16238ex) circle (1pt); \fill
(-0.46875ex,-1.13666ex) circle (1pt); \fill (1.03125ex,-1.13666ex)
circle (1pt); \fill (-0.46875ex,1.46142ex) circle (1pt); \fill
(1.03125ex,1.46142ex) circle (1pt); \fill (0.28125ex,0.16238ex) circle
(1pt); \fill (1.78125ex,0.16238ex) circle (1pt); \draw[ultra thick]
(-1.96875ex,-1.13666ex)--(-0.46875ex,-1.13666ex); \draw[ultra thick]
(-1.21875ex,0.16238ex)--(-0.46875ex,1.46142ex); \draw[ultra thick]
(0.28125ex,0.16238ex)--(1.03125ex,1.46142ex); \draw[ultra thick]
(1.03125ex,-1.13666ex)--(1.78125ex,0.16238ex); \draw[thin]
(-1.96875ex,-1.13666ex)--(-1.21875ex,0.16238ex); \draw[thin]
(-0.46875ex,-1.13666ex)--(1.03125ex,-1.13666ex); \draw[thin]
(-0.46875ex,1.46142ex)--(1.03125ex,1.46142ex); \draw[thin]
(0.28125ex,0.16238ex)--(1.78125ex,0.16238ex); } } \def\dimersLket{
\tikz[baseline=-0.5ex]{ \fill (-1.96875ex,-1.13666ex) circle (1pt);
\fill (-1.21875ex,0.16238ex) circle (1pt); \fill
(-0.46875ex,-1.13666ex) circle (1pt); \fill (1.03125ex,-1.13666ex)
circle (1pt); \fill (-0.46875ex,1.46142ex) circle (1pt); \fill
(0.28125ex,0.16238ex) circle (1pt); \fill (1.03125ex,1.46142ex) circle
(1pt); \fill (1.78125ex,0.16238ex) circle (1pt); \draw[ultra thick]
(-1.96875ex,-1.13666ex)--(-0.46875ex,-1.13666ex); \draw[ultra thick]
(-1.21875ex,0.16238ex)--(0.28125ex,0.16238ex); \draw[ultra thick]
(-0.46875ex,1.46142ex)--(1.03125ex,1.46142ex); \draw[ultra thick]
(1.03125ex,-1.13666ex)--(1.78125ex,0.16238ex); \draw[thin]
(-1.96875ex,-1.13666ex)--(-1.21875ex,0.16238ex); \draw[thin]
(-0.46875ex,-1.13666ex)--(1.03125ex,-1.13666ex); \draw[thin]
(-0.46875ex,1.46142ex)--(0.28125ex,0.16238ex); \draw[thin]
(1.03125ex,1.46142ex)--(1.78125ex,0.16238ex); } } \def\dimersLIket{
\tikz[baseline=-0.5ex]{ \fill (-2.25ex,0.ex) circle (1pt); \fill
(-1.5ex,1.29904ex) circle (1pt); \fill (-1.5ex,-1.29904ex) circle
(1pt); \fill (0.ex,-1.29904ex) circle (1pt); \fill (0.ex,1.29904ex)
circle (1pt); \fill (1.5ex,1.29904ex) circle (1pt); \fill
(1.5ex,-1.29904ex) circle (1pt); \fill (2.25ex,0.ex) circle (1pt);
\draw[ultra thick] (-2.25ex,0.ex)--(-1.5ex,-1.29904ex); \draw[ultra
thick] (-1.5ex,1.29904ex)--(0.ex,1.29904ex); \draw[ultra thick]
(0.ex,-1.29904ex)--(1.5ex,-1.29904ex); \draw[ultra thick]
(1.5ex,1.29904ex)--(2.25ex,0.ex); \draw[thin]
(-2.25ex,0.ex)--(-1.5ex,1.29904ex); \draw[thin]
(-1.5ex,-1.29904ex)--(0.ex,-1.29904ex); \draw[thin]
(0.ex,1.29904ex)--(1.5ex,1.29904ex); \draw[thin]
(1.5ex,-1.29904ex)--(2.25ex,0.ex); } } \def\dimersLIIket{
\tikz[baseline=-0.5ex]{ \fill (-2.25ex,0.ex) circle (1pt); \fill
(-1.5ex,-1.29904ex) circle (1pt); \fill (-1.5ex,1.29904ex) circle
(1pt); \fill (0.ex,1.29904ex) circle (1pt); \fill (0.ex,-1.29904ex)
circle (1pt); \fill (1.5ex,-1.29904ex) circle (1pt); \fill
(1.5ex,1.29904ex) circle (1pt); \fill (2.25ex,0.ex) circle (1pt);
\draw[ultra thick] (-2.25ex,0.ex)--(-1.5ex,1.29904ex); \draw[ultra
thick] (-1.5ex,-1.29904ex)--(0.ex,-1.29904ex); \draw[ultra thick]
(0.ex,1.29904ex)--(1.5ex,1.29904ex); \draw[ultra thick]
(1.5ex,-1.29904ex)--(2.25ex,0.ex); \draw[thin]
(-2.25ex,0.ex)--(-1.5ex,-1.29904ex); \draw[thin]
(-1.5ex,1.29904ex)--(0.ex,1.29904ex); \draw[thin]
(0.ex,-1.29904ex)--(1.5ex,-1.29904ex); \draw[thin]
(1.5ex,1.29904ex)--(2.25ex,0.ex); } } 
\def\dimersIbra{ \tikz[baseline=-0.5ex]{ \fill (-1.125ex,-0.649519ex)
circle (1pt); \fill (-0.375ex,0.649519ex) circle (1pt); \fill
(0.375ex,-0.649519ex) circle (1pt); \fill (1.125ex,0.649519ex) circle
(1pt); \draw[ultra thick]
(-1.125ex,-0.649519ex)--(0.375ex,-0.649519ex); \draw[ultra thick]
(-0.375ex,0.649519ex)--(1.125ex,0.649519ex); \draw[thin]
(-1.125ex,-0.649519ex)--(-0.375ex,0.649519ex); \draw[thin]
(0.375ex,-0.649519ex)--(1.125ex,0.649519ex); } } \def\dimersIIbra{
\tikz[baseline=-0.5ex]{ \fill (-1.125ex,-0.649519ex) circle (1pt);
\fill (0.375ex,-0.649519ex) circle (1pt); \fill (-0.375ex,0.649519ex)
circle (1pt); \fill (1.125ex,0.649519ex) circle (1pt); \draw[ultra
thick] (-1.125ex,-0.649519ex)--(-0.375ex,0.649519ex); \draw[ultra
thick] (0.375ex,-0.649519ex)--(1.125ex,0.649519ex); \draw[thin]
(-1.125ex,-0.649519ex)--(0.375ex,-0.649519ex); \draw[thin]
(-0.375ex,0.649519ex)--(1.125ex,0.649519ex); } } \def\dimersIIIbra{
\tikz[baseline=-0.5ex]{ \fill (-1.875ex,-0.649519ex) circle (1pt);
\fill (-1.125ex,0.649519ex) circle (1pt); \fill (-0.375ex,-0.649519ex)
circle (1pt); \fill (1.125ex,-0.649519ex) circle (1pt); \fill
(0.375ex,0.649519ex) circle (1pt); \fill (1.875ex,0.649519ex) circle
(1pt); \draw[ultra thick]
(-1.875ex,-0.649519ex)--(-0.375ex,-0.649519ex); \draw[ultra thick]
(-1.125ex,0.649519ex)--(0.375ex,0.649519ex); \draw[ultra thick]
(1.125ex,-0.649519ex)--(1.875ex,0.649519ex); \draw[thin]
(-1.875ex,-0.649519ex)--(-1.125ex,0.649519ex); \draw[thin]
(-0.375ex,-0.649519ex)--(1.125ex,-0.649519ex); \draw[thin]
(0.375ex,0.649519ex)--(1.875ex,0.649519ex); } } \def\dimersIVbra{
\tikz[baseline=-0.5ex]{ \fill (-1.875ex,-0.649519ex) circle (1pt);
\fill (-0.375ex,-0.649519ex) circle (1pt); \fill (-1.125ex,0.649519ex)
circle (1pt); \fill (0.375ex,0.649519ex) circle (1pt); \fill
(1.125ex,-0.649519ex) circle (1pt); \fill (1.875ex,0.649519ex) circle
(1pt); \draw[ultra thick]
(-1.875ex,-0.649519ex)--(-1.125ex,0.649519ex); \draw[ultra thick]
(-0.375ex,-0.649519ex)--(1.125ex,-0.649519ex); \draw[ultra thick]
(0.375ex,0.649519ex)--(1.875ex,0.649519ex); \draw[thin]
(-1.875ex,-0.649519ex)--(-0.375ex,-0.649519ex); \draw[thin]
(-1.125ex,0.649519ex)--(0.375ex,0.649519ex); \draw[thin]
(1.125ex,-0.649519ex)--(1.875ex,0.649519ex); } } \def\dimersVbra{
\tikz[baseline=-0.5ex]{ \fill (-1.ex,-1.29904ex) circle (1pt); \fill
(-0.25ex,0.ex) circle (1pt); \fill (-1.ex,1.29904ex) circle (1pt);
\fill (0.5ex,1.29904ex) circle (1pt); \fill (0.5ex,-1.29904ex) circle
(1pt); \fill (1.25ex,0.ex) circle (1pt); \draw[ultra thick]
(-1.ex,-1.29904ex)--(0.5ex,-1.29904ex); \draw[ultra thick]
(-1.ex,1.29904ex)--(-0.25ex,0.ex); \draw[ultra thick]
(0.5ex,1.29904ex)--(1.25ex,0.ex); \draw[thin]
(-1.ex,-1.29904ex)--(-0.25ex,0.ex); \draw[thin]
(-1.ex,1.29904ex)--(0.5ex,1.29904ex); \draw[thin]
(0.5ex,-1.29904ex)--(1.25ex,0.ex); } } \def\dimersVIbra{
\tikz[baseline=-0.5ex]{ \fill (-1.ex,-1.29904ex) circle (1pt); \fill
(0.5ex,-1.29904ex) circle (1pt); \fill (-1.ex,1.29904ex) circle (1pt);
\fill (-0.25ex,0.ex) circle (1pt); \fill (0.5ex,1.29904ex) circle
(1pt); \fill (1.25ex,0.ex) circle (1pt); \draw[ultra thick]
(-1.ex,-1.29904ex)--(-0.25ex,0.ex); \draw[ultra thick]
(-1.ex,1.29904ex)--(0.5ex,1.29904ex); \draw[ultra thick]
(0.5ex,-1.29904ex)--(1.25ex,0.ex); \draw[thin]
(-1.ex,-1.29904ex)--(0.5ex,-1.29904ex); \draw[thin]
(-1.ex,1.29904ex)--(-0.25ex,0.ex); \draw[thin]
(0.5ex,1.29904ex)--(1.25ex,0.ex); } } \def\dimersVIIbra{
\tikz[baseline=-0.5ex]{ \fill (-1.125ex,-0.649519ex) circle (1pt);
\fill (-0.375ex,0.649519ex) circle (1pt); \fill (0.375ex,-0.649519ex)
circle (1pt); \fill (1.125ex,0.649519ex) circle (1pt); \draw[ultra
thick] (-1.125ex,-0.649519ex)--(-0.375ex,0.649519ex); \draw[ultra
thick] (0.375ex,-0.649519ex)--(1.125ex,0.649519ex); \draw[thin]
(-1.125ex,-0.649519ex)--(-0.375ex,0.649519ex); \draw[thin]
(0.375ex,-0.649519ex)--(1.125ex,0.649519ex); } } \def\dimersVIIIbra{
\tikz[baseline=-0.5ex]{ \fill (-1.125ex,-0.649519ex) circle (1pt);
\fill (0.375ex,-0.649519ex) circle (1pt); \fill (-0.375ex,0.649519ex)
circle (1pt); \fill (1.125ex,0.649519ex) circle (1pt); \draw[ultra
thick] (-1.125ex,-0.649519ex)--(0.375ex,-0.649519ex); \draw[ultra
thick] (-0.375ex,0.649519ex)--(1.125ex,0.649519ex); \draw[thin]
(-1.125ex,-0.649519ex)--(0.375ex,-0.649519ex); \draw[thin]
(-0.375ex,0.649519ex)--(1.125ex,0.649519ex); } }   \def\dimersXIbra{
\tikz[baseline=-0.5ex]{ \fill (-1.96875ex,-1.13666ex) circle (1pt);
\fill (-1.21875ex,0.16238ex) circle (1pt); \fill
(-0.46875ex,-1.13666ex) circle (1pt); \fill (1.03125ex,-1.13666ex)
circle (1pt); \fill (-0.46875ex,1.46142ex) circle (1pt); \fill
(0.28125ex,0.16238ex) circle (1pt); \fill (1.03125ex,1.46142ex) circle
(1pt); \fill (1.78125ex,0.16238ex) circle (1pt); \draw[ultra thick]
(-1.96875ex,-1.13666ex)--(-0.46875ex,-1.13666ex); \draw[ultra thick]
(-1.21875ex,0.16238ex)--(-0.46875ex,1.46142ex); \draw[ultra thick]
(0.28125ex,0.16238ex)--(1.03125ex,1.46142ex); \draw[ultra thick]
(1.03125ex,-1.13666ex)--(1.78125ex,0.16238ex); \draw[thin]
(-1.96875ex,-1.13666ex)--(-1.21875ex,0.16238ex); \draw[thin]
(-0.46875ex,-1.13666ex)--(1.03125ex,-1.13666ex); \draw[thin]
(-0.46875ex,1.46142ex)--(0.28125ex,0.16238ex); \draw[thin]
(1.03125ex,1.46142ex)--(1.78125ex,0.16238ex); } } \def\dimersXIIbra{
\tikz[baseline=-0.5ex]{ \fill (-1.96875ex,-1.13666ex) circle (1pt);
\fill (-0.46875ex,-1.13666ex) circle (1pt); \fill
(-1.21875ex,0.16238ex) circle (1pt); \fill (-0.46875ex,1.46142ex)
circle (1pt); \fill (0.28125ex,0.16238ex) circle (1pt); \fill
(1.03125ex,1.46142ex) circle (1pt); \fill (1.03125ex,-1.13666ex)
circle (1pt); \fill (1.78125ex,0.16238ex) circle (1pt); \draw[ultra
thick] (-1.96875ex,-1.13666ex)--(-1.21875ex,0.16238ex); \draw[ultra
thick] (-0.46875ex,-1.13666ex)--(1.03125ex,-1.13666ex); \draw[ultra
thick] (-0.46875ex,1.46142ex)--(0.28125ex,0.16238ex); \draw[ultra
thick] (1.03125ex,1.46142ex)--(1.78125ex,0.16238ex); \draw[thin]
(-1.96875ex,-1.13666ex)--(-0.46875ex,-1.13666ex); \draw[thin]
(-1.21875ex,0.16238ex)--(-0.46875ex,1.46142ex); \draw[thin]
(0.28125ex,0.16238ex)--(1.03125ex,1.46142ex); \draw[thin]
(1.03125ex,-1.13666ex)--(1.78125ex,0.16238ex); } } \def\dimersXIIIbra{
\tikz[baseline=-0.5ex]{ \fill (-1.5ex,1.29904ex) circle (1pt); \fill
(-0.75ex,0.ex) circle (1pt); \fill (-1.5ex,-1.29904ex) circle (1pt);
\fill (0.ex,-1.29904ex) circle (1pt); \fill (0.ex,1.29904ex) circle
(1pt); \fill (1.5ex,1.29904ex) circle (1pt); \fill (0.75ex,0.ex)
circle (1pt); \fill (1.5ex,-1.29904ex) circle (1pt); \draw[ultra
thick] (-1.5ex,1.29904ex)--(0.ex,1.29904ex); \draw[ultra thick]
(-1.5ex,-1.29904ex)--(-0.75ex,0.ex); \draw[ultra thick]
(0.ex,-1.29904ex)--(1.5ex,-1.29904ex); \draw[ultra thick]
(0.75ex,0.ex)--(1.5ex,1.29904ex); \draw[thin]
(-1.5ex,1.29904ex)--(-0.75ex,0.ex); \draw[thin]
(-1.5ex,-1.29904ex)--(0.ex,-1.29904ex); \draw[thin]
(0.ex,1.29904ex)--(1.5ex,1.29904ex); \draw[thin]
(0.75ex,0.ex)--(1.5ex,-1.29904ex); } } \def\dimersXIVbra{
\tikz[baseline=-0.5ex]{ \fill (-1.5ex,1.29904ex) circle (1pt); \fill
(0.ex,1.29904ex) circle (1pt); \fill (-1.5ex,-1.29904ex) circle (1pt);
\fill (-0.75ex,0.ex) circle (1pt); \fill (0.ex,-1.29904ex) circle
(1pt); \fill (1.5ex,-1.29904ex) circle (1pt); \fill (0.75ex,0.ex)
circle (1pt); \fill (1.5ex,1.29904ex) circle (1pt); \draw[ultra thick]
(-1.5ex,1.29904ex)--(-0.75ex,0.ex); \draw[ultra thick]
(-1.5ex,-1.29904ex)--(0.ex,-1.29904ex); \draw[ultra thick]
(0.ex,1.29904ex)--(1.5ex,1.29904ex); \draw[ultra thick]
(0.75ex,0.ex)--(1.5ex,-1.29904ex); \draw[thin]
(-1.5ex,1.29904ex)--(0.ex,1.29904ex); \draw[thin]
(-1.5ex,-1.29904ex)--(-0.75ex,0.ex); \draw[thin]
(0.ex,-1.29904ex)--(1.5ex,-1.29904ex); \draw[thin]
(0.75ex,0.ex)--(1.5ex,1.29904ex); } } \def\dimersXVbra{
\tikz[baseline=-0.5ex]{ \fill (-2.34375ex,0.811899ex) circle (1pt);
\fill (-1.59375ex,-0.487139ex) circle (1pt); \fill
(-0.84375ex,0.811899ex) circle (1pt); \fill (0.65625ex,0.811899ex)
circle (1pt); \fill (-0.09375ex,-0.487139ex) circle (1pt); \fill
(0.65625ex,-1.78618ex) circle (1pt); \fill (1.40625ex,-0.487139ex)
circle (1pt); \fill (2.15625ex,0.811899ex) circle (1pt); \draw[ultra
thick] (-2.34375ex,0.811899ex)--(-0.84375ex,0.811899ex); \draw[ultra
thick] (-1.59375ex,-0.487139ex)--(-0.09375ex,-0.487139ex); \draw[ultra
thick] (0.65625ex,0.811899ex)--(2.15625ex,0.811899ex); \draw[ultra
thick] (0.65625ex,-1.78618ex)--(1.40625ex,-0.487139ex); \draw[thin]
(-2.34375ex,0.811899ex)--(-1.59375ex,-0.487139ex); \draw[thin]
(-0.84375ex,0.811899ex)--(0.65625ex,0.811899ex); \draw[thin]
(-0.09375ex,-0.487139ex)--(0.65625ex,-1.78618ex); \draw[thin]
(1.40625ex,-0.487139ex)--(2.15625ex,0.811899ex); } }
\def\dimersXVIbra{ \tikz[baseline=-0.5ex]{ \fill
(-2.34375ex,0.811899ex) circle (1pt); \fill (-0.84375ex,0.811899ex)
circle (1pt); \fill (-1.59375ex,-0.487139ex) circle (1pt); \fill
(-0.09375ex,-0.487139ex) circle (1pt); \fill (0.65625ex,0.811899ex)
circle (1pt); \fill (2.15625ex,0.811899ex) circle (1pt); \fill
(0.65625ex,-1.78618ex) circle (1pt); \fill (1.40625ex,-0.487139ex)
circle (1pt); \draw[ultra thick]
(-2.34375ex,0.811899ex)--(-1.59375ex,-0.487139ex); \draw[ultra thick]
(-0.84375ex,0.811899ex)--(0.65625ex,0.811899ex); \draw[ultra thick]
(-0.09375ex,-0.487139ex)--(0.65625ex,-1.78618ex); \draw[ultra thick]
(1.40625ex,-0.487139ex)--(2.15625ex,0.811899ex); \draw[thin]
(-2.34375ex,0.811899ex)--(-0.84375ex,0.811899ex); \draw[thin]
(-1.59375ex,-0.487139ex)--(-0.09375ex,-0.487139ex); \draw[thin]
(0.65625ex,0.811899ex)--(2.15625ex,0.811899ex); \draw[thin]
(0.65625ex,-1.78618ex)--(1.40625ex,-0.487139ex); } }
        \def\dimersXXVbra{
\tikz[baseline=-0.5ex]{ \fill (-2.34375ex,-1.13666ex) circle (1pt);
\fill (-1.59375ex,0.16238ex) circle (1pt); \fill
(-0.84375ex,-1.13666ex) circle (1pt); \fill (0.65625ex,-1.13666ex)
circle (1pt); \fill (-0.09375ex,0.16238ex) circle (1pt); \fill
(0.65625ex,1.46142ex) circle (1pt); \fill (1.40625ex,0.16238ex) circle
(1pt); \fill (2.15625ex,1.46142ex) circle (1pt); \draw[ultra thick]
(-2.34375ex,-1.13666ex)--(-0.84375ex,-1.13666ex); \draw[ultra thick]
(-1.59375ex,0.16238ex)--(-0.09375ex,0.16238ex); \draw[ultra thick]
(0.65625ex,-1.13666ex)--(1.40625ex,0.16238ex); \draw[ultra thick]
(0.65625ex,1.46142ex)--(2.15625ex,1.46142ex); \draw[thin]
(-2.34375ex,-1.13666ex)--(-1.59375ex,0.16238ex); \draw[thin]
(-0.84375ex,-1.13666ex)--(0.65625ex,-1.13666ex); \draw[thin]
(-0.09375ex,0.16238ex)--(0.65625ex,1.46142ex); \draw[thin]
(1.40625ex,0.16238ex)--(2.15625ex,1.46142ex); } } \def\dimersXXVIbra{
\tikz[baseline=-0.5ex]{ \fill (-2.34375ex,-1.13666ex) circle (1pt);
\fill (-0.84375ex,-1.13666ex) circle (1pt); \fill
(-1.59375ex,0.16238ex) circle (1pt); \fill (-0.09375ex,0.16238ex)
circle (1pt); \fill (0.65625ex,-1.13666ex) circle (1pt); \fill
(1.40625ex,0.16238ex) circle (1pt); \fill (0.65625ex,1.46142ex) circle
(1pt); \fill (2.15625ex,1.46142ex) circle (1pt); \draw[ultra thick]
(-2.34375ex,-1.13666ex)--(-1.59375ex,0.16238ex); \draw[ultra thick]
(-0.84375ex,-1.13666ex)--(0.65625ex,-1.13666ex); \draw[ultra thick]
(-0.09375ex,0.16238ex)--(0.65625ex,1.46142ex); \draw[ultra thick]
(1.40625ex,0.16238ex)--(2.15625ex,1.46142ex); \draw[thin]
(-2.34375ex,-1.13666ex)--(-0.84375ex,-1.13666ex); \draw[thin]
(-1.59375ex,0.16238ex)--(-0.09375ex,0.16238ex); \draw[thin]
(0.65625ex,-1.13666ex)--(1.40625ex,0.16238ex); \draw[thin]
(0.65625ex,1.46142ex)--(2.15625ex,1.46142ex); } } \def\dimersXXVIIbra{
\tikz[baseline=-0.5ex]{ \fill (-2.25ex,0.32476ex) circle (1pt); \fill
(-1.5ex,-0.974279ex) circle (1pt); \fill (-0.75ex,0.32476ex) circle
(1pt); \fill (0.ex,1.6238ex) circle (1pt); \fill (0.ex,-0.974279ex)
circle (1pt); \fill (1.5ex,-0.974279ex) circle (1pt); \fill
(0.75ex,0.32476ex) circle (1pt); \fill (2.25ex,0.32476ex) circle
(1pt); \draw[ultra thick] (-2.25ex,0.32476ex)--(-0.75ex,0.32476ex);
\draw[ultra thick] (-1.5ex,-0.974279ex)--(0.ex,-0.974279ex);
\draw[ultra thick] (0.ex,1.6238ex)--(0.75ex,0.32476ex); \draw[ultra
thick] (1.5ex,-0.974279ex)--(2.25ex,0.32476ex); \draw[thin]
(-2.25ex,0.32476ex)--(-1.5ex,-0.974279ex); \draw[thin]
(-0.75ex,0.32476ex)--(0.ex,1.6238ex); \draw[thin]
(0.ex,-0.974279ex)--(1.5ex,-0.974279ex); \draw[thin]
(0.75ex,0.32476ex)--(2.25ex,0.32476ex); } } \def\dimersXXVIIIbra{
\tikz[baseline=-0.5ex]{ \fill (-2.25ex,0.32476ex) circle (1pt); \fill
(-0.75ex,0.32476ex) circle (1pt); \fill (-1.5ex,-0.974279ex) circle
(1pt); \fill (0.ex,-0.974279ex) circle (1pt); \fill (0.ex,1.6238ex)
circle (1pt); \fill (0.75ex,0.32476ex) circle (1pt); \fill
(1.5ex,-0.974279ex) circle (1pt); \fill (2.25ex,0.32476ex) circle
(1pt); \draw[ultra thick] (-2.25ex,0.32476ex)--(-1.5ex,-0.974279ex);
\draw[ultra thick] (-0.75ex,0.32476ex)--(0.ex,1.6238ex); \draw[ultra
thick] (0.ex,-0.974279ex)--(1.5ex,-0.974279ex); \draw[ultra thick]
(0.75ex,0.32476ex)--(2.25ex,0.32476ex); \draw[thin]
(-2.25ex,0.32476ex)--(-0.75ex,0.32476ex); \draw[thin]
(-1.5ex,-0.974279ex)--(0.ex,-0.974279ex); \draw[thin]
(0.ex,1.6238ex)--(0.75ex,0.32476ex); \draw[thin]
(1.5ex,-0.974279ex)--(2.25ex,0.32476ex); } } \def\dimersXXIXbra{
\tikz[baseline=-0.5ex]{ \fill (-1.59375ex,-1.13666ex) circle (1pt);
\fill (-0.84375ex,0.16238ex) circle (1pt); \fill
(-1.59375ex,1.46142ex) circle (1pt); \fill (-0.09375ex,1.46142ex)
circle (1pt); \fill (-0.09375ex,-1.13666ex) circle (1pt); \fill
(1.40625ex,-1.13666ex) circle (1pt); \fill (0.65625ex,0.16238ex)
circle (1pt); \fill (2.15625ex,0.16238ex) circle (1pt); \draw[ultra
thick] (-1.59375ex,-1.13666ex)--(-0.09375ex,-1.13666ex); \draw[ultra
thick] (-1.59375ex,1.46142ex)--(-0.84375ex,0.16238ex); \draw[ultra
thick] (-0.09375ex,1.46142ex)--(0.65625ex,0.16238ex); \draw[ultra
thick] (1.40625ex,-1.13666ex)--(2.15625ex,0.16238ex); \draw[thin]
(-1.59375ex,-1.13666ex)--(-0.84375ex,0.16238ex); \draw[thin]
(-1.59375ex,1.46142ex)--(-0.09375ex,1.46142ex); \draw[thin]
(-0.09375ex,-1.13666ex)--(1.40625ex,-1.13666ex); \draw[thin]
(0.65625ex,0.16238ex)--(2.15625ex,0.16238ex); } } \def\dimersXXXbra{
\tikz[baseline=-0.5ex]{ \fill (-1.59375ex,-1.13666ex) circle (1pt);
\fill (-0.09375ex,-1.13666ex) circle (1pt); \fill
(-1.59375ex,1.46142ex) circle (1pt); \fill (-0.84375ex,0.16238ex)
circle (1pt); \fill (-0.09375ex,1.46142ex) circle (1pt); \fill
(0.65625ex,0.16238ex) circle (1pt); \fill (1.40625ex,-1.13666ex)
circle (1pt); \fill (2.15625ex,0.16238ex) circle (1pt); \draw[ultra
thick] (-1.59375ex,-1.13666ex)--(-0.84375ex,0.16238ex); \draw[ultra
thick] (-1.59375ex,1.46142ex)--(-0.09375ex,1.46142ex); \draw[ultra
thick] (-0.09375ex,-1.13666ex)--(1.40625ex,-1.13666ex); \draw[ultra
thick] (0.65625ex,0.16238ex)--(2.15625ex,0.16238ex); \draw[thin]
(-1.59375ex,-1.13666ex)--(-0.09375ex,-1.13666ex); \draw[thin]
(-1.59375ex,1.46142ex)--(-0.84375ex,0.16238ex); \draw[thin]
(-0.09375ex,1.46142ex)--(0.65625ex,0.16238ex); \draw[thin]
(1.40625ex,-1.13666ex)--(2.15625ex,0.16238ex); } } \def\dimersXXXIbra{
\tikz[baseline=-0.5ex]{ \fill (-2.625ex,-0.649519ex) circle (1pt);
\fill (-1.875ex,0.649519ex) circle (1pt); \fill (-1.125ex,-0.649519ex)
circle (1pt); \fill (0.375ex,-0.649519ex) circle (1pt); \fill
(-0.375ex,0.649519ex) circle (1pt); \fill (1.125ex,0.649519ex) circle
(1pt); \fill (1.875ex,-0.649519ex) circle (1pt); \fill
(2.625ex,0.649519ex) circle (1pt); \draw[ultra thick]
(-2.625ex,-0.649519ex)--(-1.125ex,-0.649519ex); \draw[ultra thick]
(-1.875ex,0.649519ex)--(-0.375ex,0.649519ex); \draw[ultra thick]
(0.375ex,-0.649519ex)--(1.875ex,-0.649519ex); \draw[ultra thick]
(1.125ex,0.649519ex)--(2.625ex,0.649519ex); \draw[thin]
(-2.625ex,-0.649519ex)--(-1.875ex,0.649519ex); \draw[thin]
(-1.125ex,-0.649519ex)--(0.375ex,-0.649519ex); \draw[thin]
(-0.375ex,0.649519ex)--(1.125ex,0.649519ex); \draw[thin]
(1.875ex,-0.649519ex)--(2.625ex,0.649519ex); } } \def\dimersXXXIIbra{
\tikz[baseline=-0.5ex]{ \fill (-2.625ex,-0.649519ex) circle (1pt);
\fill (-1.125ex,-0.649519ex) circle (1pt); \fill (-1.875ex,0.649519ex)
circle (1pt); \fill (-0.375ex,0.649519ex) circle (1pt); \fill
(0.375ex,-0.649519ex) circle (1pt); \fill (1.875ex,-0.649519ex) circle
(1pt); \fill (1.125ex,0.649519ex) circle (1pt); \fill
(2.625ex,0.649519ex) circle (1pt); \draw[ultra thick]
(-2.625ex,-0.649519ex)--(-1.875ex,0.649519ex); \draw[ultra thick]
(-1.125ex,-0.649519ex)--(0.375ex,-0.649519ex); \draw[ultra thick]
(-0.375ex,0.649519ex)--(1.125ex,0.649519ex); \draw[ultra thick]
(1.875ex,-0.649519ex)--(2.625ex,0.649519ex); \draw[thin]
(-2.625ex,-0.649519ex)--(-1.125ex,-0.649519ex); \draw[thin]
(-1.875ex,0.649519ex)--(-0.375ex,0.649519ex); \draw[thin]
(0.375ex,-0.649519ex)--(1.875ex,-0.649519ex); \draw[thin]
(1.125ex,0.649519ex)--(2.625ex,0.649519ex); } } \def\dimersXXXIIIbra{
\tikz[baseline=-0.5ex]{ \fill (-2.4375ex,0.32476ex) circle (1pt);
\fill (-1.6875ex,-0.974279ex) circle (1pt); \fill
(-0.9375ex,0.32476ex) circle (1pt); \fill (0.5625ex,0.32476ex) circle
(1pt); \fill (-0.1875ex,-0.974279ex) circle (1pt); \fill
(1.3125ex,-0.974279ex) circle (1pt); \fill (1.3125ex,1.6238ex) circle
(1pt); \fill (2.0625ex,0.32476ex) circle (1pt); \draw[ultra thick]
(-2.4375ex,0.32476ex)--(-0.9375ex,0.32476ex); \draw[ultra thick]
(-1.6875ex,-0.974279ex)--(-0.1875ex,-0.974279ex); \draw[ultra thick]
(0.5625ex,0.32476ex)--(1.3125ex,1.6238ex); \draw[ultra thick]
(1.3125ex,-0.974279ex)--(2.0625ex,0.32476ex); \draw[thin]
(-2.4375ex,0.32476ex)--(-1.6875ex,-0.974279ex); \draw[thin]
(-0.9375ex,0.32476ex)--(0.5625ex,0.32476ex); \draw[thin]
(-0.1875ex,-0.974279ex)--(1.3125ex,-0.974279ex); \draw[thin]
(1.3125ex,1.6238ex)--(2.0625ex,0.32476ex); } } \def\dimersXXXIVbra{
\tikz[baseline=-0.5ex]{ \fill (-2.4375ex,0.32476ex) circle (1pt);
\fill (-0.9375ex,0.32476ex) circle (1pt); \fill
(-1.6875ex,-0.974279ex) circle (1pt); \fill (-0.1875ex,-0.974279ex)
circle (1pt); \fill (0.5625ex,0.32476ex) circle (1pt); \fill
(1.3125ex,1.6238ex) circle (1pt); \fill (1.3125ex,-0.974279ex) circle
(1pt); \fill (2.0625ex,0.32476ex) circle (1pt); \draw[ultra thick]
(-2.4375ex,0.32476ex)--(-1.6875ex,-0.974279ex); \draw[ultra thick]
(-0.9375ex,0.32476ex)--(0.5625ex,0.32476ex); \draw[ultra thick]
(-0.1875ex,-0.974279ex)--(1.3125ex,-0.974279ex); \draw[ultra thick]
(1.3125ex,1.6238ex)--(2.0625ex,0.32476ex); \draw[thin]
(-2.4375ex,0.32476ex)--(-0.9375ex,0.32476ex); \draw[thin]
(-1.6875ex,-0.974279ex)--(-0.1875ex,-0.974279ex); \draw[thin]
(0.5625ex,0.32476ex)--(1.3125ex,1.6238ex); \draw[thin]
(1.3125ex,-0.974279ex)--(2.0625ex,0.32476ex); } } \def\dimersXXXVbra{
\tikz[baseline=-0.5ex]{ \fill (-1.125ex,-1.94856ex) circle (1pt);
\fill (-0.375ex,-0.649519ex) circle (1pt); \fill (-1.125ex,0.649519ex)
circle (1pt); \fill (-0.375ex,1.94856ex) circle (1pt); \fill
(0.375ex,-1.94856ex) circle (1pt); \fill (1.125ex,-0.649519ex) circle
(1pt); \fill (0.375ex,0.649519ex) circle (1pt); \fill
(1.125ex,1.94856ex) circle (1pt); \draw[ultra thick]
(-1.125ex,-1.94856ex)--(0.375ex,-1.94856ex); \draw[ultra thick]
(-1.125ex,0.649519ex)--(-0.375ex,-0.649519ex); \draw[ultra thick]
(-0.375ex,1.94856ex)--(1.125ex,1.94856ex); \draw[ultra thick]
(0.375ex,0.649519ex)--(1.125ex,-0.649519ex); \draw[thin]
(-1.125ex,-1.94856ex)--(-0.375ex,-0.649519ex); \draw[thin]
(-1.125ex,0.649519ex)--(-0.375ex,1.94856ex); \draw[thin]
(0.375ex,-1.94856ex)--(1.125ex,-0.649519ex); \draw[thin]
(0.375ex,0.649519ex)--(1.125ex,1.94856ex); } } \def\dimersXXXVIbra{
\tikz[baseline=-0.5ex]{ \fill (-1.125ex,-1.94856ex) circle (1pt);
\fill (0.375ex,-1.94856ex) circle (1pt); \fill (-1.125ex,0.649519ex)
circle (1pt); \fill (-0.375ex,-0.649519ex) circle (1pt); \fill
(-0.375ex,1.94856ex) circle (1pt); \fill (1.125ex,1.94856ex) circle
(1pt); \fill (0.375ex,0.649519ex) circle (1pt); \fill
(1.125ex,-0.649519ex) circle (1pt); \draw[ultra thick]
(-1.125ex,-1.94856ex)--(-0.375ex,-0.649519ex); \draw[ultra thick]
(-1.125ex,0.649519ex)--(-0.375ex,1.94856ex); \draw[ultra thick]
(0.375ex,-1.94856ex)--(1.125ex,-0.649519ex); \draw[ultra thick]
(0.375ex,0.649519ex)--(1.125ex,1.94856ex); \draw[thin]
(-1.125ex,-1.94856ex)--(0.375ex,-1.94856ex); \draw[thin]
(-1.125ex,0.649519ex)--(-0.375ex,-0.649519ex); \draw[thin]
(-0.375ex,1.94856ex)--(1.125ex,1.94856ex); \draw[thin]
(0.375ex,0.649519ex)--(1.125ex,-0.649519ex); } } \def\dimersXXXVIIbra{
\tikz[baseline=-0.5ex]{ \fill (-1.875ex,-0.649519ex) circle (1pt);
\fill (-1.125ex,0.649519ex) circle (1pt); \fill (-0.375ex,-0.649519ex)
circle (1pt); \fill (0.375ex,0.649519ex) circle (1pt); \fill
(1.125ex,-0.649519ex) circle (1pt); \fill (1.875ex,0.649519ex) circle
(1pt); \draw[ultra thick]
(-1.875ex,-0.649519ex)--(-1.125ex,0.649519ex); \draw[ultra thick]
(-0.375ex,-0.649519ex)--(1.125ex,-0.649519ex); \draw[ultra thick]
(0.375ex,0.649519ex)--(1.875ex,0.649519ex); \draw[thin]
(-1.875ex,-0.649519ex)--(-1.125ex,0.649519ex); \draw[thin]
(-0.375ex,-0.649519ex)--(0.375ex,0.649519ex); \draw[thin]
(1.125ex,-0.649519ex)--(1.875ex,0.649519ex); } }
\def\dimersXXXVIIIbra{ \tikz[baseline=-0.5ex]{ \fill
(-1.875ex,-0.649519ex) circle (1pt); \fill (-1.125ex,0.649519ex)
circle (1pt); \fill (-0.375ex,-0.649519ex) circle (1pt); \fill
(0.375ex,0.649519ex) circle (1pt); \fill (1.125ex,-0.649519ex) circle
(1pt); \fill (1.875ex,0.649519ex) circle (1pt); \draw[ultra thick]
(-1.875ex,-0.649519ex)--(-0.375ex,-0.649519ex); \draw[ultra thick]
(-1.125ex,0.649519ex)--(0.375ex,0.649519ex); \draw[ultra thick]
(1.125ex,-0.649519ex)--(1.875ex,0.649519ex); \draw[thin]
(-1.875ex,-0.649519ex)--(-1.125ex,0.649519ex); \draw[thin]
(-0.375ex,-0.649519ex)--(0.375ex,0.649519ex); \draw[thin]
(1.125ex,-0.649519ex)--(1.875ex,0.649519ex); } } \def\dimersXXXIXbra{
\tikz[baseline=-0.5ex]{ \fill (-1.875ex,-0.649519ex) circle (1pt);
\fill (-1.125ex,0.649519ex) circle (1pt); \fill (-0.375ex,-0.649519ex)
circle (1pt); \fill (1.125ex,-0.649519ex) circle (1pt); \fill
(0.375ex,0.649519ex) circle (1pt); \fill (1.875ex,0.649519ex) circle
(1pt); \draw[ultra thick]
(-1.875ex,-0.649519ex)--(-1.125ex,0.649519ex); \draw[ultra thick]
(-0.375ex,-0.649519ex)--(0.375ex,0.649519ex); \draw[ultra thick]
(1.125ex,-0.649519ex)--(1.875ex,0.649519ex); \draw[thin]
(-1.875ex,-0.649519ex)--(-1.125ex,0.649519ex); \draw[thin]
(-0.375ex,-0.649519ex)--(1.125ex,-0.649519ex); \draw[thin]
(0.375ex,0.649519ex)--(1.875ex,0.649519ex); } } \def\dimersXLbra{
\tikz[baseline=-0.5ex]{ \fill (-1.875ex,-0.649519ex) circle (1pt);
\fill (-0.375ex,-0.649519ex) circle (1pt); \fill (-1.125ex,0.649519ex)
circle (1pt); \fill (0.375ex,0.649519ex) circle (1pt); \fill
(1.125ex,-0.649519ex) circle (1pt); \fill (1.875ex,0.649519ex) circle
(1pt); \draw[ultra thick]
(-1.875ex,-0.649519ex)--(-1.125ex,0.649519ex); \draw[ultra thick]
(-0.375ex,-0.649519ex)--(0.375ex,0.649519ex); \draw[ultra thick]
(1.125ex,-0.649519ex)--(1.875ex,0.649519ex); \draw[thin]
(-1.875ex,-0.649519ex)--(-0.375ex,-0.649519ex); \draw[thin]
(-1.125ex,0.649519ex)--(0.375ex,0.649519ex); \draw[thin]
(1.125ex,-0.649519ex)--(1.875ex,0.649519ex); } } \def\dimersXLIbra{
\tikz[baseline=-0.5ex]{ \fill (-1.ex,-1.29904ex) circle (1pt); \fill
(-0.25ex,0.ex) circle (1pt); \fill (-1.ex,1.29904ex) circle (1pt);
\fill (0.5ex,1.29904ex) circle (1pt); \fill (0.5ex,-1.29904ex) circle
(1pt); \fill (1.25ex,0.ex) circle (1pt); \draw[ultra thick]
(-1.ex,-1.29904ex)--(0.5ex,-1.29904ex); \draw[ultra thick]
(-1.ex,1.29904ex)--(0.5ex,1.29904ex); \draw[ultra thick]
(-0.25ex,0.ex)--(1.25ex,0.ex); \draw[thin]
(-1.ex,-1.29904ex)--(-0.25ex,0.ex); \draw[thin]
(-1.ex,1.29904ex)--(0.5ex,1.29904ex); \draw[thin]
(0.5ex,-1.29904ex)--(1.25ex,0.ex); } } \def\dimersXLIIbra{
\tikz[baseline=-0.5ex]{ \fill (-1.ex,-1.29904ex) circle (1pt); \fill
(0.5ex,-1.29904ex) circle (1pt); \fill (-1.ex,1.29904ex) circle (1pt);
\fill (-0.25ex,0.ex) circle (1pt); \fill (0.5ex,1.29904ex) circle
(1pt); \fill (1.25ex,0.ex) circle (1pt); \draw[ultra thick]
(-1.ex,-1.29904ex)--(0.5ex,-1.29904ex); \draw[ultra thick]
(-1.ex,1.29904ex)--(0.5ex,1.29904ex); \draw[ultra thick]
(-0.25ex,0.ex)--(1.25ex,0.ex); \draw[thin]
(-1.ex,-1.29904ex)--(0.5ex,-1.29904ex); \draw[thin]
(-1.ex,1.29904ex)--(-0.25ex,0.ex); \draw[thin]
(0.5ex,1.29904ex)--(1.25ex,0.ex); } } \def\dimersXLIIIbra{
\tikz[baseline=-0.5ex]{ \fill (-1.ex,-1.29904ex) circle (1pt); \fill
(0.5ex,-1.29904ex) circle (1pt); \fill (-1.ex,1.29904ex) circle (1pt);
\fill (0.5ex,1.29904ex) circle (1pt); \fill (-0.25ex,0.ex) circle
(1pt); \fill (1.25ex,0.ex) circle (1pt); \draw[ultra thick]
(-1.ex,-1.29904ex)--(-0.25ex,0.ex); \draw[ultra thick]
(-1.ex,1.29904ex)--(0.5ex,1.29904ex); \draw[ultra thick]
(0.5ex,-1.29904ex)--(1.25ex,0.ex); \draw[thin]
(-1.ex,-1.29904ex)--(0.5ex,-1.29904ex); \draw[thin]
(-1.ex,1.29904ex)--(0.5ex,1.29904ex); \draw[thin]
(-0.25ex,0.ex)--(1.25ex,0.ex); } } \def\dimersXLIVbra{
\tikz[baseline=-0.5ex]{ \fill (-1.ex,-1.29904ex) circle (1pt); \fill
(0.5ex,-1.29904ex) circle (1pt); \fill (-1.ex,1.29904ex) circle (1pt);
\fill (0.5ex,1.29904ex) circle (1pt); \fill (-0.25ex,0.ex) circle
(1pt); \fill (1.25ex,0.ex) circle (1pt); \draw[ultra thick]
(-1.ex,-1.29904ex)--(0.5ex,-1.29904ex); \draw[ultra thick]
(-1.ex,1.29904ex)--(-0.25ex,0.ex); \draw[ultra thick]
(0.5ex,1.29904ex)--(1.25ex,0.ex); \draw[thin]
(-1.ex,-1.29904ex)--(0.5ex,-1.29904ex); \draw[thin]
(-1.ex,1.29904ex)--(0.5ex,1.29904ex); \draw[thin]
(-0.25ex,0.ex)--(1.25ex,0.ex); } } \def\dimersXLVbra{
\tikz[baseline=-0.5ex]{ \fill (-1.5ex,0.866025ex) circle (1pt); \fill
(-0.75ex,-0.433013ex) circle (1pt); \fill (0.ex,0.866025ex) circle
(1pt); \fill (1.5ex,0.866025ex) circle (1pt); \fill (0.ex,-1.73205ex)
circle (1pt); \fill (0.75ex,-0.433013ex) circle (1pt); \draw[ultra
thick] (-1.5ex,0.866025ex)--(0.ex,0.866025ex); \draw[ultra thick]
(-0.75ex,-0.433013ex)--(0.ex,-1.73205ex); \draw[ultra thick]
(0.75ex,-0.433013ex)--(1.5ex,0.866025ex); \draw[thin]
(-1.5ex,0.866025ex)--(-0.75ex,-0.433013ex); \draw[thin]
(0.ex,0.866025ex)--(1.5ex,0.866025ex); \draw[thin]
(0.ex,-1.73205ex)--(0.75ex,-0.433013ex); } } \def\dimersXLVIbra{
\tikz[baseline=-0.5ex]{ \fill (-1.5ex,0.866025ex) circle (1pt); \fill
(0.ex,0.866025ex) circle (1pt); \fill (-0.75ex,-0.433013ex) circle
(1pt); \fill (0.ex,-1.73205ex) circle (1pt); \fill
(0.75ex,-0.433013ex) circle (1pt); \fill (1.5ex,0.866025ex) circle
(1pt); \draw[ultra thick] (-1.5ex,0.866025ex)--(-0.75ex,-0.433013ex);
\draw[ultra thick] (0.ex,0.866025ex)--(1.5ex,0.866025ex); \draw[ultra
thick] (0.ex,-1.73205ex)--(0.75ex,-0.433013ex); \draw[thin]
(-1.5ex,0.866025ex)--(0.ex,0.866025ex); \draw[thin]
(-0.75ex,-0.433013ex)--(0.ex,-1.73205ex); \draw[thin]
(0.75ex,-0.433013ex)--(1.5ex,0.866025ex); } } \def\dimersXLVIIbra{
\tikz[baseline=-0.5ex]{ \fill (-1.96875ex,-1.13666ex) circle (1pt);
\fill (-1.21875ex,0.16238ex) circle (1pt); \fill
(-0.46875ex,-1.13666ex) circle (1pt); \fill (1.03125ex,-1.13666ex)
circle (1pt); \fill (-0.46875ex,1.46142ex) circle (1pt); \fill
(0.28125ex,0.16238ex) circle (1pt); \fill (1.03125ex,1.46142ex) circle
(1pt); \fill (1.78125ex,0.16238ex) circle (1pt); \draw[ultra thick]
(-1.96875ex,-1.13666ex)--(-0.46875ex,-1.13666ex); \draw[ultra thick]
(-1.21875ex,0.16238ex)--(0.28125ex,0.16238ex); \draw[ultra thick]
(-0.46875ex,1.46142ex)--(1.03125ex,1.46142ex); \draw[ultra thick]
(1.03125ex,-1.13666ex)--(1.78125ex,0.16238ex); \draw[thin]
(-1.96875ex,-1.13666ex)--(-1.21875ex,0.16238ex); \draw[thin]
(-0.46875ex,-1.13666ex)--(1.03125ex,-1.13666ex); \draw[thin]
(-0.46875ex,1.46142ex)--(0.28125ex,0.16238ex); \draw[thin]
(1.03125ex,1.46142ex)--(1.78125ex,0.16238ex); } }
\def\dimersXLVIIIbra{ \tikz[baseline=-0.5ex]{ \fill
(-1.96875ex,-1.13666ex) circle (1pt); \fill (-1.21875ex,0.16238ex)
circle (1pt); \fill (-0.46875ex,-1.13666ex) circle (1pt); \fill
(1.03125ex,-1.13666ex) circle (1pt); \fill (-0.46875ex,1.46142ex)
circle (1pt); \fill (1.03125ex,1.46142ex) circle (1pt); \fill
(0.28125ex,0.16238ex) circle (1pt); \fill (1.78125ex,0.16238ex) circle
(1pt); \draw[ultra thick]
(-1.96875ex,-1.13666ex)--(-0.46875ex,-1.13666ex); \draw[ultra thick]
(-1.21875ex,0.16238ex)--(-0.46875ex,1.46142ex); \draw[ultra thick]
(0.28125ex,0.16238ex)--(1.03125ex,1.46142ex); \draw[ultra thick]
(1.03125ex,-1.13666ex)--(1.78125ex,0.16238ex); \draw[thin]
(-1.96875ex,-1.13666ex)--(-1.21875ex,0.16238ex); \draw[thin]
(-0.46875ex,-1.13666ex)--(1.03125ex,-1.13666ex); \draw[thin]
(-0.46875ex,1.46142ex)--(1.03125ex,1.46142ex); \draw[thin]
(0.28125ex,0.16238ex)--(1.78125ex,0.16238ex); } } \def\dimersXLIXbra{
\tikz[baseline=-0.5ex]{ \fill (-1.96875ex,-1.13666ex) circle (1pt);
\fill (-0.46875ex,-1.13666ex) circle (1pt); \fill
(-1.21875ex,0.16238ex) circle (1pt); \fill (-0.46875ex,1.46142ex)
circle (1pt); \fill (0.28125ex,0.16238ex) circle (1pt); \fill
(1.03125ex,1.46142ex) circle (1pt); \fill (1.03125ex,-1.13666ex)
circle (1pt); \fill (1.78125ex,0.16238ex) circle (1pt); \draw[ultra
thick] (-1.96875ex,-1.13666ex)--(-1.21875ex,0.16238ex); \draw[ultra
thick] (-0.46875ex,-1.13666ex)--(1.03125ex,-1.13666ex); \draw[ultra
thick] (-0.46875ex,1.46142ex)--(1.03125ex,1.46142ex); \draw[ultra
thick] (0.28125ex,0.16238ex)--(1.78125ex,0.16238ex); \draw[thin]
(-1.96875ex,-1.13666ex)--(-0.46875ex,-1.13666ex); \draw[thin]
(-1.21875ex,0.16238ex)--(-0.46875ex,1.46142ex); \draw[thin]
(0.28125ex,0.16238ex)--(1.03125ex,1.46142ex); \draw[thin]
(1.03125ex,-1.13666ex)--(1.78125ex,0.16238ex); } } \def\dimersLbra{
\tikz[baseline=-0.5ex]{ \fill (-1.96875ex,-1.13666ex) circle (1pt);
\fill (-0.46875ex,-1.13666ex) circle (1pt); \fill
(-1.21875ex,0.16238ex) circle (1pt); \fill (0.28125ex,0.16238ex)
circle (1pt); \fill (-0.46875ex,1.46142ex) circle (1pt); \fill
(1.03125ex,1.46142ex) circle (1pt); \fill (1.03125ex,-1.13666ex)
circle (1pt); \fill (1.78125ex,0.16238ex) circle (1pt); \draw[ultra
thick] (-1.96875ex,-1.13666ex)--(-1.21875ex,0.16238ex); \draw[ultra
thick] (-0.46875ex,-1.13666ex)--(1.03125ex,-1.13666ex); \draw[ultra
thick] (-0.46875ex,1.46142ex)--(0.28125ex,0.16238ex); \draw[ultra
thick] (1.03125ex,1.46142ex)--(1.78125ex,0.16238ex); \draw[thin]
(-1.96875ex,-1.13666ex)--(-0.46875ex,-1.13666ex); \draw[thin]
(-1.21875ex,0.16238ex)--(0.28125ex,0.16238ex); \draw[thin]
(-0.46875ex,1.46142ex)--(1.03125ex,1.46142ex); \draw[thin]
(1.03125ex,-1.13666ex)--(1.78125ex,0.16238ex); } } \def\dimersLIbra{
\tikz[baseline=-0.5ex]{ \fill (-2.25ex,0.ex) circle (1pt); \fill
(-1.5ex,-1.29904ex) circle (1pt); \fill (-1.5ex,1.29904ex) circle
(1pt); \fill (0.ex,1.29904ex) circle (1pt); \fill (0.ex,-1.29904ex)
circle (1pt); \fill (1.5ex,-1.29904ex) circle (1pt); \fill
(1.5ex,1.29904ex) circle (1pt); \fill (2.25ex,0.ex) circle (1pt);
\draw[ultra thick] (-2.25ex,0.ex)--(-1.5ex,1.29904ex); \draw[ultra
thick] (-1.5ex,-1.29904ex)--(0.ex,-1.29904ex); \draw[ultra thick]
(0.ex,1.29904ex)--(1.5ex,1.29904ex); \draw[ultra thick]
(1.5ex,-1.29904ex)--(2.25ex,0.ex); \draw[thin]
(-2.25ex,0.ex)--(-1.5ex,-1.29904ex); \draw[thin]
(-1.5ex,1.29904ex)--(0.ex,1.29904ex); \draw[thin]
(0.ex,-1.29904ex)--(1.5ex,-1.29904ex); \draw[thin]
(1.5ex,1.29904ex)--(2.25ex,0.ex); } } \def\dimersLIIbra{
\tikz[baseline=-0.5ex]{ \fill (-2.25ex,0.ex) circle (1pt); \fill
(-1.5ex,1.29904ex) circle (1pt); \fill (-1.5ex,-1.29904ex) circle
(1pt); \fill (0.ex,-1.29904ex) circle (1pt); \fill (0.ex,1.29904ex)
circle (1pt); \fill (1.5ex,1.29904ex) circle (1pt); \fill
(1.5ex,-1.29904ex) circle (1pt); \fill (2.25ex,0.ex) circle (1pt);
\draw[ultra thick] (-2.25ex,0.ex)--(-1.5ex,-1.29904ex); \draw[ultra
thick] (-1.5ex,1.29904ex)--(0.ex,1.29904ex); \draw[ultra thick]
(0.ex,-1.29904ex)--(1.5ex,-1.29904ex); \draw[ultra thick]
(1.5ex,1.29904ex)--(2.25ex,0.ex); \draw[thin]
(-2.25ex,0.ex)--(-1.5ex,1.29904ex); \draw[thin]
(-1.5ex,-1.29904ex)--(0.ex,-1.29904ex); \draw[thin]
(0.ex,1.29904ex)--(1.5ex,1.29904ex); \draw[thin]
(1.5ex,-1.29904ex)--(2.25ex,0.ex); } } \def\dimersIket{
\tikz[baseline=-0.5ex]{ \fill (-1.125ex,-0.649519ex) circle (1pt);
\fill (0.375ex,-0.649519ex) circle (1pt); \fill (-0.375ex,0.649519ex)
circle (1pt); \fill (1.125ex,0.649519ex) circle (1pt); \draw[ultra
thick] (-1.125ex,-0.649519ex)--(-0.375ex,0.649519ex); \draw[ultra
thick] (0.375ex,-0.649519ex)--(1.125ex,0.649519ex); \draw[thin]
(-1.125ex,-0.649519ex)--(0.375ex,-0.649519ex); \draw[thin]
(-0.375ex,0.649519ex)--(1.125ex,0.649519ex); } } \def\dimersIIket{
\tikz[baseline=-0.5ex]{ \fill (-1.125ex,-0.649519ex) circle (1pt);
\fill (-0.375ex,0.649519ex) circle (1pt); \fill (0.375ex,-0.649519ex)
circle (1pt); \fill (1.125ex,0.649519ex) circle (1pt); \draw[ultra
thick] (-1.125ex,-0.649519ex)--(0.375ex,-0.649519ex); \draw[ultra
thick] (-0.375ex,0.649519ex)--(1.125ex,0.649519ex); \draw[thin]
(-1.125ex,-0.649519ex)--(-0.375ex,0.649519ex); \draw[thin]
(0.375ex,-0.649519ex)--(1.125ex,0.649519ex); } } \def\dimersIIIket{
\tikz[baseline=-0.5ex]{ \fill (-1.875ex,-0.649519ex) circle (1pt);
\fill (-0.375ex,-0.649519ex) circle (1pt); \fill (-1.125ex,0.649519ex)
circle (1pt); \fill (0.375ex,0.649519ex) circle (1pt); \fill
(1.125ex,-0.649519ex) circle (1pt); \fill (1.875ex,0.649519ex) circle
(1pt); \draw[ultra thick]
(-1.875ex,-0.649519ex)--(-1.125ex,0.649519ex); \draw[ultra thick]
(-0.375ex,-0.649519ex)--(1.125ex,-0.649519ex); \draw[ultra thick]
(0.375ex,0.649519ex)--(1.875ex,0.649519ex); \draw[thin]
(-1.875ex,-0.649519ex)--(-0.375ex,-0.649519ex); \draw[thin]
(-1.125ex,0.649519ex)--(0.375ex,0.649519ex); \draw[thin]
(1.125ex,-0.649519ex)--(1.875ex,0.649519ex); } } \def\dimersIVket{
\tikz[baseline=-0.5ex]{ \fill (-1.875ex,-0.649519ex) circle (1pt);
\fill (-1.125ex,0.649519ex) circle (1pt); \fill (-0.375ex,-0.649519ex)
circle (1pt); \fill (1.125ex,-0.649519ex) circle (1pt); \fill
(0.375ex,0.649519ex) circle (1pt); \fill (1.875ex,0.649519ex) circle
(1pt); \draw[ultra thick]
(-1.875ex,-0.649519ex)--(-0.375ex,-0.649519ex); \draw[ultra thick]
(-1.125ex,0.649519ex)--(0.375ex,0.649519ex); \draw[ultra thick]
(1.125ex,-0.649519ex)--(1.875ex,0.649519ex); \draw[thin]
(-1.875ex,-0.649519ex)--(-1.125ex,0.649519ex); \draw[thin]
(-0.375ex,-0.649519ex)--(1.125ex,-0.649519ex); \draw[thin]
(0.375ex,0.649519ex)--(1.875ex,0.649519ex); } } \def\dimersVket{
\tikz[baseline=-0.5ex]{ \fill (-1.ex,-1.29904ex) circle (1pt); \fill
(0.5ex,-1.29904ex) circle (1pt); \fill (-1.ex,1.29904ex) circle (1pt);
\fill (-0.25ex,0.ex) circle (1pt); \fill (0.5ex,1.29904ex) circle
(1pt); \fill (1.25ex,0.ex) circle (1pt); \draw[ultra thick]
(-1.ex,-1.29904ex)--(-0.25ex,0.ex); \draw[ultra thick]
(-1.ex,1.29904ex)--(0.5ex,1.29904ex); \draw[ultra thick]
(0.5ex,-1.29904ex)--(1.25ex,0.ex); \draw[thin]
(-1.ex,-1.29904ex)--(0.5ex,-1.29904ex); \draw[thin]
(-1.ex,1.29904ex)--(-0.25ex,0.ex); \draw[thin]
(0.5ex,1.29904ex)--(1.25ex,0.ex); } } \def\dimersVIket{
\tikz[baseline=-0.5ex]{ \fill (-1.ex,-1.29904ex) circle (1pt); \fill
(-0.25ex,0.ex) circle (1pt); \fill (-1.ex,1.29904ex) circle (1pt);
\fill (0.5ex,1.29904ex) circle (1pt); \fill (0.5ex,-1.29904ex) circle
(1pt); \fill (1.25ex,0.ex) circle (1pt); \draw[ultra thick]
(-1.ex,-1.29904ex)--(0.5ex,-1.29904ex); \draw[ultra thick]
(-1.ex,1.29904ex)--(-0.25ex,0.ex); \draw[ultra thick]
(0.5ex,1.29904ex)--(1.25ex,0.ex); \draw[thin]
(-1.ex,-1.29904ex)--(-0.25ex,0.ex); \draw[thin]
(-1.ex,1.29904ex)--(0.5ex,1.29904ex); \draw[thin]
(0.5ex,-1.29904ex)--(1.25ex,0.ex); } } \def\dimersVIIket{
\tikz[baseline=-0.5ex]{ \fill (-1.125ex,-0.649519ex) circle (1pt);
\fill (-0.375ex,0.649519ex) circle (1pt); \fill (0.375ex,-0.649519ex)
circle (1pt); \fill (1.125ex,0.649519ex) circle (1pt); \draw[ultra
thick] (-1.125ex,-0.649519ex)--(-0.375ex,0.649519ex); \draw[ultra
thick] (0.375ex,-0.649519ex)--(1.125ex,0.649519ex); \draw[thin]
(-1.125ex,-0.649519ex)--(-0.375ex,0.649519ex); \draw[thin]
(0.375ex,-0.649519ex)--(1.125ex,0.649519ex); } } \def\dimersVIIIket{
\tikz[baseline=-0.5ex]{ \fill (-1.125ex,-0.649519ex) circle (1pt);
\fill (0.375ex,-0.649519ex) circle (1pt); \fill (-0.375ex,0.649519ex)
circle (1pt); \fill (1.125ex,0.649519ex) circle (1pt); \draw[ultra
thick] (-1.125ex,-0.649519ex)--(0.375ex,-0.649519ex); \draw[ultra
thick] (-0.375ex,0.649519ex)--(1.125ex,0.649519ex); \draw[thin]
(-1.125ex,-0.649519ex)--(0.375ex,-0.649519ex); \draw[thin]
(-0.375ex,0.649519ex)--(1.125ex,0.649519ex); } }   \def\dimersXIket{
\tikz[baseline=-0.5ex]{ \fill (-1.96875ex,-1.13666ex) circle (1pt);
\fill (-0.46875ex,-1.13666ex) circle (1pt); \fill
(-1.21875ex,0.16238ex) circle (1pt); \fill (-0.46875ex,1.46142ex)
circle (1pt); \fill (0.28125ex,0.16238ex) circle (1pt); \fill
(1.03125ex,1.46142ex) circle (1pt); \fill (1.03125ex,-1.13666ex)
circle (1pt); \fill (1.78125ex,0.16238ex) circle (1pt); \draw[ultra
thick] (-1.96875ex,-1.13666ex)--(-1.21875ex,0.16238ex); \draw[ultra
thick] (-0.46875ex,-1.13666ex)--(1.03125ex,-1.13666ex); \draw[ultra
thick] (-0.46875ex,1.46142ex)--(0.28125ex,0.16238ex); \draw[ultra
thick] (1.03125ex,1.46142ex)--(1.78125ex,0.16238ex); \draw[thin]
(-1.96875ex,-1.13666ex)--(-0.46875ex,-1.13666ex); \draw[thin]
(-1.21875ex,0.16238ex)--(-0.46875ex,1.46142ex); \draw[thin]
(0.28125ex,0.16238ex)--(1.03125ex,1.46142ex); \draw[thin]
(1.03125ex,-1.13666ex)--(1.78125ex,0.16238ex); } } \def\dimersXIIket{
\tikz[baseline=-0.5ex]{ \fill (-1.96875ex,-1.13666ex) circle (1pt);
\fill (-1.21875ex,0.16238ex) circle (1pt); \fill
(-0.46875ex,-1.13666ex) circle (1pt); \fill (1.03125ex,-1.13666ex)
circle (1pt); \fill (-0.46875ex,1.46142ex) circle (1pt); \fill
(0.28125ex,0.16238ex) circle (1pt); \fill (1.03125ex,1.46142ex) circle
(1pt); \fill (1.78125ex,0.16238ex) circle (1pt); \draw[ultra thick]
(-1.96875ex,-1.13666ex)--(-0.46875ex,-1.13666ex); \draw[ultra thick]
(-1.21875ex,0.16238ex)--(-0.46875ex,1.46142ex); \draw[ultra thick]
(0.28125ex,0.16238ex)--(1.03125ex,1.46142ex); \draw[ultra thick]
(1.03125ex,-1.13666ex)--(1.78125ex,0.16238ex); \draw[thin]
(-1.96875ex,-1.13666ex)--(-1.21875ex,0.16238ex); \draw[thin]
(-0.46875ex,-1.13666ex)--(1.03125ex,-1.13666ex); \draw[thin]
(-0.46875ex,1.46142ex)--(0.28125ex,0.16238ex); \draw[thin]
(1.03125ex,1.46142ex)--(1.78125ex,0.16238ex); } } \def\dimersXIIIket{
\tikz[baseline=-0.5ex]{ \fill (-1.5ex,1.29904ex) circle (1pt); \fill
(0.ex,1.29904ex) circle (1pt); \fill (-1.5ex,-1.29904ex) circle (1pt);
\fill (-0.75ex,0.ex) circle (1pt); \fill (0.ex,-1.29904ex) circle
(1pt); \fill (1.5ex,-1.29904ex) circle (1pt); \fill (0.75ex,0.ex)
circle (1pt); \fill (1.5ex,1.29904ex) circle (1pt); \draw[ultra thick]
(-1.5ex,1.29904ex)--(-0.75ex,0.ex); \draw[ultra thick]
(-1.5ex,-1.29904ex)--(0.ex,-1.29904ex); \draw[ultra thick]
(0.ex,1.29904ex)--(1.5ex,1.29904ex); \draw[ultra thick]
(0.75ex,0.ex)--(1.5ex,-1.29904ex); \draw[thin]
(-1.5ex,1.29904ex)--(0.ex,1.29904ex); \draw[thin]
(-1.5ex,-1.29904ex)--(-0.75ex,0.ex); \draw[thin]
(0.ex,-1.29904ex)--(1.5ex,-1.29904ex); \draw[thin]
(0.75ex,0.ex)--(1.5ex,1.29904ex); } } \def\dimersXIVket{
\tikz[baseline=-0.5ex]{ \fill (-1.5ex,1.29904ex) circle (1pt); \fill
(-0.75ex,0.ex) circle (1pt); \fill (-1.5ex,-1.29904ex) circle (1pt);
\fill (0.ex,-1.29904ex) circle (1pt); \fill (0.ex,1.29904ex) circle
(1pt); \fill (1.5ex,1.29904ex) circle (1pt); \fill (0.75ex,0.ex)
circle (1pt); \fill (1.5ex,-1.29904ex) circle (1pt); \draw[ultra
thick] (-1.5ex,1.29904ex)--(0.ex,1.29904ex); \draw[ultra thick]
(-1.5ex,-1.29904ex)--(-0.75ex,0.ex); \draw[ultra thick]
(0.ex,-1.29904ex)--(1.5ex,-1.29904ex); \draw[ultra thick]
(0.75ex,0.ex)--(1.5ex,1.29904ex); \draw[thin]
(-1.5ex,1.29904ex)--(-0.75ex,0.ex); \draw[thin]
(-1.5ex,-1.29904ex)--(0.ex,-1.29904ex); \draw[thin]
(0.ex,1.29904ex)--(1.5ex,1.29904ex); \draw[thin]
(0.75ex,0.ex)--(1.5ex,-1.29904ex); } } \def\dimersXVket{
\tikz[baseline=-0.5ex]{ \fill (-2.34375ex,0.811899ex) circle (1pt);
\fill (-0.84375ex,0.811899ex) circle (1pt); \fill
(-1.59375ex,-0.487139ex) circle (1pt); \fill (-0.09375ex,-0.487139ex)
circle (1pt); \fill (0.65625ex,0.811899ex) circle (1pt); \fill
(2.15625ex,0.811899ex) circle (1pt); \fill (0.65625ex,-1.78618ex)
circle (1pt); \fill (1.40625ex,-0.487139ex) circle (1pt); \draw[ultra
thick] (-2.34375ex,0.811899ex)--(-1.59375ex,-0.487139ex); \draw[ultra
thick] (-0.84375ex,0.811899ex)--(0.65625ex,0.811899ex); \draw[ultra
thick] (-0.09375ex,-0.487139ex)--(0.65625ex,-1.78618ex); \draw[ultra
thick] (1.40625ex,-0.487139ex)--(2.15625ex,0.811899ex); \draw[thin]
(-2.34375ex,0.811899ex)--(-0.84375ex,0.811899ex); \draw[thin]
(-1.59375ex,-0.487139ex)--(-0.09375ex,-0.487139ex); \draw[thin]
(0.65625ex,0.811899ex)--(2.15625ex,0.811899ex); \draw[thin]
(0.65625ex,-1.78618ex)--(1.40625ex,-0.487139ex); } }
\def\dimersXVIket{ \tikz[baseline=-0.5ex]{ \fill
(-2.34375ex,0.811899ex) circle (1pt); \fill (-1.59375ex,-0.487139ex)
circle (1pt); \fill (-0.84375ex,0.811899ex) circle (1pt); \fill
(0.65625ex,0.811899ex) circle (1pt); \fill (-0.09375ex,-0.487139ex)
circle (1pt); \fill (0.65625ex,-1.78618ex) circle (1pt); \fill
(1.40625ex,-0.487139ex) circle (1pt); \fill (2.15625ex,0.811899ex)
circle (1pt); \draw[ultra thick]
(-2.34375ex,0.811899ex)--(-0.84375ex,0.811899ex); \draw[ultra thick]
(-1.59375ex,-0.487139ex)--(-0.09375ex,-0.487139ex); \draw[ultra thick]
(0.65625ex,0.811899ex)--(2.15625ex,0.811899ex); \draw[ultra thick]
(0.65625ex,-1.78618ex)--(1.40625ex,-0.487139ex); \draw[thin]
(-2.34375ex,0.811899ex)--(-1.59375ex,-0.487139ex); \draw[thin]
(-0.84375ex,0.811899ex)--(0.65625ex,0.811899ex); \draw[thin]
(-0.09375ex,-0.487139ex)--(0.65625ex,-1.78618ex); \draw[thin]
(1.40625ex,-0.487139ex)--(2.15625ex,0.811899ex); } }

       \def\dimersXXVket{
\tikz[baseline=-0.5ex]{ \fill (-2.34375ex,-1.13666ex) circle (1pt);
\fill (-0.84375ex,-1.13666ex) circle (1pt); \fill
(-1.59375ex,0.16238ex) circle (1pt); \fill (-0.09375ex,0.16238ex)
circle (1pt); \fill (0.65625ex,-1.13666ex) circle (1pt); \fill
(1.40625ex,0.16238ex) circle (1pt); \fill (0.65625ex,1.46142ex) circle
(1pt); \fill (2.15625ex,1.46142ex) circle (1pt); \draw[ultra thick]
(-2.34375ex,-1.13666ex)--(-1.59375ex,0.16238ex); \draw[ultra thick]
(-0.84375ex,-1.13666ex)--(0.65625ex,-1.13666ex); \draw[ultra thick]
(-0.09375ex,0.16238ex)--(0.65625ex,1.46142ex); \draw[ultra thick]
(1.40625ex,0.16238ex)--(2.15625ex,1.46142ex); \draw[thin]
(-2.34375ex,-1.13666ex)--(-0.84375ex,-1.13666ex); \draw[thin]
(-1.59375ex,0.16238ex)--(-0.09375ex,0.16238ex); \draw[thin]
(0.65625ex,-1.13666ex)--(1.40625ex,0.16238ex); \draw[thin]
(0.65625ex,1.46142ex)--(2.15625ex,1.46142ex); } } \def\dimersXXVIket{
\tikz[baseline=-0.5ex]{ \fill (-2.34375ex,-1.13666ex) circle (1pt);
\fill (-1.59375ex,0.16238ex) circle (1pt); \fill
(-0.84375ex,-1.13666ex) circle (1pt); \fill (0.65625ex,-1.13666ex)
circle (1pt); \fill (-0.09375ex,0.16238ex) circle (1pt); \fill
(0.65625ex,1.46142ex) circle (1pt); \fill (1.40625ex,0.16238ex) circle
(1pt); \fill (2.15625ex,1.46142ex) circle (1pt); \draw[ultra thick]
(-2.34375ex,-1.13666ex)--(-0.84375ex,-1.13666ex); \draw[ultra thick]
(-1.59375ex,0.16238ex)--(-0.09375ex,0.16238ex); \draw[ultra thick]
(0.65625ex,-1.13666ex)--(1.40625ex,0.16238ex); \draw[ultra thick]
(0.65625ex,1.46142ex)--(2.15625ex,1.46142ex); \draw[thin]
(-2.34375ex,-1.13666ex)--(-1.59375ex,0.16238ex); \draw[thin]
(-0.84375ex,-1.13666ex)--(0.65625ex,-1.13666ex); \draw[thin]
(-0.09375ex,0.16238ex)--(0.65625ex,1.46142ex); \draw[thin]
(1.40625ex,0.16238ex)--(2.15625ex,1.46142ex); } } \def\dimersXXVIIket{
\tikz[baseline=-0.5ex]{ \fill (-2.25ex,0.32476ex) circle (1pt); \fill
(-0.75ex,0.32476ex) circle (1pt); \fill (-1.5ex,-0.974279ex) circle
(1pt); \fill (0.ex,-0.974279ex) circle (1pt); \fill (0.ex,1.6238ex)
circle (1pt); \fill (0.75ex,0.32476ex) circle (1pt); \fill
(1.5ex,-0.974279ex) circle (1pt); \fill (2.25ex,0.32476ex) circle
(1pt); \draw[ultra thick] (-2.25ex,0.32476ex)--(-1.5ex,-0.974279ex);
\draw[ultra thick] (-0.75ex,0.32476ex)--(0.ex,1.6238ex); \draw[ultra
thick] (0.ex,-0.974279ex)--(1.5ex,-0.974279ex); \draw[ultra thick]
(0.75ex,0.32476ex)--(2.25ex,0.32476ex); \draw[thin]
(-2.25ex,0.32476ex)--(-0.75ex,0.32476ex); \draw[thin]
(-1.5ex,-0.974279ex)--(0.ex,-0.974279ex); \draw[thin]
(0.ex,1.6238ex)--(0.75ex,0.32476ex); \draw[thin]
(1.5ex,-0.974279ex)--(2.25ex,0.32476ex); } } \def\dimersXXVIIIket{
\tikz[baseline=-0.5ex]{ \fill (-2.25ex,0.32476ex) circle (1pt); \fill
(-1.5ex,-0.974279ex) circle (1pt); \fill (-0.75ex,0.32476ex) circle
(1pt); \fill (0.ex,1.6238ex) circle (1pt); \fill (0.ex,-0.974279ex)
circle (1pt); \fill (1.5ex,-0.974279ex) circle (1pt); \fill
(0.75ex,0.32476ex) circle (1pt); \fill (2.25ex,0.32476ex) circle
(1pt); \draw[ultra thick] (-2.25ex,0.32476ex)--(-0.75ex,0.32476ex);
\draw[ultra thick] (-1.5ex,-0.974279ex)--(0.ex,-0.974279ex);
\draw[ultra thick] (0.ex,1.6238ex)--(0.75ex,0.32476ex); \draw[ultra
thick] (1.5ex,-0.974279ex)--(2.25ex,0.32476ex); \draw[thin]
(-2.25ex,0.32476ex)--(-1.5ex,-0.974279ex); \draw[thin]
(-0.75ex,0.32476ex)--(0.ex,1.6238ex); \draw[thin]
(0.ex,-0.974279ex)--(1.5ex,-0.974279ex); \draw[thin]
(0.75ex,0.32476ex)--(2.25ex,0.32476ex); } } \def\dimersXXIXket{
\tikz[baseline=-0.5ex]{ \fill (-1.59375ex,-1.13666ex) circle (1pt);
\fill (-0.09375ex,-1.13666ex) circle (1pt); \fill
(-1.59375ex,1.46142ex) circle (1pt); \fill (-0.84375ex,0.16238ex)
circle (1pt); \fill (-0.09375ex,1.46142ex) circle (1pt); \fill
(0.65625ex,0.16238ex) circle (1pt); \fill (1.40625ex,-1.13666ex)
circle (1pt); \fill (2.15625ex,0.16238ex) circle (1pt); \draw[ultra
thick] (-1.59375ex,-1.13666ex)--(-0.84375ex,0.16238ex); \draw[ultra
thick] (-1.59375ex,1.46142ex)--(-0.09375ex,1.46142ex); \draw[ultra
thick] (-0.09375ex,-1.13666ex)--(1.40625ex,-1.13666ex); \draw[ultra
thick] (0.65625ex,0.16238ex)--(2.15625ex,0.16238ex); \draw[thin]
(-1.59375ex,-1.13666ex)--(-0.09375ex,-1.13666ex); \draw[thin]
(-1.59375ex,1.46142ex)--(-0.84375ex,0.16238ex); \draw[thin]
(-0.09375ex,1.46142ex)--(0.65625ex,0.16238ex); \draw[thin]
(1.40625ex,-1.13666ex)--(2.15625ex,0.16238ex); } } \def\dimersXXXket{
\tikz[baseline=-0.5ex]{ \fill (-1.59375ex,-1.13666ex) circle (1pt);
\fill (-0.84375ex,0.16238ex) circle (1pt); \fill
(-1.59375ex,1.46142ex) circle (1pt); \fill (-0.09375ex,1.46142ex)
circle (1pt); \fill (-0.09375ex,-1.13666ex) circle (1pt); \fill
(1.40625ex,-1.13666ex) circle (1pt); \fill (0.65625ex,0.16238ex)
circle (1pt); \fill (2.15625ex,0.16238ex) circle (1pt); \draw[ultra
thick] (-1.59375ex,-1.13666ex)--(-0.09375ex,-1.13666ex); \draw[ultra
thick] (-1.59375ex,1.46142ex)--(-0.84375ex,0.16238ex); \draw[ultra
thick] (-0.09375ex,1.46142ex)--(0.65625ex,0.16238ex); \draw[ultra
thick] (1.40625ex,-1.13666ex)--(2.15625ex,0.16238ex); \draw[thin]
(-1.59375ex,-1.13666ex)--(-0.84375ex,0.16238ex); \draw[thin]
(-1.59375ex,1.46142ex)--(-0.09375ex,1.46142ex); \draw[thin]
(-0.09375ex,-1.13666ex)--(1.40625ex,-1.13666ex); \draw[thin]
(0.65625ex,0.16238ex)--(2.15625ex,0.16238ex); } } \def\dimersXXXIket{
\tikz[baseline=-0.5ex]{ \fill (-2.625ex,-0.649519ex) circle (1pt);
\fill (-1.125ex,-0.649519ex) circle (1pt); \fill (-1.875ex,0.649519ex)
circle (1pt); \fill (-0.375ex,0.649519ex) circle (1pt); \fill
(0.375ex,-0.649519ex) circle (1pt); \fill (1.875ex,-0.649519ex) circle
(1pt); \fill (1.125ex,0.649519ex) circle (1pt); \fill
(2.625ex,0.649519ex) circle (1pt); \draw[ultra thick]
(-2.625ex,-0.649519ex)--(-1.875ex,0.649519ex); \draw[ultra thick]
(-1.125ex,-0.649519ex)--(0.375ex,-0.649519ex); \draw[ultra thick]
(-0.375ex,0.649519ex)--(1.125ex,0.649519ex); \draw[ultra thick]
(1.875ex,-0.649519ex)--(2.625ex,0.649519ex); \draw[thin]
(-2.625ex,-0.649519ex)--(-1.125ex,-0.649519ex); \draw[thin]
(-1.875ex,0.649519ex)--(-0.375ex,0.649519ex); \draw[thin]
(0.375ex,-0.649519ex)--(1.875ex,-0.649519ex); \draw[thin]
(1.125ex,0.649519ex)--(2.625ex,0.649519ex); } } \def\dimersXXXIIket{
\tikz[baseline=-0.5ex]{ \fill (-2.625ex,-0.649519ex) circle (1pt);
\fill (-1.875ex,0.649519ex) circle (1pt); \fill (-1.125ex,-0.649519ex)
circle (1pt); \fill (0.375ex,-0.649519ex) circle (1pt); \fill
(-0.375ex,0.649519ex) circle (1pt); \fill (1.125ex,0.649519ex) circle
(1pt); \fill (1.875ex,-0.649519ex) circle (1pt); \fill
(2.625ex,0.649519ex) circle (1pt); \draw[ultra thick]
(-2.625ex,-0.649519ex)--(-1.125ex,-0.649519ex); \draw[ultra thick]
(-1.875ex,0.649519ex)--(-0.375ex,0.649519ex); \draw[ultra thick]
(0.375ex,-0.649519ex)--(1.875ex,-0.649519ex); \draw[ultra thick]
(1.125ex,0.649519ex)--(2.625ex,0.649519ex); \draw[thin]
(-2.625ex,-0.649519ex)--(-1.875ex,0.649519ex); \draw[thin]
(-1.125ex,-0.649519ex)--(0.375ex,-0.649519ex); \draw[thin]
(-0.375ex,0.649519ex)--(1.125ex,0.649519ex); \draw[thin]
(1.875ex,-0.649519ex)--(2.625ex,0.649519ex); } } \def\dimersXXXIIIket{
\tikz[baseline=-0.5ex]{ \fill (-2.4375ex,0.32476ex) circle (1pt);
\fill (-0.9375ex,0.32476ex) circle (1pt); \fill
(-1.6875ex,-0.974279ex) circle (1pt); \fill (-0.1875ex,-0.974279ex)
circle (1pt); \fill (0.5625ex,0.32476ex) circle (1pt); \fill
(1.3125ex,1.6238ex) circle (1pt); \fill (1.3125ex,-0.974279ex) circle
(1pt); \fill (2.0625ex,0.32476ex) circle (1pt); \draw[ultra thick]
(-2.4375ex,0.32476ex)--(-1.6875ex,-0.974279ex); \draw[ultra thick]
(-0.9375ex,0.32476ex)--(0.5625ex,0.32476ex); \draw[ultra thick]
(-0.1875ex,-0.974279ex)--(1.3125ex,-0.974279ex); \draw[ultra thick]
(1.3125ex,1.6238ex)--(2.0625ex,0.32476ex); \draw[thin]
(-2.4375ex,0.32476ex)--(-0.9375ex,0.32476ex); \draw[thin]
(-1.6875ex,-0.974279ex)--(-0.1875ex,-0.974279ex); \draw[thin]
(0.5625ex,0.32476ex)--(1.3125ex,1.6238ex); \draw[thin]
(1.3125ex,-0.974279ex)--(2.0625ex,0.32476ex); } } \def\dimersXXXIVket{
\tikz[baseline=-0.5ex]{ \fill (-2.4375ex,0.32476ex) circle (1pt);
\fill (-1.6875ex,-0.974279ex) circle (1pt); \fill
(-0.9375ex,0.32476ex) circle (1pt); \fill (0.5625ex,0.32476ex) circle
(1pt); \fill (-0.1875ex,-0.974279ex) circle (1pt); \fill
(1.3125ex,-0.974279ex) circle (1pt); \fill (1.3125ex,1.6238ex) circle
(1pt); \fill (2.0625ex,0.32476ex) circle (1pt); \draw[ultra thick]
(-2.4375ex,0.32476ex)--(-0.9375ex,0.32476ex); \draw[ultra thick]
(-1.6875ex,-0.974279ex)--(-0.1875ex,-0.974279ex); \draw[ultra thick]
(0.5625ex,0.32476ex)--(1.3125ex,1.6238ex); \draw[ultra thick]
(1.3125ex,-0.974279ex)--(2.0625ex,0.32476ex); \draw[thin]
(-2.4375ex,0.32476ex)--(-1.6875ex,-0.974279ex); \draw[thin]
(-0.9375ex,0.32476ex)--(0.5625ex,0.32476ex); \draw[thin]
(-0.1875ex,-0.974279ex)--(1.3125ex,-0.974279ex); \draw[thin]
(1.3125ex,1.6238ex)--(2.0625ex,0.32476ex); } } \def\dimersXXXVket{
\tikz[baseline=-0.5ex]{ \fill (-1.125ex,-1.94856ex) circle (1pt);
\fill (0.375ex,-1.94856ex) circle (1pt); \fill (-1.125ex,0.649519ex)
circle (1pt); \fill (-0.375ex,-0.649519ex) circle (1pt); \fill
(-0.375ex,1.94856ex) circle (1pt); \fill (1.125ex,1.94856ex) circle
(1pt); \fill (0.375ex,0.649519ex) circle (1pt); \fill
(1.125ex,-0.649519ex) circle (1pt); \draw[ultra thick]
(-1.125ex,-1.94856ex)--(-0.375ex,-0.649519ex); \draw[ultra thick]
(-1.125ex,0.649519ex)--(-0.375ex,1.94856ex); \draw[ultra thick]
(0.375ex,-1.94856ex)--(1.125ex,-0.649519ex); \draw[ultra thick]
(0.375ex,0.649519ex)--(1.125ex,1.94856ex); \draw[thin]
(-1.125ex,-1.94856ex)--(0.375ex,-1.94856ex); \draw[thin]
(-1.125ex,0.649519ex)--(-0.375ex,-0.649519ex); \draw[thin]
(-0.375ex,1.94856ex)--(1.125ex,1.94856ex); \draw[thin]
(0.375ex,0.649519ex)--(1.125ex,-0.649519ex); } } \def\dimersXXXVIket{
\tikz[baseline=-0.5ex]{ \fill (-1.125ex,-1.94856ex) circle (1pt);
\fill (-0.375ex,-0.649519ex) circle (1pt); \fill (-1.125ex,0.649519ex)
circle (1pt); \fill (-0.375ex,1.94856ex) circle (1pt); \fill
(0.375ex,-1.94856ex) circle (1pt); \fill (1.125ex,-0.649519ex) circle
(1pt); \fill (0.375ex,0.649519ex) circle (1pt); \fill
(1.125ex,1.94856ex) circle (1pt); \draw[ultra thick]
(-1.125ex,-1.94856ex)--(0.375ex,-1.94856ex); \draw[ultra thick]
(-1.125ex,0.649519ex)--(-0.375ex,-0.649519ex); \draw[ultra thick]
(-0.375ex,1.94856ex)--(1.125ex,1.94856ex); \draw[ultra thick]
(0.375ex,0.649519ex)--(1.125ex,-0.649519ex); \draw[thin]
(-1.125ex,-1.94856ex)--(-0.375ex,-0.649519ex); \draw[thin]
(-1.125ex,0.649519ex)--(-0.375ex,1.94856ex); \draw[thin]
(0.375ex,-1.94856ex)--(1.125ex,-0.649519ex); \draw[thin]
(0.375ex,0.649519ex)--(1.125ex,1.94856ex); } } \def\dimersXXXVIIket{
\tikz[baseline=-0.5ex]{ \fill (-1.875ex,-0.649519ex) circle (1pt);
\fill (-1.125ex,0.649519ex) circle (1pt); \fill (-0.375ex,-0.649519ex)
circle (1pt); \fill (1.125ex,-0.649519ex) circle (1pt); \fill
(0.375ex,0.649519ex) circle (1pt); \fill (1.875ex,0.649519ex) circle
(1pt); \draw[ultra thick]
(-1.875ex,-0.649519ex)--(-1.125ex,0.649519ex); \draw[ultra thick]
(-0.375ex,-0.649519ex)--(0.375ex,0.649519ex); \draw[ultra thick]
(1.125ex,-0.649519ex)--(1.875ex,0.649519ex); \draw[thin]
(-1.875ex,-0.649519ex)--(-1.125ex,0.649519ex); \draw[thin]
(-0.375ex,-0.649519ex)--(1.125ex,-0.649519ex); \draw[thin]
(0.375ex,0.649519ex)--(1.875ex,0.649519ex); } } \def\dimersXXXVIIIket{
\tikz[baseline=-0.5ex]{ \fill (-1.875ex,-0.649519ex) circle (1pt);
\fill (-0.375ex,-0.649519ex) circle (1pt); \fill (-1.125ex,0.649519ex)
circle (1pt); \fill (0.375ex,0.649519ex) circle (1pt); \fill
(1.125ex,-0.649519ex) circle (1pt); \fill (1.875ex,0.649519ex) circle
(1pt); \draw[ultra thick]
(-1.875ex,-0.649519ex)--(-1.125ex,0.649519ex); \draw[ultra thick]
(-0.375ex,-0.649519ex)--(0.375ex,0.649519ex); \draw[ultra thick]
(1.125ex,-0.649519ex)--(1.875ex,0.649519ex); \draw[thin]
(-1.875ex,-0.649519ex)--(-0.375ex,-0.649519ex); \draw[thin]
(-1.125ex,0.649519ex)--(0.375ex,0.649519ex); \draw[thin]
(1.125ex,-0.649519ex)--(1.875ex,0.649519ex); } } \def\dimersXXXIXket{
\tikz[baseline=-0.5ex]{ \fill (-1.875ex,-0.649519ex) circle (1pt);
\fill (-1.125ex,0.649519ex) circle (1pt); \fill (-0.375ex,-0.649519ex)
circle (1pt); \fill (0.375ex,0.649519ex) circle (1pt); \fill
(1.125ex,-0.649519ex) circle (1pt); \fill (1.875ex,0.649519ex) circle
(1pt); \draw[ultra thick]
(-1.875ex,-0.649519ex)--(-1.125ex,0.649519ex); \draw[ultra thick]
(-0.375ex,-0.649519ex)--(1.125ex,-0.649519ex); \draw[ultra thick]
(0.375ex,0.649519ex)--(1.875ex,0.649519ex); \draw[thin]
(-1.875ex,-0.649519ex)--(-1.125ex,0.649519ex); \draw[thin]
(-0.375ex,-0.649519ex)--(0.375ex,0.649519ex); \draw[thin]
(1.125ex,-0.649519ex)--(1.875ex,0.649519ex); } } \def\dimersXLket{
\tikz[baseline=-0.5ex]{ \fill (-1.875ex,-0.649519ex) circle (1pt);
\fill (-1.125ex,0.649519ex) circle (1pt); \fill (-0.375ex,-0.649519ex)
circle (1pt); \fill (0.375ex,0.649519ex) circle (1pt); \fill
(1.125ex,-0.649519ex) circle (1pt); \fill (1.875ex,0.649519ex) circle
(1pt); \draw[ultra thick]
(-1.875ex,-0.649519ex)--(-0.375ex,-0.649519ex); \draw[ultra thick]
(-1.125ex,0.649519ex)--(0.375ex,0.649519ex); \draw[ultra thick]
(1.125ex,-0.649519ex)--(1.875ex,0.649519ex); \draw[thin]
(-1.875ex,-0.649519ex)--(-1.125ex,0.649519ex); \draw[thin]
(-0.375ex,-0.649519ex)--(0.375ex,0.649519ex); \draw[thin]
(1.125ex,-0.649519ex)--(1.875ex,0.649519ex); } } \def\dimersXLIket{
\tikz[baseline=-0.5ex]{ \fill (-1.ex,-1.29904ex) circle (1pt); \fill
(0.5ex,-1.29904ex) circle (1pt); \fill (-1.ex,1.29904ex) circle (1pt);
\fill (0.5ex,1.29904ex) circle (1pt); \fill (-0.25ex,0.ex) circle
(1pt); \fill (1.25ex,0.ex) circle (1pt); \draw[ultra thick]
(-1.ex,-1.29904ex)--(-0.25ex,0.ex); \draw[ultra thick]
(-1.ex,1.29904ex)--(0.5ex,1.29904ex); \draw[ultra thick]
(0.5ex,-1.29904ex)--(1.25ex,0.ex); \draw[thin]
(-1.ex,-1.29904ex)--(0.5ex,-1.29904ex); \draw[thin]
(-1.ex,1.29904ex)--(0.5ex,1.29904ex); \draw[thin]
(-0.25ex,0.ex)--(1.25ex,0.ex); } } \def\dimersXLIIket{
\tikz[baseline=-0.5ex]{ \fill (-1.ex,-1.29904ex) circle (1pt); \fill
(0.5ex,-1.29904ex) circle (1pt); \fill (-1.ex,1.29904ex) circle (1pt);
\fill (0.5ex,1.29904ex) circle (1pt); \fill (-0.25ex,0.ex) circle
(1pt); \fill (1.25ex,0.ex) circle (1pt); \draw[ultra thick]
(-1.ex,-1.29904ex)--(0.5ex,-1.29904ex); \draw[ultra thick]
(-1.ex,1.29904ex)--(-0.25ex,0.ex); \draw[ultra thick]
(0.5ex,1.29904ex)--(1.25ex,0.ex); \draw[thin]
(-1.ex,-1.29904ex)--(0.5ex,-1.29904ex); \draw[thin]
(-1.ex,1.29904ex)--(0.5ex,1.29904ex); \draw[thin]
(-0.25ex,0.ex)--(1.25ex,0.ex); } } \def\dimersXLIIIket{
\tikz[baseline=-0.5ex]{ \fill (-1.ex,-1.29904ex) circle (1pt); \fill
(-0.25ex,0.ex) circle (1pt); \fill (-1.ex,1.29904ex) circle (1pt);
\fill (0.5ex,1.29904ex) circle (1pt); \fill (0.5ex,-1.29904ex) circle
(1pt); \fill (1.25ex,0.ex) circle (1pt); \draw[ultra thick]
(-1.ex,-1.29904ex)--(0.5ex,-1.29904ex); \draw[ultra thick]
(-1.ex,1.29904ex)--(0.5ex,1.29904ex); \draw[ultra thick]
(-0.25ex,0.ex)--(1.25ex,0.ex); \draw[thin]
(-1.ex,-1.29904ex)--(-0.25ex,0.ex); \draw[thin]
(-1.ex,1.29904ex)--(0.5ex,1.29904ex); \draw[thin]
(0.5ex,-1.29904ex)--(1.25ex,0.ex); } } \def\dimersXLIVket{
\tikz[baseline=-0.5ex]{ \fill (-1.ex,-1.29904ex) circle (1pt); \fill
(0.5ex,-1.29904ex) circle (1pt); \fill (-1.ex,1.29904ex) circle (1pt);
\fill (-0.25ex,0.ex) circle (1pt); \fill (0.5ex,1.29904ex) circle
(1pt); \fill (1.25ex,0.ex) circle (1pt); \draw[ultra thick]
(-1.ex,-1.29904ex)--(0.5ex,-1.29904ex); \draw[ultra thick]
(-1.ex,1.29904ex)--(0.5ex,1.29904ex); \draw[ultra thick]
(-0.25ex,0.ex)--(1.25ex,0.ex); \draw[thin]
(-1.ex,-1.29904ex)--(0.5ex,-1.29904ex); \draw[thin]
(-1.ex,1.29904ex)--(-0.25ex,0.ex); \draw[thin]
(0.5ex,1.29904ex)--(1.25ex,0.ex); } } \def\dimersXLVket{
\tikz[baseline=-0.5ex]{ \fill (-1.5ex,0.866025ex) circle (1pt); \fill
(0.ex,0.866025ex) circle (1pt); \fill (-0.75ex,-0.433013ex) circle
(1pt); \fill (0.ex,-1.73205ex) circle (1pt); \fill
(0.75ex,-0.433013ex) circle (1pt); \fill (1.5ex,0.866025ex) circle
(1pt); \draw[ultra thick] (-1.5ex,0.866025ex)--(-0.75ex,-0.433013ex);
\draw[ultra thick] (0.ex,0.866025ex)--(1.5ex,0.866025ex); \draw[ultra
thick] (0.ex,-1.73205ex)--(0.75ex,-0.433013ex); \draw[thin]
(-1.5ex,0.866025ex)--(0.ex,0.866025ex); \draw[thin]
(-0.75ex,-0.433013ex)--(0.ex,-1.73205ex); \draw[thin]
(0.75ex,-0.433013ex)--(1.5ex,0.866025ex); } } \def\dimersXLVIket{
\tikz[baseline=-0.5ex]{ \fill (-1.5ex,0.866025ex) circle (1pt); \fill
(-0.75ex,-0.433013ex) circle (1pt); \fill (0.ex,0.866025ex) circle
(1pt); \fill (1.5ex,0.866025ex) circle (1pt); \fill (0.ex,-1.73205ex)
circle (1pt); \fill (0.75ex,-0.433013ex) circle (1pt); \draw[ultra
thick] (-1.5ex,0.866025ex)--(0.ex,0.866025ex); \draw[ultra thick]
(-0.75ex,-0.433013ex)--(0.ex,-1.73205ex); \draw[ultra thick]
(0.75ex,-0.433013ex)--(1.5ex,0.866025ex); \draw[thin]
(-1.5ex,0.866025ex)--(-0.75ex,-0.433013ex); \draw[thin]
(0.ex,0.866025ex)--(1.5ex,0.866025ex); \draw[thin]
(0.ex,-1.73205ex)--(0.75ex,-0.433013ex); } } \def\dimersXLVIIket{
\tikz[baseline=-0.5ex]{ \fill (-1.96875ex,-1.13666ex) circle (1pt);
\fill (-0.46875ex,-1.13666ex) circle (1pt); \fill
(-1.21875ex,0.16238ex) circle (1pt); \fill (0.28125ex,0.16238ex)
circle (1pt); \fill (-0.46875ex,1.46142ex) circle (1pt); \fill
(1.03125ex,1.46142ex) circle (1pt); \fill (1.03125ex,-1.13666ex)
circle (1pt); \fill (1.78125ex,0.16238ex) circle (1pt); \draw[ultra
thick] (-1.96875ex,-1.13666ex)--(-1.21875ex,0.16238ex); \draw[ultra
thick] (-0.46875ex,-1.13666ex)--(1.03125ex,-1.13666ex); \draw[ultra
thick] (-0.46875ex,1.46142ex)--(0.28125ex,0.16238ex); \draw[ultra
thick] (1.03125ex,1.46142ex)--(1.78125ex,0.16238ex); \draw[thin]
(-1.96875ex,-1.13666ex)--(-0.46875ex,-1.13666ex); \draw[thin]
(-1.21875ex,0.16238ex)--(0.28125ex,0.16238ex); \draw[thin]
(-0.46875ex,1.46142ex)--(1.03125ex,1.46142ex); \draw[thin]
(1.03125ex,-1.13666ex)--(1.78125ex,0.16238ex); } }
\def\dimersXLVIIIket{ \tikz[baseline=-0.5ex]{ \fill
(-1.96875ex,-1.13666ex) circle (1pt); \fill (-0.46875ex,-1.13666ex)
circle (1pt); \fill (-1.21875ex,0.16238ex) circle (1pt); \fill
(-0.46875ex,1.46142ex) circle (1pt); \fill (0.28125ex,0.16238ex)
circle (1pt); \fill (1.03125ex,1.46142ex) circle (1pt); \fill
(1.03125ex,-1.13666ex) circle (1pt); \fill (1.78125ex,0.16238ex)
circle (1pt); \draw[ultra thick]
(-1.96875ex,-1.13666ex)--(-1.21875ex,0.16238ex); \draw[ultra thick]
(-0.46875ex,-1.13666ex)--(1.03125ex,-1.13666ex); \draw[ultra thick]
(-0.46875ex,1.46142ex)--(1.03125ex,1.46142ex); \draw[ultra thick]
(0.28125ex,0.16238ex)--(1.78125ex,0.16238ex); \draw[thin]
(-1.96875ex,-1.13666ex)--(-0.46875ex,-1.13666ex); \draw[thin]
(-1.21875ex,0.16238ex)--(-0.46875ex,1.46142ex); \draw[thin]
(0.28125ex,0.16238ex)--(1.03125ex,1.46142ex); \draw[thin]
(1.03125ex,-1.13666ex)--(1.78125ex,0.16238ex); } } \def\dimersXLIXket{
\tikz[baseline=-0.5ex]{ \fill (-1.96875ex,-1.13666ex) circle (1pt);
\fill (-1.21875ex,0.16238ex) circle (1pt); \fill
(-0.46875ex,-1.13666ex) circle (1pt); \fill (1.03125ex,-1.13666ex)
circle (1pt); \fill (-0.46875ex,1.46142ex) circle (1pt); \fill
(1.03125ex,1.46142ex) circle (1pt); \fill (0.28125ex,0.16238ex) circle
(1pt); \fill (1.78125ex,0.16238ex) circle (1pt); \draw[ultra thick]
(-1.96875ex,-1.13666ex)--(-0.46875ex,-1.13666ex); \draw[ultra thick]
(-1.21875ex,0.16238ex)--(-0.46875ex,1.46142ex); \draw[ultra thick]
(0.28125ex,0.16238ex)--(1.03125ex,1.46142ex); \draw[ultra thick]
(1.03125ex,-1.13666ex)--(1.78125ex,0.16238ex); \draw[thin]
(-1.96875ex,-1.13666ex)--(-1.21875ex,0.16238ex); \draw[thin]
(-0.46875ex,-1.13666ex)--(1.03125ex,-1.13666ex); \draw[thin]
(-0.46875ex,1.46142ex)--(1.03125ex,1.46142ex); \draw[thin]
(0.28125ex,0.16238ex)--(1.78125ex,0.16238ex); } } \def\dimersLket{
\tikz[baseline=-0.5ex]{ \fill (-1.96875ex,-1.13666ex) circle (1pt);
\fill (-1.21875ex,0.16238ex) circle (1pt); \fill
(-0.46875ex,-1.13666ex) circle (1pt); \fill (1.03125ex,-1.13666ex)
circle (1pt); \fill (-0.46875ex,1.46142ex) circle (1pt); \fill
(0.28125ex,0.16238ex) circle (1pt); \fill (1.03125ex,1.46142ex) circle
(1pt); \fill (1.78125ex,0.16238ex) circle (1pt); \draw[ultra thick]
(-1.96875ex,-1.13666ex)--(-0.46875ex,-1.13666ex); \draw[ultra thick]
(-1.21875ex,0.16238ex)--(0.28125ex,0.16238ex); \draw[ultra thick]
(-0.46875ex,1.46142ex)--(1.03125ex,1.46142ex); \draw[ultra thick]
(1.03125ex,-1.13666ex)--(1.78125ex,0.16238ex); \draw[thin]
(-1.96875ex,-1.13666ex)--(-1.21875ex,0.16238ex); \draw[thin]
(-0.46875ex,-1.13666ex)--(1.03125ex,-1.13666ex); \draw[thin]
(-0.46875ex,1.46142ex)--(0.28125ex,0.16238ex); \draw[thin]
(1.03125ex,1.46142ex)--(1.78125ex,0.16238ex); } } \def\dimersLIket{
\tikz[baseline=-0.5ex]{ \fill (-2.25ex,0.ex) circle (1pt); \fill
(-1.5ex,1.29904ex) circle (1pt); \fill (-1.5ex,-1.29904ex) circle
(1pt); \fill (0.ex,-1.29904ex) circle (1pt); \fill (0.ex,1.29904ex)
circle (1pt); \fill (1.5ex,1.29904ex) circle (1pt); \fill
(1.5ex,-1.29904ex) circle (1pt); \fill (2.25ex,0.ex) circle (1pt);
\draw[ultra thick] (-2.25ex,0.ex)--(-1.5ex,-1.29904ex); \draw[ultra
thick] (-1.5ex,1.29904ex)--(0.ex,1.29904ex); \draw[ultra thick]
(0.ex,-1.29904ex)--(1.5ex,-1.29904ex); \draw[ultra thick]
(1.5ex,1.29904ex)--(2.25ex,0.ex); \draw[thin]
(-2.25ex,0.ex)--(-1.5ex,1.29904ex); \draw[thin]
(-1.5ex,-1.29904ex)--(0.ex,-1.29904ex); \draw[thin]
(0.ex,1.29904ex)--(1.5ex,1.29904ex); \draw[thin]
(1.5ex,-1.29904ex)--(2.25ex,0.ex); } } \def\dimersLIIket{
\tikz[baseline=-0.5ex]{ \fill (-2.25ex,0.ex) circle (1pt); \fill
(-1.5ex,-1.29904ex) circle (1pt); \fill (-1.5ex,1.29904ex) circle
(1pt); \fill (0.ex,1.29904ex) circle (1pt); \fill (0.ex,-1.29904ex)
circle (1pt); \fill (1.5ex,-1.29904ex) circle (1pt); \fill
(1.5ex,1.29904ex) circle (1pt); \fill (2.25ex,0.ex) circle (1pt);
\draw[ultra thick] (-2.25ex,0.ex)--(-1.5ex,1.29904ex); \draw[ultra
thick] (-1.5ex,-1.29904ex)--(0.ex,-1.29904ex); \draw[ultra thick]
(0.ex,1.29904ex)--(1.5ex,1.29904ex); \draw[ultra thick]
(1.5ex,-1.29904ex)--(2.25ex,0.ex); \draw[thin]
(-2.25ex,0.ex)--(-1.5ex,-1.29904ex); \draw[thin]
(-1.5ex,1.29904ex)--(0.ex,1.29904ex); \draw[thin]
(0.ex,-1.29904ex)--(1.5ex,-1.29904ex); \draw[thin]
(1.5ex,1.29904ex)--(2.25ex,0.ex); } }

\def\dimerseqcoef{-t\Big(\left|\dimersIket\right\rangle\left\langle\dimersIbra\right|
+ \left|\dimersIIket\right\rangle\left\langle\dimersIIbra\right|\Big)
+ v \Big(\left|\dimersVIIket\right\rangle\left\langle\dimersVIIbra\right|
+ \left|\dimersVIIIket\right\rangle\left\langle\dimersVIIIbra\right|\Big) \nonumber \\
&-&t_{6,a}\Big(\left|\dimersIIIket\right\rangle\left\langle\dimersIIIbra\right|
+ \left|\dimersIVket\right\rangle\left\langle\dimersIVbra\right| +
\left|\dimersVket\right\rangle\left\langle\dimersVbra\right| +
\left|\dimersVIket\right\rangle\left\langle\dimersVIbra\right|\Big) \nonumber \\
&-&t_{6,b} \Big(\left|\dimersXLVket\right\rangle\left\langle\dimersXLVbra\right|
+ \left|\dimersXLVIket\right\rangle\left\langle\dimersXLVIbra\right|\Big)\nonumber \\
&+&u\Big(\left|\dimersXXXVIIket\right\rangle\left\langle\dimersXXXVIIbra\right|
+ \left|\dimersXXXVIIIket\right\rangle\left\langle\dimersXXXVIIIbra\right|
+ \left|\dimersXXXIXket\right\rangle\left\langle\dimersXXXIXbra\right|
+ \left|\dimersXLket\right\rangle\left\langle\dimersXLbra\right| \nonumber \\
&& \hspace{1.0in}+
\left|\dimersXLIket\right\rangle\left\langle\dimersXLIbra\right| +
\left|\dimersXLIIket\right\rangle\left\langle\dimersXLIIbra\right| +
\left|\dimersXLIIIket\right\rangle\left\langle\dimersXLIIIbra\right| +
\left|\dimersXLIVket\right\rangle\left\langle\dimersXLIVbra\right|\Big)\nonumber \\
&-&t_{8,a}\Big(\left|\dimersLIket\right\rangle\left\langle\dimersLIbra\right|
+\left|\dimersLIIket\right\rangle\left\langle\dimersLIIbra\right|\Big)
-t_{8,b}\Big(\left|\dimersXIket\right\rangle\left\langle\dimersXIbra\right| +
\left|\dimersXIIket\right\rangle\left\langle\dimersXIIbra\right|\Big)\nonumber\\
&-&t_{8,c}\Big(\left|\dimersXIIIket\right\rangle\left\langle\dimersXIIIbra\right|
+ \left|\dimersXIVket\right\rangle\left\langle\dimersXIVbra\right|\Big)
-t_{8,d}\Big(\left|\dimersXVket\right\rangle\left\langle\dimersXVbra\right| +
\left|\dimersXVIket\right\rangle\left\langle\dimersXVIbra\right|\Big) \nonumber \\
&-&t_{8,e}\Big(\left|\dimersXXVket\right\rangle\left\langle\dimersXXVbra\right|
+ \left|\dimersXXVIket\right\rangle\left\langle\dimersXXVIbra\right| +
\left|\dimersXXVIIket\right\rangle\left\langle\dimersXXVIIbra\right| +
\left|\dimersXXVIIIket\right\rangle\left\langle\dimersXXVIIIbra\right| \nonumber \\
&& \hspace{1.0in}
+ \left|\dimersXXIXket\right\rangle\left\langle\dimersXXIXbra\right| +
\left|\dimersXXXket\right\rangle\left\langle\dimersXXXbra\right|\Big)\nonumber \\
&-&t_{8,f}\Big(\left|\dimersXXXIket\right\rangle\left\langle\dimersXXXIbra\right|
+ \left|\dimersXXXIIket\right\rangle\left\langle\dimersXXXIIbra\right|
+ \left|\dimersXXXIIIket\right\rangle\left\langle\dimersXXXIIIbra\right|
+ \left|\dimersXXXIVket\right\rangle\left\langle\dimersXXXIVbra\right|\nonumber \\
&& \hspace{1.0in}
+ \left|\dimersXXXVket\right\rangle\left\langle\dimersXXXVbra\right| +
\left|\dimersXXXVIket\right\rangle\left\langle\dimersXXXVIbra\right|\Big)\nonumber \\
&-&t_{8,g}\Big(\left|\dimersXLVIIket\right\rangle\left\langle\dimersXLVIIbra\right|
+ \left|\dimersXLVIIIket\right\rangle\left\langle\dimersXLVIIIbra\right|
+ \left|\dimersXLIXket\right\rangle\left\langle\dimersXLIXbra\right| +
\left|\dimersLket\right\rangle\left\langle\dimersLbra\right|\Big)}

\begin{document}

\begin{center}
{\Large \textbf{Dimer description of the SU(4) antiferromagnet on the triangular lattice}}
\end{center}

\begin{center}
Anna Keselman\textsuperscript{1*,2},
Lucile Savary\textsuperscript{3,1},
Leon Balents\textsuperscript{1,4}
\end{center}

\begin{center}
{\bf 1} Kavli Institute for Theoretical Physics, University of California, Santa Barbara, CA 93106-4030
\\
{\bf 2} Station Q, Microsoft Corporation, Santa Barbara, California 93106-6105, USA
\\
{\bf 3} Universit\'e de Lyon, \'{E}cole Normale Sup\'{e}rieure de Lyon, Universit\'e Claude Bernard Lyon I, CNRS, Laboratoire de physique, 46, all\'{e}e d'Italie, 69007 Lyon
\\
{\bf 4} Canadian Institute for Advanced Research, Toronto, Ontario, Canada\\
* akeselman@kitp.ucsb.edu
\end{center}

\begin{center}
\today
\end{center}

% For convenience during refereeing: line numbers
%\linenumbers

\section*{Abstract}
{\bf
In systems with many local degrees of freedom, high-symmetry points in the phase diagram can provide an important starting point for the investigation of their properties throughout the phase diagram. 
In systems with both spin and orbital (or valley) degrees of freedom such a starting point gives rise to SU(4)-symmetric models.
Here we consider SU(4)-symmetric ``spin'' models, corresponding to Mott phases at half-filling, i.e.\ the six-dimensional representation of SU(4). 
This may be relevant to twisted multilayer graphene.
In particular, we study the SU(4) antiferromagnetic ``Heisenberg'' model on the triangular lattice, both in the classical limit and in the quantum regime. 
Carrying out a numerical study using the density matrix renormalization group (DMRG), we argue that the ground state is non-magnetic.
We then derive a dimer expansion of the SU(4) spin model.   An exact diagonalization (ED) study of the effective dimer model suggests that the ground state breaks translation invariance, forming a valence bond solid (VBS) with a 12-site unit cell.
Finally, we consider the effect of SU(4)-symmetry breaking interactions due to Hund's coupling, and argue for a possible phase transition between a VBS and a magnetically ordered state.
}

\vspace{10pt}
\noindent\rule{\textwidth}{1pt}
\tableofcontents\thispagestyle{fancy}
\noindent\rule{\textwidth}{1pt}
\vspace{10pt}

\section{Introduction}
\label{sec:intro}

Frustrated quantum antiferromagnets may possess non-magnetic ground
states, avoiding spin order through short or long range entanglement
of spins.  In the late 1980s and early 1990s, a dominant approach to
this physics was based on generalizing the SU(2) group of spin
rotations to SU(N) or Sp(2N)
\cite{read1991large,read1989valence,chung2001large,PhysRevB.42.2526}.
In the limit $N\rightarrow \infty$, models with such enlarged symmetry
may be solved exactly by a fully symmetric saddle point of a path
integral representation of the partition function, and consequently
possess non-magnetic ground states.  More recently, it has become
possible to study models with SU(N) symmetry for finite $N$ using
computational methods.
In particular, models describing SU(N) fermions at half filling were shown to host a variety of non-magnetic states
\cite{harada2003neel,Assaad2005,lou2009antiferromagnetic,kaul2011quantum,Lang2013,kim2019},
indicating that features of the $N\rightarrow\infty$ solutions survive to $N$ of order one.

Interest in models of this type has also been stimulated by their
possible experimental realization in cold
atoms \cite{cazalilla2014ultracold,gorshkov2010two}, in materials with
orbital degeneracy \cite{kugel1973crystal,li19984},
and, more recently, in moir\'e superlattices with valley degeneracy \cite{Cao2018,yankowitz2018tuning,Chen2019,xu2018topological}.
Here we investigate in particular the SU(4) antiferromagnet in the
self-conjugate six dimensional representation.  This is the
representation corresponding to two electrons distributed amongst four
degenerate spin/orbital states on each site. This representation thus
occurs naturally in systems with spin and a two-fold orbital or valley
degeneracy.  Like the familiar S=1/2 SU(2) spins, pairs of spins in
this representation may form a singlet ``valence bond'', so that there
is a natural ``dimer'' picture upon which non-magnetic states may be
based.  

Indeed, it has been shown by Rokhsar \cite{PhysRevB.42.2526} that at
$N=\infty$, dimerized  states (i.e. products of singlet bonds) are the
ground states in the self-conjugate representation for
a very wide class of lattices and exchange interactions (including
almost all those of physical interest).  For $N$ large but finite, it
is therefore expected that a quantum dimer
model \cite{PhysRevB.81.214413,PhysRevB.80.184427,Ralko2005,Moessner2001,PhysRevLett.61.2376},
which describes the projection of the Hamiltonian to the singlet subspace,
should capture the physics of the problem.  A dimer model consists of
a Hilbert space in which dimers, which stand in for singlets, realize a
covering of the lattice, with each site covered by one,
and only one dimer. Such models are known to predict quantum spin liquids and various
valence bond solid orders, depending on the exact model,
dimensionality, and lattice.  Notably, the simplest quantum dimer model on the
triangular lattice was argued to possess a $\mathbb{Z}_2$ spin liquid ground
state \cite{Moessner2001}.

With these motivations, we study the aforementioned SU(4) model on the
triangular lattice both analytically and numerically.  In contrast to
the previous numerical works mentioned above, which focused on
bipartite lattices, the triangular lattice considered here does not to
our knowledge admit a sign-free Monte Carlo approach for generic
parameters (though for a specific choice of parameters a sign-free
algorithm exists\cite{PhysRevLett.115.157202}).  Hence we attack the
problem differently, using the density matrix renormalization group
(DMRG), exact diagonalization, and an analytic dimer expansion.
First, we describe a reformulation which takes advantage of the fact
that SU(4) is a double cover of SO(6), through a mapping of the
self-conjugate representation of SU(4) to the vector representation of
SO(6), which we define.  We determine the classical ground states of
the model, and find that they generalize the three-sublattice coplanar
states of the S=1/2 Heisenberg model on the triangular lattice.
However, carrying out a numerical study of the model using the DMRG
method, we argue that this order is absent in the quantum limit.  We
note that this result is in agreement with a recent pseudo-fermion
functional renormalization group study \cite{kiese2019} of the same
model.  We then extend the ``overlap expansion'' approach --
originally developed by Rokhsar and Kivelson
\cite{PhysRevLett.61.2376} to obtain the quantum dimer model
Hamiltonian from the SU(2) spin Hamiltonian -- to derive an analytic
overlap expansion for the quantum dimer model relevant to the SU(4)
case.  We show that the parameter for the overlap expansion is $x=1/6$
in the SU(4) case, which should be compared to $x=1/2$ for SU(2)
spins: hence the expansion is expected to be much more accurate for
the present problem.  Our exact diagonalization (ED) study of the
effective dimer model suggests that the ground state is a twelve-site
valence bond solid (VBS), although larger system sizes are required to
conclusively rule out a proximate quantum spin liquid state.  We
remark that the related model of Ref.\cite{PhysRevLett.115.157202} 
shows such VBS order when generalized from SO(6) to SO(N) with N$>$12,
which provides some further indications for this VBS state in this family
of model Hamiltonians.  

In addition, we study a generalization of the model which breaks SU(4)
symmetry down to SU(2)$\times$SU(2), by including an additional
``atomic'' Hund's coupling $J_H$ on the sites.  In the limit of large
$J_H$, the model reduces to the SU(2) symmetric spin S=1 Heisenberg
Hamiltonian, and the ground state has long-range three sublattice
order.  The generalized model thus exhibits a quantum phase transition
from a paramagnetic to antiferromagnetically ordered state at zero
temperature by varying $J_H$.  We present signs for this transition in numerics.

\section{Model}

\subsection{From Hubbard to Heisenberg}
\label{sec:from-hubb-heis}

In the limit of strong interactions, electrons localize and
the appropriate description of their physics becomes that of their
spins, localized at lattice sites. Relevant models may be derived from
Hubbard models. Here we proceed with this approach and consider two
electrons per site hopping on the triangular lattice with hopping
parameter $t$ and subjected to an on-site Hubbard interaction $U$ as well as a Hund's coupling $J_{H}$, which tends to enforce an alignment of the spin degrees of
freedom and thus breaks SU(4) symmetry. More precisely, we consider the following Hamiltonian:
\begin{equation}
  \label{eq:46}
  H=-t\sum_{\langle ij\rangle}\sum_{a={\sf 1}}^{\sf 4}(c_{ia}^\dagger
  c_{ja}^{\vphantom{\dagger}}+{\rm h.c.})+U\sum_in_i(n_i-2)-J_{H}\sum_i\left(c_{i}^\dagger
  \boldsymbol{\sigma}c_{i}^{\vphantom{\dagger}}\right)^2,
\end{equation}
where $c_{i,a}^\dagger,c_{i,a}^{\vphantom{\dagger}}$ are flavor $a$ electron
creation and annihilation operators at site $i$,
$n_i=\sum_{a={\sf 1}}^{\sf 4}c_{i,a}^\dagger c_{i,a}^{\vphantom{\dagger}}$.  In
the final term, we used a notation based on the physical origin of the
four electronic states from the spin-1/2 of the electron $s^z=\pm 1/2$ and a two-fold
orbital degeneracy $\tau^z=\pm 1/2$.  We may associate $a={\sf
  1,2,3,4}$ (note the use of sans-serif font for this purpose) to the states
with $(2s^z,2\tau^z) = (1,1),(-1,1),(1,-1),(-1,-1)$, 
respectively. 
The  $\sigma^\mu=\sigma^\mu \otimes \mathbb{Id}_2$ ($\mu=x,y,z$) Pauli
matrices act in spin space. 
We specialize to the case of filling $n_i=2$, and carry out the
standard degenerate perturbation theory in $t/U$ to derive
an effective ``spin'' model.  A basis for the states with
$n_i=2$ on a single site is:
\begin{align}
  \label{eq:54}
  |1\rangle & =|{\sf 12}\rangle, \qquad  |2\rangle  =|{\sf 13}\rangle, \qquad
                                               |3\rangle =|{\sf 14}\rangle,
                                                           \nonumber \\
  |4\rangle &=|{\sf 23}\rangle, \qquad  |5\rangle =|{\sf 24}\rangle,
              \qquad  |6\rangle =|{\sf 34}\rangle,
\end{align}
where, on the right-hand-side of the equations,
$|ab\rangle=c_a^\dagger c_b^{\dagger}|0\rangle$, where $|0\rangle$ is
the vacuum. 

We start by analyzing the SU(4) symmetric model, i.e. we take $J_{H}=0$, and return to the effects of a finite $J_H$ later in Sec.~\ref{sec:probing-mag-order}. 
At second order in small $t/U$,
Eq.~\eqref{eq:46} becomes
\begin{equation}
  \label{eq:1}
  \hat{H}=J\sum_{\langle ij\rangle}\left( \sum_{a,b={\sf 1}}^{\sf 4}
  \hat{T}_i^{ab}\hat{T}_j^{ba}- \hat{\mathrm{Id}}_6\right),
\end{equation}
with $J\sim t^2/U$ and  $\hat{T}^{ab}={\rm
  P}_6c^\dagger_ac^{\vphantom{\dagger}}_b{\rm P}_6$, with ${\rm P}_6$  the
projection operator onto the six-dimensional vector space defined
above.   The $\hat{T}^{ab}$ are 6$\times$6 matrices which are related
to the generators of SU(4).  In Eq.~\eqref{eq:1} we extracted a
constant which sets the zero of energy at a convenient value.  To see
this, we bring out the analogy to SU(2) spins by extracting the trace from
the $\hat{T}^{ab}$ matrices:
\begin{equation}
  \label{eq:30}
  \tilde{T}^{ab} = \hat{T}^{ab} - \frac{1}{2} \delta^{ab}\hat{\mathrm{Id}}_6.
\end{equation}
With this definition ${\rm Tr}\, \tilde{T}^{ab}=0$ (note that the trace here is over
the 6-dimensional SU(4) space). The Hamiltonian in Eq.~\eqref{eq:1} can now be written as
\begin{equation}
  \label{eq:31}
  \hat{H}=J\sum_{\langle ij\rangle} \sum_{a,b={\sf 1}}^{\sf 4}
  \tilde{T}_i^{ab}\tilde{T}_j^{ba}.
\end{equation}
One can also check that $\sum_a \tilde{T}^{aa} = 0$, so that there are
clearly only 15 such independent SU(4) matrices, which comprise a
basis for the generators of SU(4) in the 6-dimensional representation.

Now, regardless of the precise microscopic Hamiltonian $H$, we may
consider a spin model, determined on the basis of, and constrained by
symmetry. To proceed to the derivation of the most general SU(4)
model, it is useful to make use of the following.

\subsection{Map to the vector representation of SO(6)}
\label{sec:map-vect-repr}

SU(4) is a double
cover of SO(6) and there exists a convenient map (which is faithful) from the
six-dimensional representation of SU(4) to the
fundamental (vector) representation of SO(6) \cite{Wang2009}. By using the following basis,
where each basis state transforms under the vector representation of
SO(6), i.e.\ $O:|\hat{n}\rangle\mapsto O|\hat{n}\rangle$, where $O$ is
an SO(6) matrix,
\begin{align}
  \label{eq:55}
|\hat{1}\rangle& =\frac{1}{\sqrt{2}}\left(|2\rangle-|5\rangle\right),\qquad 
|\hat{2}\rangle =\frac{-i}{\sqrt{2}}\left(|2\rangle+|5\rangle\right), \nonumber\\
|\hat{3}\rangle& =\frac{1}{\sqrt{2}}\left(|3\rangle+|4\rangle\right),\qquad
|\hat{4}\rangle=\frac{-i}{\sqrt{2}}\left(|3\rangle-|4\rangle\right), \nonumber\\
|\hat{5}\rangle& =\frac{1}{\sqrt{2}}\left(|1\rangle+|6\rangle\right), \qquad
|\hat{6}\rangle=\frac{-i}{\sqrt{2}}\left(|1\rangle-|6\rangle\right), 
\end{align}
it is straightforward to write all the SU(4) invariant two-site operators:
\begin{eqnarray}
  \label{eq:56}
  \hat{{\rm Id}}_{ij}&=&\sum_{n,m=1}^6\left(|\hat{n}\rangle\langle
                \hat{n}|\right)_i\left(|\hat{m}\rangle\langle\hat{m}|\right)_j
                         ,\\
\label{eq:56b}
6\hat{P}_{ij}=\hat{Q}_{ij}&=&\sum_{n,m=1}^6\left(|\hat{n}\rangle\langle
           \hat{m}|\right)_i\left(|\hat{n}\rangle\langle\hat{m}|\right)_j ,\\
\hat{\Pi}_{ij}&=&\sum_{n,m=1}^6\left(|\hat{n}\rangle\langle
            \hat{m}|\right)_i\left(|\hat{m}\rangle\langle\hat{n}|\right)_j .
\end{eqnarray}
Here $\hat{P}_{ij}$ is the singlet projector over sites $ij$, where a
(normalized) singlet over sites $ij$ is written
\begin{equation}
  \label{eq:3}
  |s\rangle_{ij}=\frac{1}{\sqrt{6}}\sum_{n=1}^6|\hat{n}\hat{n}\rangle_{ij},
\end{equation}
while $\hat{\Pi}_{ij}$ is
the permutation operator over sites $ij$ and $\hat{{\rm Id}}_{ij}$ is the
identity. Then the general SO(6) invariant Hamiltonian with nearest-neighbor
interactions is a sum of these terms:
\begin{equation}
  \label{eq:57}
  \hat{H}_{\rm gen}=\sum_{\langle ij\rangle} \left(\alpha
    \hat{Q}_{ij}+\beta\hat{\Pi}_{ij}+\gamma\hat{{\rm Id}}_{ij}\right),
\end{equation}
for $\alpha,\beta,\gamma\in\mathbb{R}$. The ``Heisenberg'' model
Eq.~\eqref{eq:1} is realized for $-\alpha=\beta=J$, $\gamma=0$. 
One can readily check then that for two sites with the
SU(4) (or SO(6)) singlet in Eq.~\eqref{eq:3},
\begin{equation}
  \label{eq:32}
  \hat{H}_{ij} |s\rangle_{ij} = J\left( -\hat{Q}_{ij} +
    \hat{\Pi}_{ij}\right) |s\rangle_{ij} = - 5J |s\rangle_{ij} .
\end{equation}

In this SO(6) basis, we may also define the symmetric $\hat{\mathcal{S}}^{mn}$ and
antisymmetric operators $\hat{\mathcal{A}}^{mn}$, as well as the
Hermitian (and still traceless),
versions of the latter, $\hat{A}^{mn}=i\hat{\mathcal{A}}^{mn}$
\begin{eqnarray}
  \label{eq:7} \hat{\mathcal{S}}^{mn}&=&\frac{1}{\sqrt{2}}\left(|\hat{m}\rangle\langle\hat{n}|+|\hat{n}\rangle\langle\hat{m}|\right),\\
\hat{\mathcal{A}}^{mn}&=&\frac{1}{\sqrt{2}}\left(|\hat{m}\rangle\langle\hat{n}|-|\hat{n}\rangle\langle\hat{m}|\right).\\
\hat{A}^{mn}&=&i \hat{\mathcal{A}}^{mn}.
\end{eqnarray}
The $\hat{A}^{mn}$ operators can be considered as the generators of SO(6), and their square, ${\rm
  Tr}[\hat{\mathbf{A}}\cdot\hat{\mathbf{A}}^T]=5\hat{{\rm Id}}_6$ is
the quadratic Casimir operator, up to a normalization constant. Here
we have {\em defined} the matrix of operators $\hat{\mathbf{A}}$
such that $(\hat{\mathbf{A}})_{mn}=\hat{A}^{mn}$.  Using
these operators, the ``Heisenberg'' Hamiltonian $\hat{H}$ becomes
\begin{equation}
  \label{eq:9}
  \hat{H}=J\sum_{\langle
    ij\rangle}\sum_{m,n=1}^6\hat{A}_i^{mn}\hat{A}_j^{mn}=J\sum_{\langle
  ij\rangle}{\rm Tr}\,\hat{\mathbf{A}}^{\vphantom{T}}_i\cdot\hat{\mathbf{A}}^T_j,
\end{equation}
where the trace, $\cdot$ and transpose operations act on the
superscripts of the $\hat{\mathbf{A}}_l$ matrices of $\hat{A}_l^{mn}$ operators.

\section{Magnetic order}

In this section, we first examine the classical ground states of the
SU(4) model, which are ``magnetically'' ordered, i.e.\ they break the SU(4)
symmetry and have a non-zero expectation value of the ``spin''
operator matrix $\tilde{T}_i^{ab}$ or $\hat{A}_i^{mn}$ on each site.
Having identified the {\em type} of magnetic order which is most
favored, we next describe numerical studies which search for it.  We
find that this magnetic order is in fact absent, and that the ground
state appears to lack any form of SU(4) symmetry breaking, i.e.\ is
non-magnetic.

\subsection{Classical limit}

In order to look for a product ground
state we first ask about the definition of the classical limit
of SU(4) (SO(6)) spins.  Like for SU(2) spins, we should replace, in the
Hamiltonian, each of the fifteen SO(6)  generators
$\hat{A}^{mn}$, which are 6$\times$6 matrices, by a single
classical number.

To do so, we interpret the classical limit as a variational problem in the subspace
of states consisting of direct products of single-site wavefunctions.
Within any such state, the expectation value of any product of
$\hat{A}^{mn}$ is replaced by a product of expectation values of each
$\hat{A}^{mn}$, which are c-numbers, as desired.  A general
single-site wavefunction is given by
$|\psi\rangle_i=\sum_{p=1}^6v^i_p|\hat{p}\rangle_i$, where
$\mathbf{v}^i$ is a complex six-dimensional unit
vector, i.e. $\sum_p |v_p^i|^2=1$, so that $|\psi\rangle_i$ is
normalized.  Upon going to the classical limit,
\begin{equation}
  \label{eq:19}
  \hat{A}^{mn}\rightarrow\mathsf{A}_{mn}=\langle\psi|\hat{A}^{mn}|\psi\rangle=\sqrt{2}{\rm Im}[v_m^{\vphantom{*}} v^*_n].
\end{equation}
The matrix $\mathsf{A}$ is now an antisymmetric $6\times6$ matrix of scalar matrix
elements $\mathsf{A}_{mn}$, and the Heisenberg Hamiltonian becomes
\begin{equation}
  \label{eq:20}
  \hat{H}=J\sum_{\langle ij\rangle} \sum_{m,n=1}^6\hat{A}_i^{mn}
  \hat{A}^{mn}_j\rightarrow J\sum_{\langle ij\rangle}
  \sum_{m,n=1}^6\mathsf{A}^i_{mn} \mathsf{A}^j_{mn}=J\sum_{\langle
    ij\rangle} {\rm Tr}\, \mathsf{A}^i (\mathsf{A}^j)^T.
\end{equation}
Note that, while each {\em operator} $\hat{A}^{mn}$ for fixed $m,n$ is Hermitian, the
matrix $\mathsf{A}$ is real and antisymmetric. Moreover, while
$\sum_{m,n=1}^6\hat{A}^{mn}\hat{A}^{mn}=5\hat{{\rm Id}}_6$, one can
show that $0\leq{\rm Tr}\mathsf{A}\mathsf{A}^T\leq 1$ (see
Appendix~\ref{app:classical}). In solving a classical SO(6) in this representation model, one
should find matrices $\mathsf{A}$ which verify the above constraints
(much like SU(2) S=1/2 (resp.\ S=1) classical spins are described by a
three-dimensional vector with unit norm $|\mathbf{S}|=1$ (resp.\ with $0\leq|\mathbf{S}|\leq1$).

\subsection{Product variational states}
\label{sec:prod-vari-stat}

We now specialize to the triangular lattice and nearest-neighbor
``Heisenberg'' Hamiltonian, and look for the ground state of the
corresponding classical model. The SU(2)-invariant spin-1/2 model on the triangular lattice is one of
the best-studied models of frustrated magnetism. Its ground states are
the so-called $120^\circ$-ordered states. They triple the unit cell and
each elementary triangular unit is such that the spins point at 120
degrees of one another. 

We may rewrite the classical version of the Heisenberg Hamiltonian
in Eq.~\eqref{eq:20}, as
\begin{equation}
  \label{eq:21}
  \mathsf{H}=
  \frac{J}{4} \sum_{t\;triangle} \left[ {\rm Tr}\, \left(\sum_{i \in
        t} \mathsf{A}_i\right)\left(\sum_{i \in
        t} \mathsf{A}_i\right)^T - {\rm Tr} \left( \sum_{i\in t}
      \mathsf{A}^{\vphantom{T}}_i \mathsf{A}_i^T\right)\right],
\end{equation}
where the sum runs over all unit triangles of the triangular
lattice. The energy is clearly minimized when the first term in the
square brackets vanishes and the second one is maximized.  
The magnitude of the second term is maximized when the upper bound on ${\rm Tr}\mathsf{A}_i \mathsf{A}_i^T$ is saturated for all $i\in t$.
As shown in Appendix~\ref{app:classical}, the upper bound is saturated 
when the complex vector $\mathbf{v}$ describing the single-site wavefunction is given by 
$\mathbf{v}=(\mathbf{x}+i\mathbf{y})/\sqrt{2}$, with
$\mathbf{x},\mathbf{y}$ real, six-dimensional orthogonal unit
vectors, i.e.\ $|\mathbf{x}|=|\mathbf{y}|=1$ and $\mathbf{x}\cdot\mathbf{y}=0$.  The first term vanishes for three-sublattice states that satisfy
$\mathsf{A}_{1}+\mathsf{A}_{2}+\mathsf{A}_{3}=0_6$.

To minimize the first term in the square brackets in
Eq.~\eqref{eq:21}, and simultaneously maximize the magnitude of the second term, we can choose
\begin{equation}
  \label{eq:23}
  \mathbf{v}_{l=1,2,3}=\mathbf{v}\left(\tfrac{2\pi l}{3}\right),\qquad\mbox{with}\qquad
\mathbf{v}(\theta)=\frac{1}{\sqrt{2}}\left(\mathbf{x}+i(\cos\theta\mathbf{y}+\sin\theta\mathbf{z})\right),
\end{equation}
corresponding to
\begin{equation}
  \label{eq:24} 
    \mathbf{A}_{l=1,2,3}=\mathbf{A}\left(\tfrac{2\pi l}{3}\right),\quad\mbox{with}\quad
    \mathsf{A}(\theta)=\frac{1}{\sqrt{2}}\left[(\cos\theta\mathbf{y}+\sin\theta\mathbf{z})\mathbf{x}^T-\mathbf{x}(\cos\theta\mathbf{y}+\sin\theta\mathbf{z})^T\right],
\end{equation}
where  $\mathbf{x},\mathbf{y},\mathbf{z}$ are three orthonormal unit vectors such
that $\mathbf{x}\cdot\mathbf{y}=\mathbf{x}\cdot\mathbf{z}=\mathbf{y}\cdot\mathbf{z}=0$. 

For $\mathbf{x},\mathbf{y},\mathbf{z}$ along each of the first three basis
vectors of $\mathbb{R}^6$, we get for example
$\mathsf{A}(\theta)=\cos\theta \mathcal{A}^{21}+\sin\theta \mathcal{A}^{31}$, 
where
$(\mathcal{A}^{\mu\nu})_{pq}=\frac{1}{\sqrt{2}}(\delta_{\mu
  p}\delta_{\nu q}-\delta_{\mu q}\delta_{\nu p})$.  Note that once the
spins on two nearest-neighbor sites are fixed, the remainder are fully
determined by the condition
$\mathsf{A}_{1}+\mathsf{A}_{2}+\mathsf{A}_{3}=0_6$, which can be
successively applied to the spins on triangles sharing two of the sites which
have already been fixed, to cover the entire lattice.  This implies
that all classical ground states are of the three-sublattice type.

\subsection{Numerical analysis using DMRG}
\label{sec:DMRG}

To probe the presence of magnetic order in the system we study the model numerically, using DMRG~\cite{White1992,Schollwoeck2005}. 
To this end, we consider finite cylinders in a geometry that allows for the formation of a $120^\circ$-ordered state. 
Denoting the basis vectors of the triangular lattice by
$\vec{a}_1=(1,0)$, $\vec{a}_2=(1/2,\sqrt{3}/2)$, we consider cylinders
such that the sites of the lattice modulo $\vec{R}=N_y \vec{a}_2$ are
identified, and $N_y$ is a multiple of three. This geometry is
depicted in Fig.~\ref{fig:dmrgMagneticOrder}(c,d). Due to the large
single-site Hilbert space dimension in this problem we are limited to
narrow cylinders with $N_y=3$ and $N_y=6$, and we only study very
short cylinders for the latter. We note that a different geometry,
namely one in which lattice sites modulo $\vec{R}'=N_y \left(
  \vec{a}_2-\vec{a}_1/2 \right)$ are identified, is also compatible
with a $120^\circ$-ordered state for $N_y=4$. However, in this case,
we find indications that the system behaves as a quasi-1D system with
localized modes at the ends of the cylinder. We thus leave out these
results from the discussion of the 2D limit presented here.

In the following, we first discuss the flavor gap in the system, and show that it remains finite, suggesting the absence of a low-energy Goldstone mode that would be expected if the system formed a magnetically ordered state.
We then probe the presence of long range order in the ground state by looking at the static response of the system to polarizing fields applied at its boundary. We show that the expectation value of the magnetization decays rapidly away from the boundary in the presence of SU(4) symmetry, implying a lack of long range order.

Our DMRG simulations were performed using the ITensor library~\cite{ITensor}. 

\subsubsection{Flavor gap}

We calculate the flavor gap only for cylinders of width $N_y=3$, as extracting the gap requires finite length scaling, and we are limited to very short systems for cylinders of width $N_y=6$, as mentioned above. Note that for $N_y=3$, the length of the system $N_x$ has to be even to allow for an $SU(4)$-singlet ground state.

Similarly to the conservation of $s^z$ which is often used when
studying SU(2) spins, we can employ the conservation of three U(1)
quantum numbers for the SU(4) case: $t_3 \equiv n_{\sf 1}-n_{\sf 2}$, $t_8 \equiv n_{\sf 1}+n_{\sf 2}-2n_{\sf 3}$, and $t_{15} \equiv n_{\sf 1}+n_{\sf 2}+n_{\sf 3}-3n_{\sf 4}$, where $n_{a={\sf 1,..,4}}$ denote the occupations of the four flavors as before.
We calculate the gap of a $\Delta_{t_3}=2$ excitation. To this end, we first obtain the ground state, which we expect to be an $SU(4)$ singlet, and hence lie in the $\left(t_3,t_8,t_{15}\right)=\left(0,0,0\right)$ sector, and then calculate the lowest energy state in the $\left(t_3,t_8,t_{15}\right)=\left(2,0,0\right)$ sector. The latter state belongs to the 15 dimensional irreducible representation of SU(4), as can be verified by calculating the quadratic Casimir operator $\sum_{a,b=\sf{1}}^{\sf{4}}\tilde{T}^{ab}\tilde{T}^{ba}$. The resulting energy gap is plotted in Fig.~\ref{fig:dmrgMagneticOrder}(a) as function of inverse system length. 
Even though we present data for relatively short systems, it is clear that the gap remains finite in the infinite system size limit, and we can estimate it to be larger than $2.5J$.
The maximal bond dimension in our simulations was $M=4000$, resulting in a truncation error of $10^{-5}$ ($10^{-4}$) for the largest system size in the $t_3=0$ ($t_3=2$) sector.

\subsubsection{Probing long range magnetic order in the ground state}\label{sec:probing-mag-order}

\begin{figure}
\begin{subfigure}[t]{.5\linewidth}
\includegraphics[width=\linewidth]{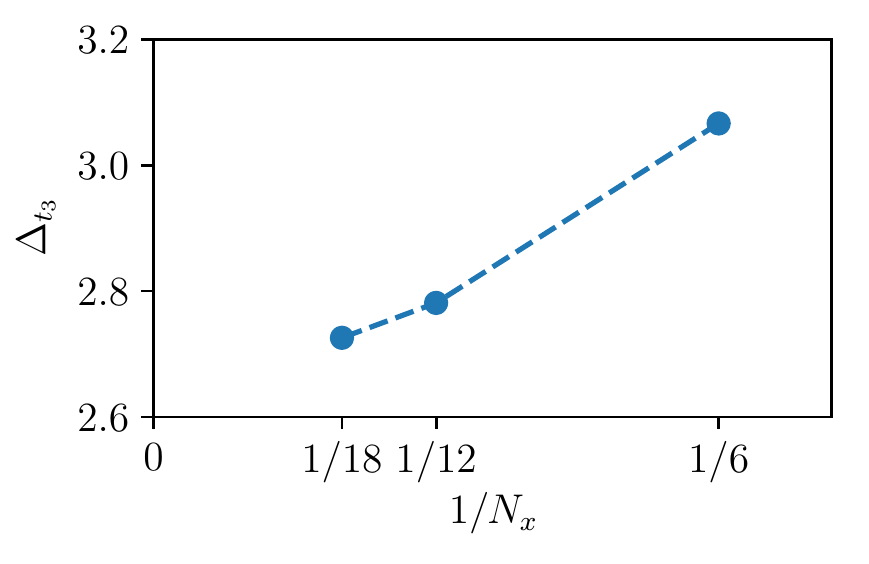}
\captionsetup{skip=-2in}
\caption{}
\end{subfigure}
\begin{subfigure}[t]{.5\linewidth}
\includegraphics[width=\linewidth]{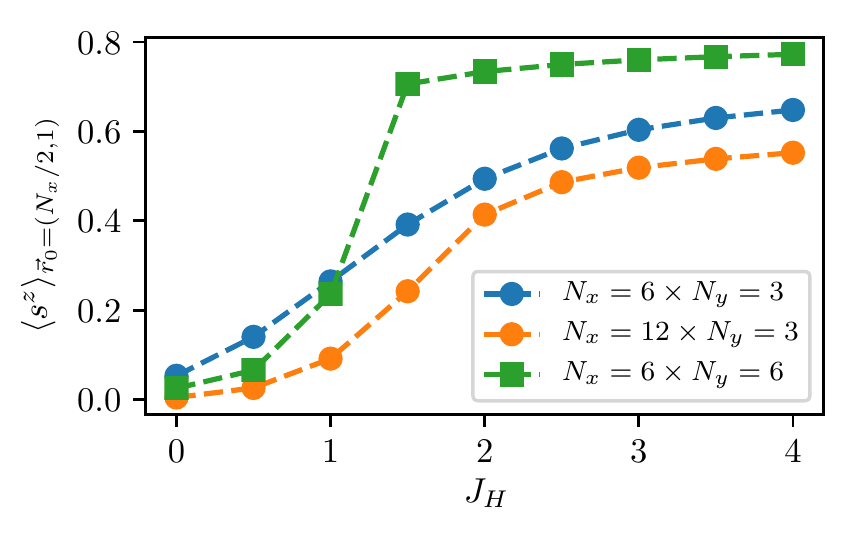}
\captionsetup{skip=-2in}
\caption{}
\end{subfigure} \\
\begin{subfigure}[t]{.5\linewidth}
\includegraphics[width=\linewidth]{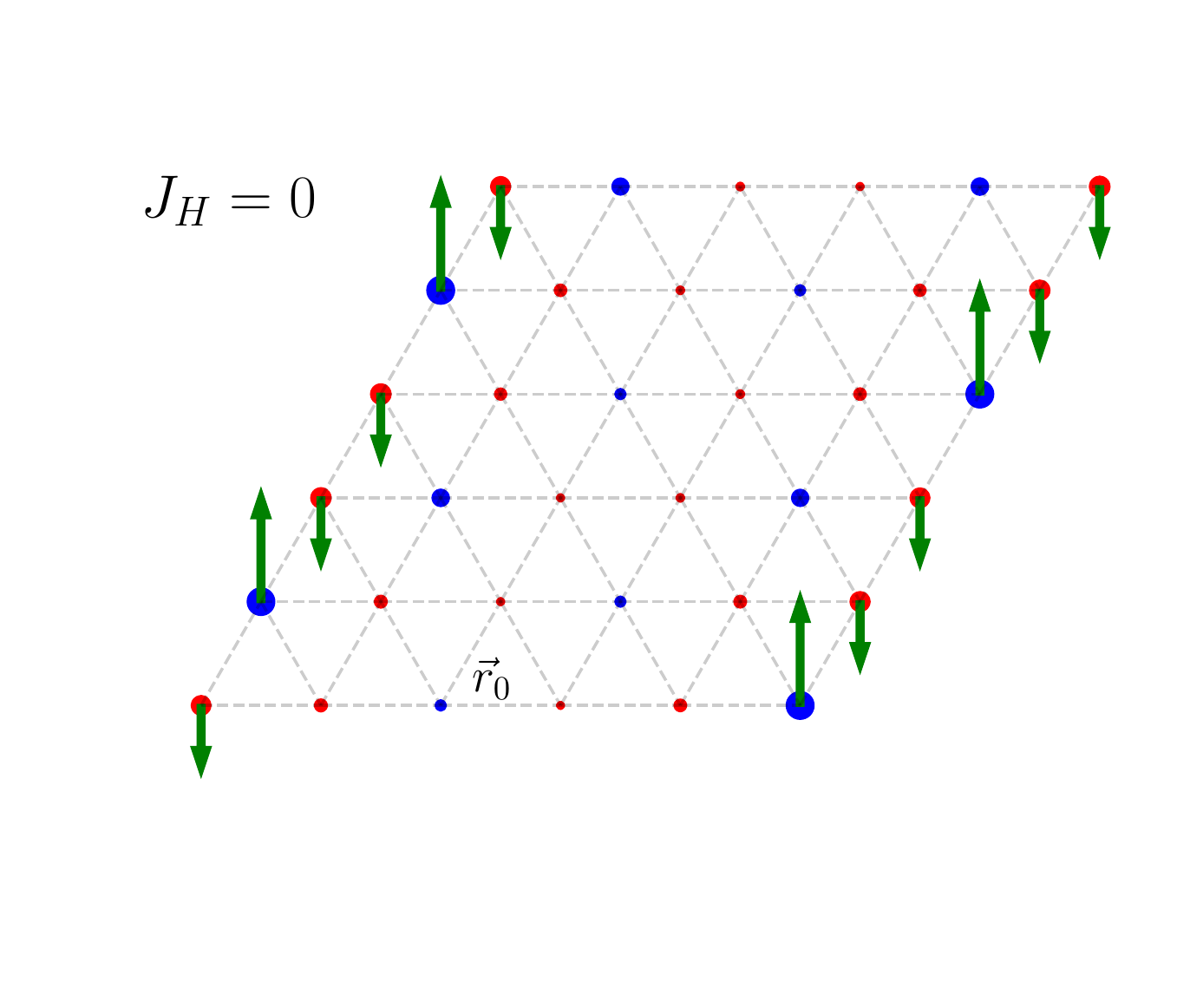}
\captionsetup{skip=-2.3in}
\caption{}
\end{subfigure}
\begin{subfigure}[t]{.5\linewidth}
\includegraphics[width=\linewidth]{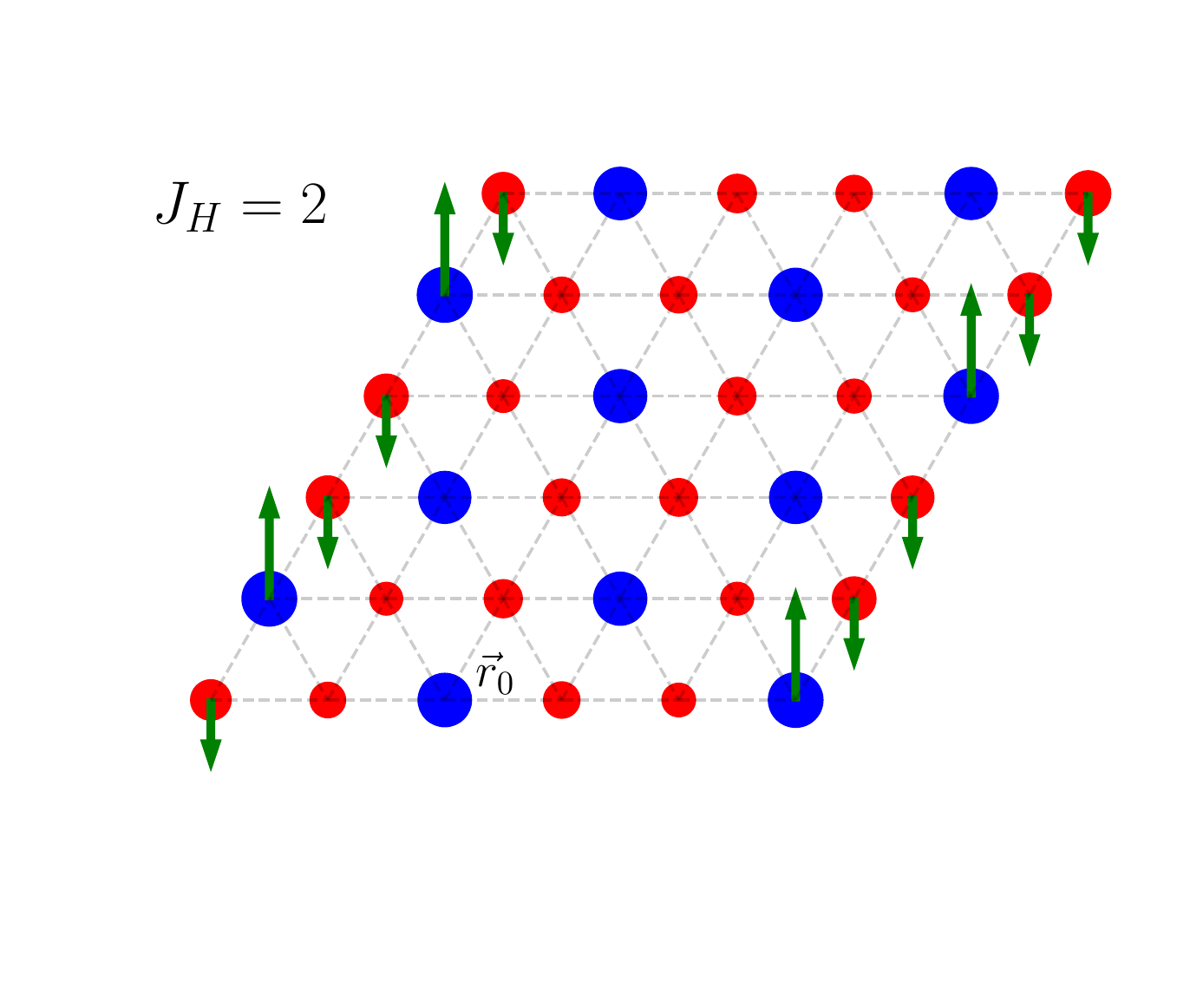}
\captionsetup{skip=-2.3in}
\caption{}
\end{subfigure}
\caption{Here we set $J=1$. (a) Flavor gap, $\Delta_{t_3}$ for $J_H=0$, as function of inverse system size, obtained for cylinders of width $N_y=3$. (b) Magnetization on a site in the middle of the system (at position $\vec{r}_0=(x_0, y_0)$ with $x_0=N_x/2, y_0=1$), as function of the Hund's coupling $J_H$ for different system sizes. Pinning fields are applied along the $z$ axis, on the sites at the boundaries of the cylinder as indicated by the green arrows in (c) and (d), where the local magnetization in a $N_x=6 \times N_y=6$ cylinder is shown for $J_H=0$ and $J_H=2$ respectively. The area of the circles is proportional to the expectation value of $s^z$ with blue (red) corresponding to a positive (negative) value.}
 \label{fig:dmrgMagneticOrder}
\end{figure}

For the analysis of long range magnetic order in the ground state, it
is instructive to consider the effect of $SU(4)$ symmetry breaking by a Hund's coupling term as introduced in Eq.~\eqref{eq:46}.
More specifically, the Hamiltonian we consider is
\begin{equation}
\label{eq:H_hund}
\hat{H} = J \sum_{\left<i,j\right>}\sum_{a,b} \tilde{T}^{ab}_i \tilde{T}^{ba}_j  - J_H \sum_i \hat{\mathbf{S}}_i^2.
\end{equation}
As was mentioned previously, a finite $J_H>0$ breaks the $SU(4)$ symmetry down to $SU(2)\times SU(2)$, pairing the two electrons on each site into a spin-triplet state. 
Using the definitions Eq.~(\ref{eq:54},\ref{eq:55}), the projection on the on-site spin-triplet and spin-singlet subspaces is given by
\begin{equation}
  \label{eq:singlet-triplet}
  \hat{\mathcal{P}}_{i,S=1} = \sum_{n=1}^3 \left(|\hat{n}\rangle\langle\hat{n}|\right)_i, \qquad
  \hat{\mathcal{P}}_{i,S=0} = \sum_{n=4}^6 \left(|\hat{n}\rangle\langle\hat{n}|\right)_i,
\end{equation}
so that the Hund's coupling term can be written simply as $\hat{\mathbf{S}}_i^2 = \sqrt{2}\hat{\mathcal{P}}_{i,S=1}$.

In the large $J_H$ limit, 
the spin model Eq.~\eqref{eq:H_hund} reduces to an SU(2) spin-1 Heisenberg model, $H=J\sum_{\langle ij\rangle}\mathbf{S}_i\cdot\mathbf{S}_j$, where $S_i^\mu$ ($\mu=x,y,z$) are $S=1$ operators (see Appendix~\ref{app:LargeJH} for further details). The latter is known to form a $120^\circ$-ordered state on the triangular lattice~\cite{GOTZE2016333}. Below, we study the model in Eq.~\eqref{eq:H_hund} as $J_H$ is increased from $J_H=0$ (the SU(4) symmetric point), where a three-sublattice order is predicted by our classical analysis, to a large $J_H \gg J$, where the $120^\circ$ order is known to form also in the quantum limit.

Once $SU(4)$ symmetry is broken down to SU(2)$\times$SU(2), only two U(1) quantum numbers are conserved: the $z$ components of the spin and valley degrees of freedom, namely $2s^{z}=n_{\sf 1}+n_{\sf 3}-n_{\sf 2}-n_{\sf 4}$ and $2\tau^{z}=n_{\sf 1}+n_{\sf 2}-n_{\sf 3}-n_{\sf 4}$. Employing the conservation of these two quantum numbers, we now look for the ground state in the sector $\left(s^z, \tau^z \right)=\left(0,0\right)$.
Once again, we consider finite cylinders of geometry and size compatible with the 3-sublattice order of the 120$^\circ$ state. In particular, we consider cylinders of width $N_y=3$ and $N_y=6$ in the same geometry as before.

To facilitate the formation of a long range ordered state, we follow the approach introduced in Ref.~\cite{White2007} and apply pinning fields at the boundaries of the cylinder. We then calculate the expectation value of the spin component parallel to the field in the bulk, far from the boundary for different ratios of the length of the cylinder to its circumference. A complementary analysis, where we calculate the spin-spin correlations in the absence of pinning fields is presented in Appendix~\ref{app:DMRG} and gives similar results.
To retain the conservation of $s^z$, we apply the pinning fields only
along the $s^z$ axis. More specifically, the field applied is $-s^z$
on the A sublattice, and $+s^z/2$ on the B and C sublattices as depicted in Fig.~\ref{fig:dmrgMagneticOrder}(c,d). Note that in the SU(4)-symmetric case this corresponds to a field along $\mathsf{A}=\mathcal{A}^{21}$ (see Sec.~\ref{sec:prod-vari-stat}).

The expectation value of $s^z$ on a site in the middle of the system, as function of $J_H$, for different system sizes is shown in Fig.~\ref{fig:dmrgMagneticOrder}(b). The expectation value of $s^z$ remains small close to $J_H=0$, even when the ratio of the length of the cylinder to its circumference is unity, suggesting the absence of magnetic order in this case. As $J_H$ is increased, a finite expectation value develops as expected. 
The range of system sizes accessible by our simulations is not large enough to perform finite size scaling, but a relatively sharp increase in the magnetization around $J_H/J\simeq1$ suggests a phase transition occurs in the vicinity of this value.
In Figs.~\ref{fig:dmrgMagneticOrder}(c,d) we plot the expectation
values of $s^z$ on all the sites of a $6 \times 6$ cylinder, for
$J_H=0$ and $J_H/J=2$ respectively. While in the former case, the magnetization decays rapidly away from the boundary where the pinning fields are applied, in the latter case the magnetization is finite and uniform across the system.

In these simulations the maximal bond dimension for cylinders of width $N_y=3$ was $M=2000$, resulting in a truncation error smaller than $5\cdot10^{-5}$. For cylinders of width $N_y=6$ the maximal bond dimension was $M=8000$ for $J_H=0$ and $M=4000$ for $J_H>0$, resulting in a truncation error of $\sim 2\cdot10^{-3}$ for values of $J_H/J <1.5 $ at which no long-range ordering is observed, and a truncation error of $ \sim 5\cdot 10^{-4}$ or smaller for $J_H/J\geq 1.5$ at which a $120^\circ$ order develops.

To summarize, our numerical study suggests that the $SU(4)$-symmetric Heisenberg model does not have magnetic long-range order. A transition into a $120^\circ$-ordered state can be driven by a Hund's coupling term which breaks $SU(4)$ symmetry.

\section{Singlet projection}

The short-range nature of the spin correlations observed in
DMRG motivates an approach focusing on SU(4) singlets.  As we saw
explicitly in Eq.~\eqref{eq:3},  the six-dimensional representation of
SU(4) considered in this work allows for the formation of a singlet on a
pair of sites.  Hence we can build many singlet states for the entire
system by partitioning the sites into pairs, and placing each pair of
corresponding spins into a singlet state.  Following the pioneering work of Rokhsar and
Kivelson \cite{PhysRevLett.61.2376} who considered the projection of
the usual SU(2) Heisenberg model (in the S=1/2 representation) to a
nearest-neighbor singlet manifold, we study the projection of the
SU(4) Heisenberg model onto the subspace of nearest-neighbor SU(4)
singlet ``dimer'' coverings of the lattice.

In this section, we start with a simple analytic comparison between energies of the
singlet states and those of the classical ones discussed earlier, showing that
the dimer states are superior in a variational sense.  Then we provide
further numerical justification for the projection to the singlet
subspace.  We next discuss the projection of the SU(4) Hamiltonian to
the nearest-neighbor singlet coverings subspace and derive an
effective dimer model. Finally we study the resulting dimer model using exact diagonalization.

\subsection{Crude estimate of energy competition between singlet and
  ordered states}
\label{sec:crude-estimate-dimer}

We first estimate the energy of such a singlet state, and compare to
that of an ordered state.   The optimal
ordered product states were found in Sec.~\ref{sec:prod-vari-stat}.
They comprise 3-sublattice ordered states which spontaneously break
SU(4) symmetry analogously to the 120$^\circ$ ordered states for
classical SU(2) Heisenberg spins.  In those states, the energy per
bond is the same for all bonds and is equal to
\begin{equation}
  \label{eq:33}
  \langle H_{ij}\rangle = J {\rm Tr} \, \mathsf{A}^i (\mathsf{A}^j)^T
  = E_{\rm bond}^{\rm class}=-\frac{1}{2}J.
\end{equation}
Hence
\begin{equation}
  \label{eq:13}
  E^{\rm class}  = - \frac{1}{2}J N_{\rm bonds} = -\frac{3}{2}J N_{\rm sites}.
\end{equation}

Now we consider a singlet state which is the product of two-site
singlet ``dimers''.    Specifically, a singlet covering is given by a partition of the set of $N$
sites $i$ into pairs
$C=\{(i_1j_1),(i_2j_2),\cdots(i_{N/2},j_{N/2})\}$, where $(i,j)$
denotes a pair of nearest-neighbor sites.  Such a state can be
visualized by drawing a dimer -- a colored bond -- between the pairs
of sites $(i_aj_a)$.     We define
\begin{equation}
  \label{eq:14}
  |\mathfrak{C}\rangle = \bigotimes_{(ij) \in C}|s\rangle_{ij},
\end{equation}
using normalized singlets $|s\rangle_{ij}$ as in Eq.~\eqref{eq:10}.
Note that in contrast to the SU(2) case, in the SO(6) representation
the singlet state has a purely positive wavefunction, and is without
any sign ambiguity.  Thus there is no need to define the
directionality of a singlet which is required to determine the sign
of the wavefunction in the SU(2) case.

For a crude estimate, we consider the variational energy of a single
dimer covering,
\begin{equation}
  \label{eq:2}
 E_{\rm dimer}= \langle \mathfrak{C}|H|\mathfrak{C}\rangle = \sum_{\langle
    ij\rangle}  \langle \mathfrak{C}|H_{ij}|\mathfrak{C}\rangle.
\end{equation}
Unlike for the classical state, all the bond expectation values are
not equal.  As shown in Eq.~\eqref{eq:32}, the singlet
$|s\rangle_{ij}$ is an eigenstate of $H_{ij}$, with energy $-5J$.
Hence $ \langle \mathfrak{C}|H_{ij}|\mathfrak{C}\rangle=-5J$ for those
bonds covered by dimers.  For bonds that are not covered by singlets,
the two spins on the bond are uncorrelated, and one has $ \langle
\mathfrak{C}|H_{ij}|\mathfrak{C}\rangle=0$ for those bonds.  Hence the
variational energy of the dimer state is $-5J$ per bond {\em times}
the fraction of bonds occupied by singlets, which is $1/6$.  Thus  
Fixed $5/3 \to 5/2$ below, since $N_{\rm bonds} = 3 N_{\rm sites}$
\begin{equation}
  \label{eq:22}
  E_{\rm dimer} = - \frac{5}{6} J N_{\rm bonds} = -\frac{5}{2} J
  N_{\rm sites}.
\end{equation}
Comparing Eq.~\eqref{eq:22} and Eq.~\eqref{eq:13}, we see that the
dimer state has lower energy.  This gives some simple understanding of
the avoidance of magnetic order.

It is instructive to compare to the SU(2) case, with spin $S$ spins.
In this case for the usual Heisenberg model the classical product
ground state with 120$^\circ$ order has $J\langle
\bm{S}_i\cdot\bm{S}_j\rangle_{\rm class} = - JS^2/2$, so the classical
energy is
\begin{equation}
  \label{eq:10}
  E^{\rm class}_{SU(2)} = - \frac{J S^2}{2} N_{\rm bonds}.
\end{equation}
For a spin singlet bond, we can write
$\bm{S}_i\cdot\bm{S}_j = \frac{1}{2}[(\bm{S}_i+\bm{S}_j)^2 -
\bm{S}_i^2-\bm{S}_j^2]$, so that
$\langle J \bm{S}_i\cdot\bm{S}_j \rangle_{\rm singlet} = -JS(S+1)$.
Thus the dimer energy is
\begin{equation}
  \label{eq:35}
  E^{\rm dimer}_{SU(2)} = -\frac{JS(S+1)}{6} N_{\rm bonds}.
\end{equation}
Comparing Eq.~\eqref{eq:10} and Eq.~\eqref{eq:35}, we see that the
energies are {\em equal} for $S=1/2$ ($E= - J N_{\rm bonds}/8$), with
the classical state superior for all larger $S$.

In summary the simplest possible variational dimer state of a single
singlet covering is already better than a classically ordered state
for the SU(4) problem, which is distinctly different from the SU(2)
case.  In the following sections we will refine the approach to
singlet states, and consider superpositions of many terms, each with
the form of Eq.~\eqref{eq:14}.

\subsection{Numerical justifications for the projection onto the singlets subspace}

We define a nearest-neighbor singlet subspace as the Hilbert space
spanned by superpositions of all nearest-neighbor singlet coverings of
the form of Eq.~\eqref{eq:14}.  In this subsection we compare the low
energy spectrum of the Hamiltonian in the full Hilbert space with that
of its projection onto this nearest-neighbor singlet space.  To this
end we perform a numerical study on systems with size of up to 18
sites, using ED for systems with less than 12 sites, and Matrix
Product State (MPS)-based simulations for larger systems, as described
in detail in Appendix~\ref{app:ProjectionWithMPS}.

We first compare the flavor gap in the $SU(4)$ spin model with the gap
in the projected problem. For cylinder of width $N_y=3$, the flavor gap
was discussed in Sec.~\ref{sec:DMRG} and estimated to be larger than
$2.5J$ for an infinitely long cylinder. The gap obtained for the
projected problem is $0.281J$, $0.203J$ for system sizes of $N_x=4$
and $N_x=6$ respectively. For cylinders of width $N_y=4$, in the same
geometry, we find the flavor gap to be very weakly dependent on system
size already for small system sizes, and larger than $3.8J$. The gap
in the projected problem is $1.738J$, $1.724J$ for system sizes of
$N_x=3$ and $N_x=4$ respectively. Thus, we find that in both cases the
gap of the projected Hamiltonian is smaller than the flavor gap,
suggesting that the low energy physics is governed by the singlets.

In addition, we calculate the overlaps between the ground state of the $SU(4)$ spin model and that of the projected Hamiltonian. These are summarized in Table~\ref{tab:GSOverlaps} for a number of system sizes and different boundary conditions. 
We find that the overlaps  decrease with increasing system size as expected. However, given the immense reduction in the dimension of the Hilbert space upon the projection, we find surprisingly large overlaps even for systems with $N\simeq 10-20$ sites.

\begin{table}
\centering
\begin{tabular}{l | c | c | c | c | c | c | c} 
              & $2\times 2$ & $2\times3$ & $4\times3$ & $6\times3$ & $2\times4$ & $3\times4$ & $4\times4$ \\
 \hline \hline 
OBC      & 0.976          &     0.946      &     0.85(2)    &    0.76(2)    &     0.921      &    0.85(2)    &     0.80(4)     \\
\hline
Cylinder & -                 &     0.875      &     0.70(1)    &   0.55(1)     &    0.918       &    0.87(1)    &     0.82(1)      \\ 
\hline
\end{tabular}
\caption{Wavefunction overlaps between the ground state of the $SU(4)$
  spin model and the ground state of the Hamiltonian projected onto
  the subspace of nearest-neighbor singlet coverings for different
  system sizes ($N_x \times N_y$). ``OBC'' indicates open boundary
  conditions along both $x$ and $y$, while ``Cylinder'' indicates
  periodic boundary conditions along $y$ and open boundary conditions
  along $x$. The error indicated in brackets is estimated from the DMRG truncation error for the ground state of the spin model. Values for which no error is indicated were obtained using ED. }
\label{tab:GSOverlaps}
\end{table}

\subsection{Derivation of the effective dimer model}
\label{sec:proj-near-neighb}

We now turn to the analytic derivation of the projected Hamiltonian.
Rokhsar and Kivelson \cite{PhysRevLett.61.2376} constructed an expansion to
express the effective projected Hamiltonian as a sum of local terms of
increasing length of dimer re-arrangements.  We obtain a similar
expansion here for the SU(4) $\sim$ SO(6) case.  We follow specifically
a reformulation of the expansion by Ralko {\em et
  al.}\ \cite{PhysRevB.80.184427}.

We seek the best
variational state of the form
\begin{equation}
  \label{eq:4}
  |\psi\rangle = \sum_C \psi_C |\mathfrak{C}\rangle.
\end{equation}
The wavefunction $\psi_C$ is required to minimize
\begin{equation}
  \label{eq:5}
  E(\psi) =\frac{\langle \psi|H|\psi\rangle}{\langle
    \psi|\psi\rangle} =  \frac{\psi^\dagger
    \mathsf{H}\psi^{\vphantom\dagger}}{\psi^\dagger \mathsf{S}\psi^{\vphantom\dagger}},
\end{equation}
where
\begin{equation}
  \label{eq:6}
  \mathsf{H}_{C'C} = \langle \mathfrak{C}'|H|\mathfrak{C}\rangle,
  \qquad \mathsf{S}_{C'C}=\langle \mathfrak{C}'|\mathfrak{C}\rangle.
\end{equation}
The minimum of the variational energy is
given by the condition $\partial E/\partial \psi^*=0$.  This gives
\begin{equation}
\label{eq:8}
  \mathsf{H} \psi = E_0 \mathsf{S} \psi,
\end{equation}
where $E_0 = \textrm{min}_\psi E(\psi)$ is the best variational energy.
This is a generalized eigenvalue problem for $E_0$.  We can convert it
to a conventional one by defining $\Psi = \mathsf{S}^{1/2} \psi$,
which leads to
\begin{equation}
\label{eq:11}
  \mathsf{H}_{\rm eff} \Psi = E_0 \Psi,
\end{equation}
with the effective Hamiltonian 
\begin{equation}
\label{eq:12}
  \mathsf{H}_{\rm eff} = \mathsf{S}^{-1/2} \mathsf{H} \mathsf{S}^{-1/2} .
\end{equation}
Therefore the variational ground state energy (and from it ultimately
the variational ground state wavefunction) is obtained from the ground
state of $\mathsf{H}_{\rm eff}$, which is the desired effective
quantum dimer Hamiltonian.

To obtain $\mathsf{H}_{\rm eff}$, we expand both $\mathsf{H}$ and
$\mathsf{S}$ in a series of increasingly small terms, which are
related to the number of dimer rearrangements forming ``loops''.  The
small parameter of this expansion is the overlap $x$ in the
smallest such non-trivial loop: two dimers cyclically permuted on four sites.
More generally, the inner product of
a sequence of
dimers pairing sites $C=\{(i_1j_1),(i_2j_2),\cdots(i_{N/2},j_{N/2})\}$
and $C'= \{(i_1j_2),(i_2j_3),\cdots(i_{N/2},j_{1})\}$ is
$\langle \mathfrak{C}'|\mathfrak{C}\rangle = x^{N/2-1}$, with $x=1/6$.  
In a full calculation of $\mathsf{H}$ and $\mathsf{S}$, products of such
overlaps appear, resulting in multiple factors of $x$, which
determines the order of these terms in the expansion.
Details of this quite technical procedure, which we formulate for
SU(4) on a general lattice, will be presented in a separate
publication.  Starting from the general SO(6) invariant Hamiltonian in Eq.~\eqref{eq:57}, carrying out this expansion, and then calculating
$\mathsf{H}_{\rm eff}$ consistently to a given order gives the final
result for the quantum dimer model Hamiltonian:
\begin{eqnarray}
  \label{eq:bigugly}
  &&   \mathsf{H}_{\rm eff} =\sum' \dimerseqcoef ,
\end{eqnarray}
where the prime on the sum indicates a sum over all the
symmetry-equivalent plaquettes shown
in the bras and kets, throughout the lattice. All the coefficients are given in terms of $\alpha$, $\beta$, and $x$ and are summarized in Table~\ref{tab:DimerCoefs}.

\begin{table}
\centering
\begin{tabular}{l | l | l}
              & expression & numerical value\\
 \hline \hline 
$t$  & $-(2\alpha+\beta)-x(\alpha+2\beta)+x^2(\alpha+2\beta)$ & 31/36 \\
\hline
$v,\ t_{6,a}$ & $-x(2\alpha+\beta)-x^2(\alpha+2\beta)$ & 5/36 \\
\hline
$t_{6,b}$ & $-3x(\alpha+\beta)-3x^2(\alpha+\beta)$ & 0 \\
\hline
$u$ & $-\frac{1}{2} x^2 (2 \alpha +\beta)$ & 1/72 \\
\hline
$t_{8,a}$ & $-4 \alpha x^2$ & 1/9 \\
\hline
$t_{8,b}$ & $-\frac{1}{2} x^2 (5 \alpha +6 \beta)$ & -1/72 \\
\hline
$t_{8,c}$ & $-x^2 (4 \alpha +5 \beta)$ & -1/36 \\
\hline
$t_{8,d}$ & $-\frac{1}{2} x^2 (5 \alpha +4 \beta)$ & 1/72 \\
\hline
$t_{8,e}$ & $-x^2 (2 \alpha +\beta)$ & 1/36 \\
\hline
$t_{8,f}$ & $-\frac{1}{2} x^2 (2 \alpha +\beta)$ & 1/72 \\
\hline
$t_{8,g}$ & $-2 x^2 (\alpha +\beta)$ & 0 \\
\hline
\end{tabular}
\caption{Coefficients of the terms in the effective dimer model given in Eq.~\eqref{eq:bigugly}. The numerical value in the last column is calculated for the parameters corresponding to the SU(4) Heisenberg model, namely $\alpha=-1,\beta=1,x=1/6$. }
\label{tab:DimerCoefs}
\end{table}

\subsection{Numerical study of the dimer model}\label{sec:DimersED}

We now turn to an analysis of the dimer model obtained in the previous
section, Eq.~\eqref{eq:bigugly}, and taking $\alpha=-1,\beta=1$, corresponding to the Heisenberg model.

To zeroth order in the expansion parameter $x$, the dimer model obtained is the standard dimer model considered by Rokhsar and Kivelson~\cite{PhysRevLett.61.2376}, i.e.\ $H=-t(|\plaquettev\rangle\langle\plaquetteh|+|\plaquetteh\rangle\langle\plaquettev|)+v (|\plaquettev\rangle\langle\plaquettev|+|\plaquetteh\rangle\langle\plaquetteh|)$,
with $v=0$ and $t=1$ -- for these values the Hamiltonian contains only
the ``flip'' term, consistent with the truncation of
Eq.~\eqref{eq:bigugly} to zeroth order in $x$.  Previous studies of this model on the triangular lattice~\cite{Moessner2001,Ralko2005,Ralko2006,Ralko2007} found that the ground state for $v/t=0$ is a $\sqrt{12}\times\sqrt{12}$ VBS state. At large enough negative $v/t$, the ground state is a columnar ordered state, while for positive $v/t$, first a phase transition into the $\mathbb{Z}_2$ RVB spin liquid phase occurs at $v/t\simeq0.83$, followed by a transition at $v/t=1$ into a staggered ordered phase.

\subsubsection{Geometry and sectors}
\label{sec:geometry-sectors}

\begin{figure}[ht]
\includegraphics[width=0.5\linewidth,valign=c]{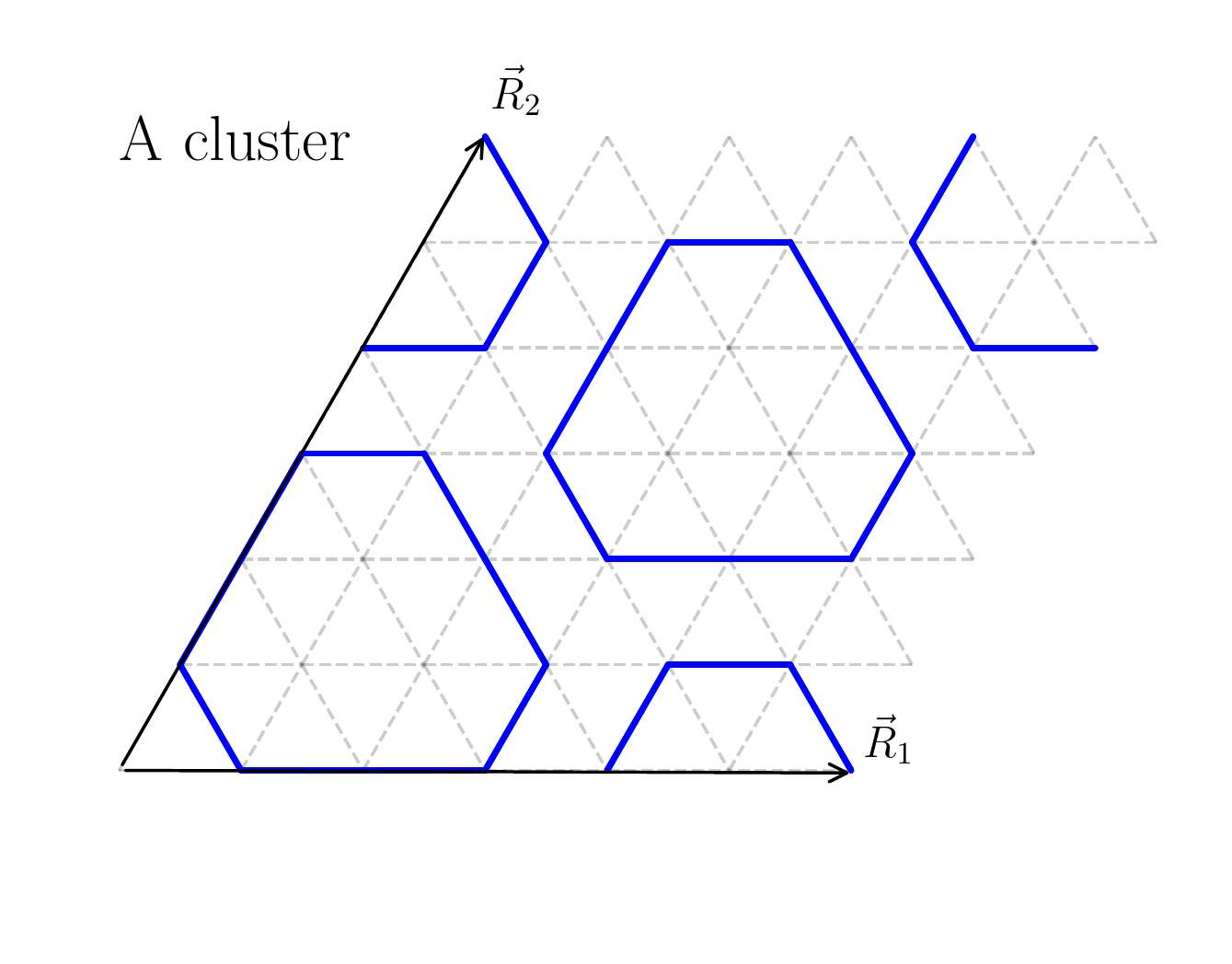}
\includegraphics[width=0.5\linewidth,valign=c]{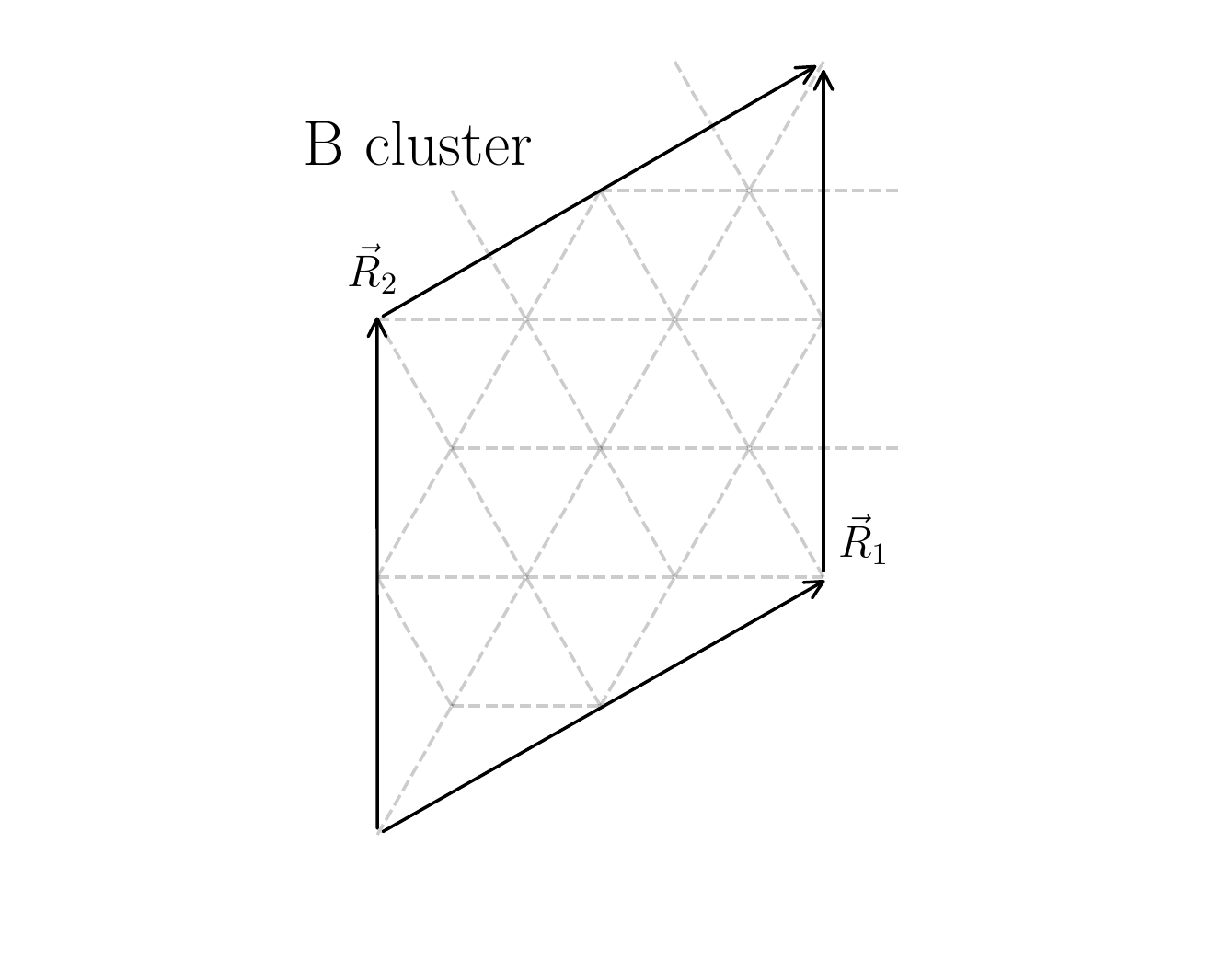}
\caption{The two types of clusters considered in the numerical study of the dimer model. A schematic representation of the $\sqrt{12}\times\sqrt{12}$ order, depicting the 12-site unit cell is shown for the $6\times6$ type A cluster.}
 \label{fig:clusters}
\end{figure}

To understand how higher order terms in the expansion affect the
ground state, we now study the model in Eq.~\eqref{eq:bigugly} numerically, using ED.
We consider systems with periodic boundary conditions along both directions (i.e.\ systems on a torus), and focus on two types of clusters which keep all the symmetries of the infinite lattice, following Ref.~\cite{Ralko2005}.
Denoting the basis vectors of the triangular lattice by
$\vec{a}_1=(1,0)$, $\vec{a}_2=(1/2,\sqrt{3}/2)$, these clusters are
defined by identifying the sites of the lattice modulo the vectors
$(\vec{R}_1,\vec{R}_2)$, where
$(\vec{R}_1,\vec{R}_2)=m(\vec{a}_1,\vec{a}_2)$ for clusters of type A,
and $(\vec{R}_1,\vec{R}_2)=(m\vec{a}_1+m\vec{a}_2,
-m\vec{a}_1+2m\vec{a}_2)$ for clusters of type B. These two types of
clusters are shown in Fig.~\ref{fig:clusters}. Note that the number of
sites in cluster of type A (B) is $m^2$ ($3m^2$).

On a torus, the Hilbert space of dimer coverings breaks up into four
distinct topological sectors defined by the parities of the number of
dimers intersected by closed loops winding around the torus along the
two axes. We will denote these sectors by ${\rm TS}(p_x,p_y)$, with
$p_x,p_y=0 (1)$ for even (odd) parity along $x$ and $y$
respectively. As pointed out in Ref.~\cite{Ralko2005}, on a cluster
with $C_6$ symmetry, three of these
topological sectors are always degenerate since they can be related by
$C_6$ rotations of the lattice. Which three  sectors are degenerate
depends on the parity of $m/2$, but in order to understand the
spectrum of the problem it is enough to consider the two sectors
$\rm{TS}(0,0)$ and $\rm{TS}(1,1)$, as these two sectors are never
related by $C_6$ rotations.

We consider system sizes of up to 36 sites, i.e.\ clusters of type B
with $m=2$ (12 sites) and clusters of type A with $m=4$ and $m=6$ (16
and 36 sites respectively). We note that to allow for a
$\sqrt{12}\times\sqrt{12}$ order, the number of sites in the system
must be a multiple of $12$. In addition, for $m/2$ odd, only the
topological sector $\rm{TS}(1,1)$ can accommodate this ordering
without defects (see Fig.~\ref{fig:clusters}).

\subsubsection{Exact diagonalization results}
\label{sec:exact-diag-results}

\begin{figure}
\begin{subfigure}[t]{.52\linewidth}
\caption{}
\includegraphics[width=\linewidth,valign=t]{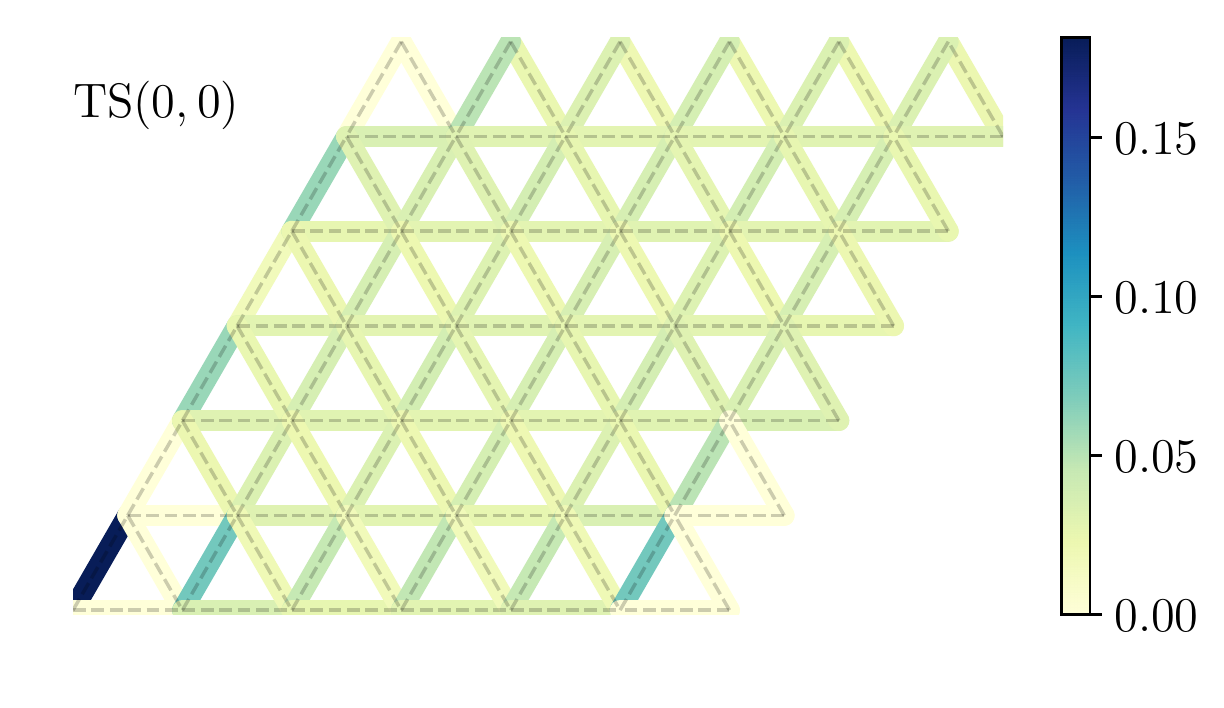}
\end{subfigure}
\begin{subfigure}[t]{.48\linewidth}
\caption{}
\includegraphics[width=\linewidth,valign=t]{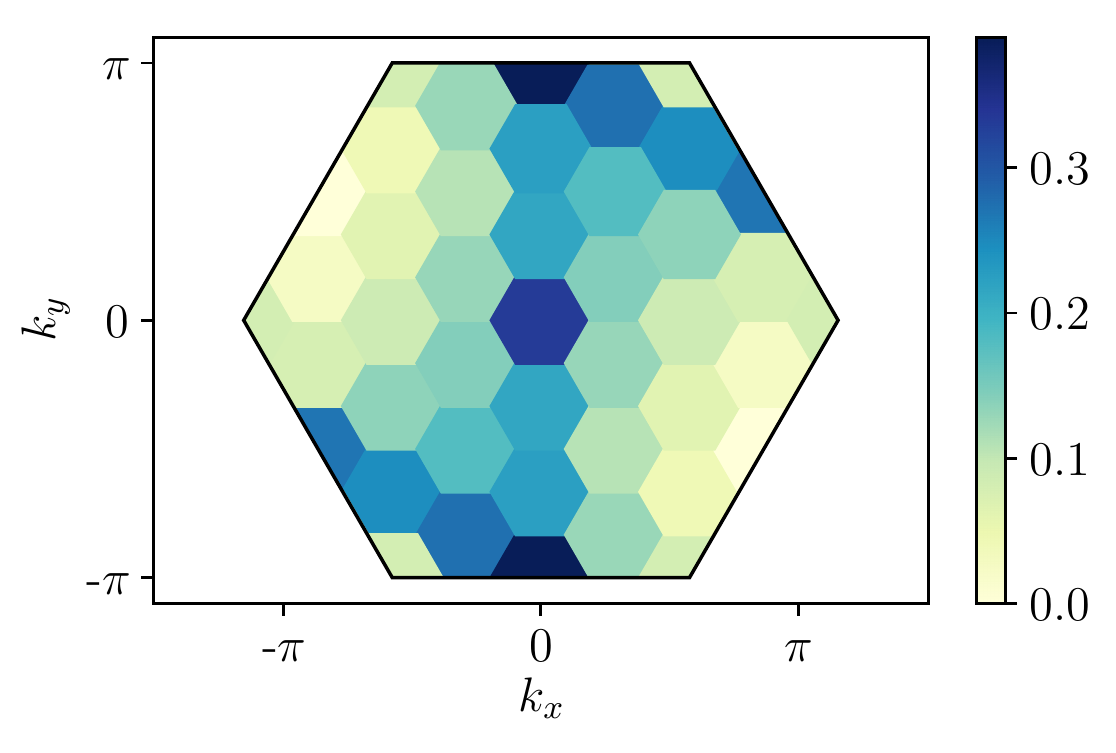}
\end{subfigure} \\
\begin{subfigure}[t]{.52\linewidth}
\caption{}
\includegraphics[width=\linewidth,valign=t]{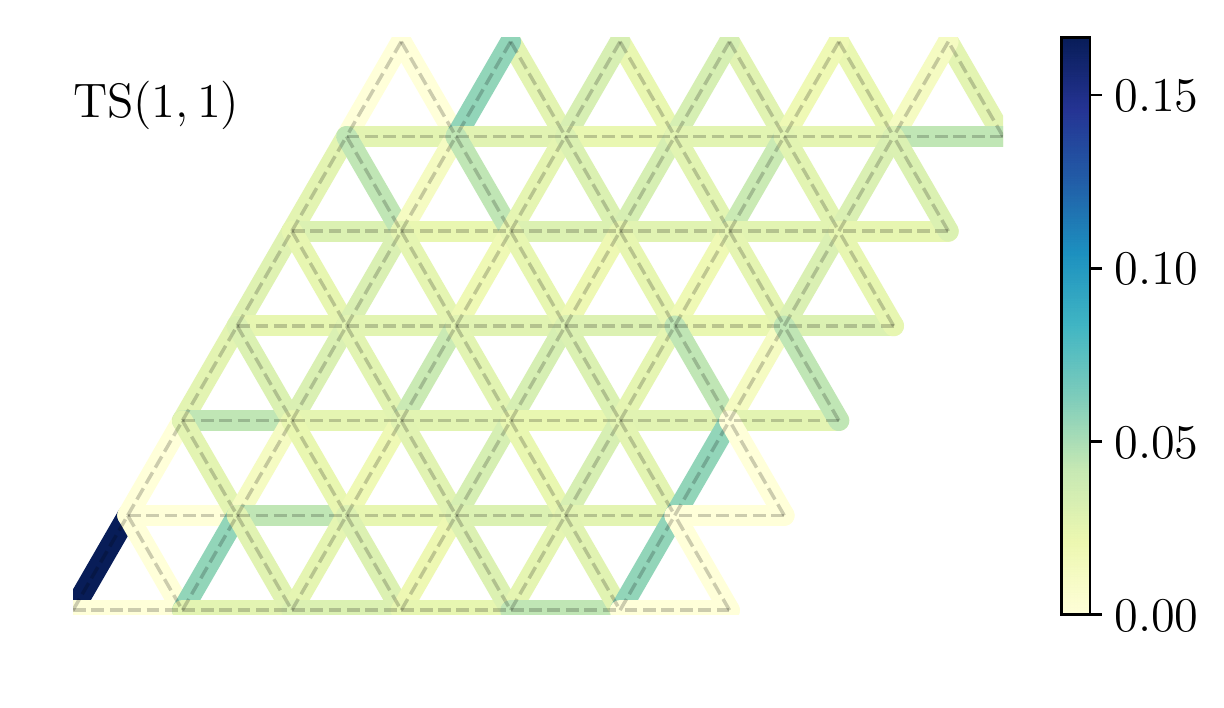}
\end{subfigure}
\begin{subfigure}[t]{.48\linewidth}
\caption{}
\includegraphics[width=\linewidth,valign=t]{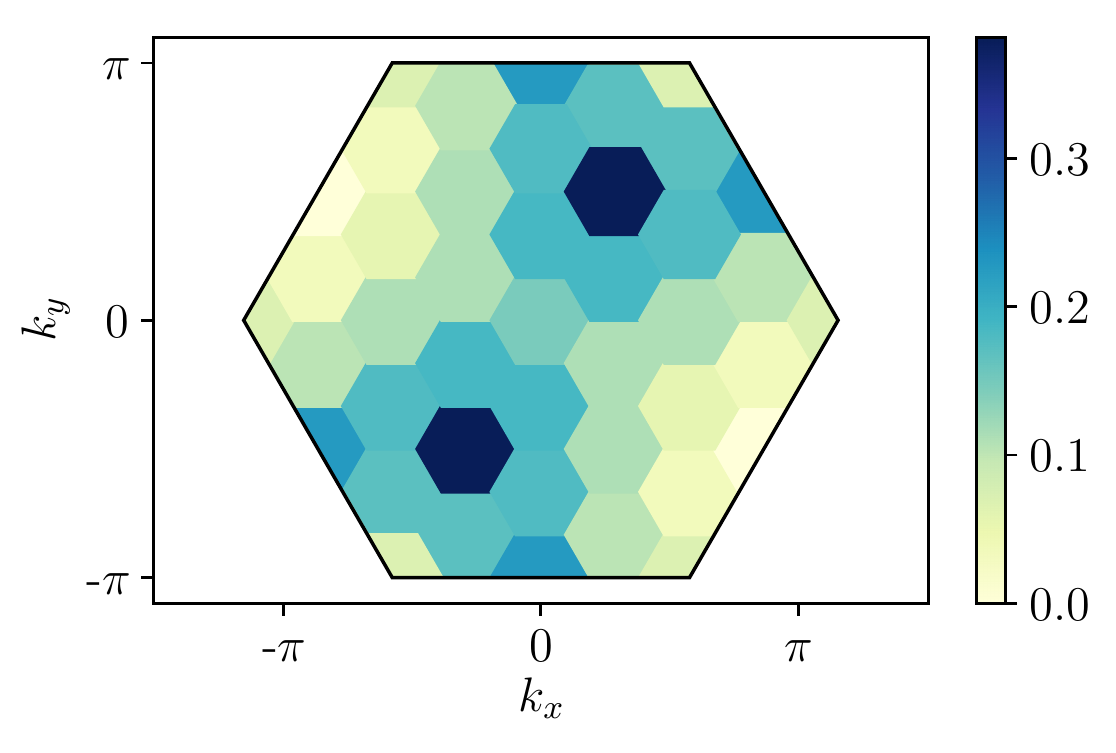}
\end{subfigure}
\caption{(a,b) Dimer-dimer
  correlations $\langle b_1 b_i \rangle$ (where $b_i=1$ if the $i$th bond is occupied by a dimer and $b_i=0$ otherwise) in the lowest energy state
  in the topological sectors $\rm{TS}(0,0)$ and $\rm{TS}(1,1)$ respectively of the
  extended dimer model $H_2$ on a $6\times6$ lattice obtained using ED.  (c,d) Fourier transform of $\langle b_1
  b_i \rangle-\langle b_1\rangle\langle b_i \rangle$ for the vertical
  bonds in the two topological sectors respectively.}
 \label{fig:edBondBond}
\end{figure}

We study the successive approximations obtained by working to
increasingly higher order in $x$, denoting by $H_n$ the sum of all
terms in the effective Hamiltonian up to and including $O(x^n)$.   
More explicitly, $H_0$ consists solely of the kinetic term on a plaquette with $t=1$, while $H_1$ contains in addition the potential energy term $v$ as well as the kinetic terms corresponding to hopping on loops of length six. The values of the coefficients in $H_1$ are given by $t=5/6$, and $v=t_{6,a}=1/6$ (note that $t_{6,b}=0$). The Hamiltonian $H_2$ contains all the terms in Eq.~\ref{eq:bigugly} with the corresponding values given in Table~\ref{tab:DimerCoefs}.
We calculate the lowest energy state in each topological sector of
$H_{n=0,1,2}$, for the physical situation $x=1/6$. The values obtained
are summarized in Appendix~\ref{app:DimerEDEn}. We find that the
correction due to second order terms is indeed small compared to the
first order ones.

We then interpolate between the Hamiltonians $H_0$ and $H_2$,
calculating the low energy spectrum of
$H(\eta)=(1-\eta) H_0+\eta H_2$. We find that there are no level
crossings in the low energy spectrum, and the ground state remains in
the topological sector $\rm{TS}(1,1)$ for system sizes which can
accommodate the $\sqrt{12}\times\sqrt{12}$ order (see
Fig.~\ref{fig:edSpectrumInterpolation} in
Appendix~\ref{app:DimerEDEn}).  The smooth continuity suggests that
$H_0$ and $H_2$ describe the same phase of matter.  Furthermore, we
find that, in each topological sector, the wavefunction overlap
between the lowest energy state of $H_0$ and that of $H_2$ is very
close to one, in particular in the topological sector
$\rm{TS}(1,1)$. More specifically, for the $6\times6$ system, the
overlaps are $0.88$ and $0.97$ for $\rm{TS}(0,0)$ and $\rm{TS}(1,1)$
respectively.  

In addition, we compare the dimer-dimer correlations in these
states. We find that the correlations in the lowest energy states of
$H_2$ become slightly more uniform compared to those in the lowest energy states of $H_0$, but overall display the same features (see Appendix~\ref{app:DimerEDCorr}).
In Fig.~\ref{fig:edBondBond} we plot the real space dimer-dimer correlations, as well as their Fourier transform, calculated in the lowest energy state of $H_2$ in the two topological sectors for a $6\times6$ lattice.
As can be clearly seen, for the state in $\rm{TS}(1,1)$ sharp peaks at
$\vec{k}=\pm(\pi/(2\sqrt{3}),\pi/2)$ are present, suggesting breaking
of translational invariance compatible with the formation of a 12-site
unit cell. We note that the six-fold rotational symmetry expected in
the ground state is broken in Figs.~\ref{fig:edBondBond}(b,d) by the
choice of the set of bonds used in the calculation of the dimer-dimer
correlation function. The structure factor shown in
Figs.~\ref{fig:edBondBond}(b,d) is obtained for the bonds parallel to
the lattice basis vector $\vec{a}_2$. When the correlation function is
calculated with respect to a set of bonds related by a $\pi/3$
rotation on each site, the peaks in the structure factor appear at
momenta related by the corresponding $\pi/3$ rotation.

Although a better finite size scaling analysis is required to make a
conclusive statement regarding the nature of the ground state of the
dimer model, we believe that these observations -- (i) the similarity of the ground state
correlations to those of the ``standard'' dimer model at $v/t=0$  for
small system sizes, and (ii) the smooth evolution of the spectrum upon
interpolation between the two models -- strongly suggest that the
ground state remains a $\sqrt{12}\times\sqrt{12}$ VBS ordered state.

\section{Conclusion}

In this work, we considered SU(4) spins in the six-dimensional
(self-conjugate) representation, on the triangular lattice, with
nearest-neighbor antiferromagnetic interactions.  Our DMRG study
suggests that the ground state is non-magnetic, but remains
inconclusive as to the exact nature of the ground state.   We
developed and carried out a dimer expansion, which we argued is
capable of capturing the low energy properties of the model.  The study of the
the associated dimer model led us to conjecture  that the ground state of
the SU(4) model may be a 12-site valence bond solid (VBS).  

As the mapping to the dimer model involves an uncontrolled projection,
we do not know how to systematically improve it.  Hence, a fully
conclusive study should return to the original SU(4) spin
model.  This, however, remains numerically challenging due to the
large on-site Hilbert space dimension.  As a first step in this
direction, we carried out preliminary calculations in addition to
those reported in this paper, using the infinite DMRG (iDMRG) method
on width-four cylinders.  By choosing appropriate boundary conditions,
this geometry is compatible with the 12-site VBS order.  However, we
did not find signatures of this order in our iDMRG simulations.  One
possible interpretation is that the non-observation of VBS order
is simply due to the effects of finite size or finite bond dimension.
Another possibility is that the VBS order is truly absent, indicating
some type of spin liquid state without broken symmetries.  The
proximity of a $\mathbb{Z}_2$ spin liquid phase in the effective dimer
model suggests this as an intriguing possibility.  Regardless, this
conundrum highlights the challenges of a direct simulation of the
original SU(4) problem.  

In the study of the effective dimer model, we focused on the
parameters corresponding to the SU(4) Heisenberg model,
$\alpha=-1,\beta=1$.  In the future it would be interesting to explore
the full phase diagram of the general dimer model derived, understand
if it can realize the $\mathbb{Z}_2$ spin liquid phase, and identify
the nature of the interactions in terms of the SU(4) spins required
for this.

In addition, it would be desirable to study in more detail the
evolution of the ground state with increasing $J_H$.  If the ground
state of the spin model at $J_H=0$ is indeed a 12-site VBS, and if
there is, as suggested by our numerics, a direct transition to a
three-sublattice ordered state with increasing $J_H$, then this is a
Landau-forbidden quantum phase transition.  If this is realized via a
continuous quantum critical point, then it must be an example of
deconfined quantum criticality.  It would be interesting to understand
the nature of this critical point and test it in numerics.

\section*{Acknowledgements}

A.K. would like to thank Bela Bauer for valuable discussions regarding the numerical studies presented in this work. 
We thank Chao-Ming Jian for discussions regarding the irreducible representations of SU(4).
This research is funded in part by the Gordon and Betty Moore Foundation through Grant GBMF8690 to UCSB to support the work of A.K.
Use was made of the computational facilities administered by the Center for Scientific Computing at the CNSI 
and MRL (an NSF MRSEC; DMR-1720256) and purchased through NSF CNS-1725797.
L.S.\ was supported by the Agence Nationale de la Recherche through
Grant ANR-18-ERC2-0003-01 (QUANTEM), and in part by the National
Science Foundation under Grant No.\ NSF PHY-1748958.  L.B.  was supported
by the DOE, Office of Science, Basic Energy Sciences under Award
No. DE-FG02-08ER46524.
\begin{appendix}

\section{Classical limit}
\label{app:classical}

\subsection{SO(6) formulation}
\label{sec:so6-formulation}

The classical limit is taken by replacing the 15 generators $\hat{A}^{mn}$ by their expectation values in a given state, i.e.:
\begin{equation}
  \label{eq:app7}
  \hat{A}^{mn}\rightarrow\mathsf{A}_{mn}=\langle\psi|\hat{A}^{mn}|\psi\rangle.
\end{equation}
Since the operators $\hat{A}^{mn}$ are hermitian and satisfy $\hat{A}^{mn}=-\hat{A}^{nm}$ , the matrix $\mathsf{A}$ is real and anti-symmetric, i.e. $\mathsf{A}^T= -\mathsf{A}$. Note that since $\mathsf{A}$ is real, ${\rm Tr}\,\mathsf{A}\mathsf{A}^T>0$, and therefore ${\rm Tr}\,\mathsf{A}^2 = - {\rm Tr}\,\mathsf{A}\mathsf{A}^T <0$.

Writing the quantum state explicitly as $|\psi\rangle=\sum_{p=1}^6 v_p |\hat{p}\rangle$, the matrix elements of $\mathsf{A}$ are given by
\begin{equation}
  \label{eq:app9}
  \mathsf{A}_{mn}=\sum_{p,p'}\frac{iv_p^*v_{p'}}{\sqrt{2}}\langle\hat{p}|\left(|\hat{m}\rangle\langle\hat{n}|-|\hat{n}\rangle\langle\hat{m}|\right)|\hat{p}'\rangle = \frac{i}{\sqrt{2}}\left(v_m^*v_{n}-v_n^*v_{m}\right) = \sqrt{2}\,{\rm Im}[v_n^*v_{m}].
\end{equation}
or in matrix notations $\mathsf{A}=i (\mathbf{v}^*\mathbf{v}^T-\mathbf{v}\mathbf{v}^\dagger)/\sqrt{2} = \sqrt{2}\,{\rm Im}[\mathbf{v}\mathbf{v}^\dagger]$.
We next note that 
\begin{eqnarray}
  \label{eq:app11}
  {\rm Tr}\,\mathsf{A}\mathsf{A}^T &=& \sum_{m,n=1}^6\mathsf{A}_{mn}\mathsf{A}_{mn} = 1-\left|\sum_{n=1}^6v_{n}^2\right|^2 \leq 1.
\end{eqnarray}
Further let $\mathbf{v}=(\mathbf{x}+i\mathbf{y})/\sqrt{2}$ with $\mathbf{x}$, $\mathbf{y}$ real six-dimensional vectors with unit norm.  Then ${\rm Tr}\,\mathsf{A}\mathsf{A}^T=1-(\mathbf{x}\cdot\mathbf{y})^2$.
It is now easy to see that the upper bound on ${\rm Tr}\,\mathsf{A}\mathsf{A}^T$ is reached when $\mathbf{x}\perp\mathbf{y}$.

In the classical limit the Hamiltonian is given by:
\begin{eqnarray}
  \label{eq:app12}
  \mathsf{H}_{\rm Heis}=J\sum_{\langle ij\rangle}\sum_{m,n=1}^6\mathsf{A}_{mn}^i\mathsf{A}_{mn}^j=J\sum_{\langle ij\rangle}{\rm Tr}[\mathsf{A}_i\mathsf{A}_j^T].
\end{eqnarray}
On the triangular lattice we can rewrite:
\begin{equation}
  \label{eq:app13}
  \mathsf{H}=
  \frac{J}{4} \sum_{t\;triangle} \left[ {\rm Tr}\, \left(\sum_{i \in
        t} \mathsf{A}_i\right)\left(\sum_{i \in
        t} \mathsf{A}_i\right)^T - {\rm Tr} \left( \sum_{i\in t}
      \mathsf{A}^{\vphantom{T}}_i \mathsf{A}_i^T\right)\right].
\end{equation}
For antiferromagnetic coupling, $J>0$, to minimize the energy, we would like the first term
to vanish, and the second to be as negative as possible. 
Let us denote by $\mathbf{x},\mathbf{y},\mathbf{z}$ three real, orthogonal, six-dimensional unit vectors, and define
\begin{eqnarray}
  \label{eq:app18}
  \mathbf{v}(\theta) & = & \frac{1}{\sqrt{2}}\left(\mathbf{y}+i(\cos\theta\mathbf{x}+\sin\theta\mathbf{z})\right), \\
  \mathsf{A}(\theta) & = &\sqrt{2}\left({\rm Im}[\mathbf{v}(\theta)]{\rm Re}[\mathbf{v}(\theta)]^T-{\rm Re}[\mathbf{v}(\theta)]{\rm Im}[\mathbf{v}(\theta)]^T\right).
\end{eqnarray}
Then, the matrices $\mathsf{A}_{l=1,2,3} =\mathsf{A}\left(\frac{2\pi
    l}{3}\right)$ satisfy $\sum_{l=1}^3 \mathsf{A}_l=0$ and ${\rm
  Tr}\,\mathsf{A}_l\mathsf{A}_l^T=1$, thus minimizing the energy on a
triangle.

Note that choosing $\mathbf{x}=\mathbf{e}_1$, $\mathbf{y}=\mathbf{e}_2$, $\mathbf{z}=\mathbf{e}_3$, with $\mathbf{e}_n$ denoting the unit vector along the $n$th dimension in $\mathbb{R}^6$ we obtain a state $\mathbf{v}(\theta)$ that belongs to the spin-triplet valley-singlet subspace on a given site (see also Appendix~\ref{app:LargeJH} below and in particular Eq.~\eqref{eq:app32} therein). The classical ground state corresponding to the states $\mathbf{v}_{l=1,2,3}=\mathbf{v}(\frac{2\pi l}{3})$ on each triangle of the lattice is then, in this case, exactly the $120^{\circ}$ ordered state of the SU(2) spin-ones.

\subsection{SU(4) formulation}
\label{sec:su4-formulation}

Here we derive the classical energy function and constraints using the
SU(4) formulation, i.e.\ starting from the Hamiltonian Eq.~\eqref{eq:31}, and using
the basis states on the right-hand-sides of the equalities in
Eq.~\eqref{eq:54}.  

We start by writing the quantum state on a single site explicitly as
$|\psi\rangle=\sum_{a,b={\sf 1}}^{\sf 4}\psi_{ab}c_a^\dagger
c_b^\dagger|0\rangle$. The $4\times4$ matrix $\psi$ must be
antisymmetric, $\psi^T=-\psi$, and the normalization constraint
$\langle\psi|\psi\rangle=1$ imposes ${\rm Tr}\psi^\dagger\psi=1/2$.
The classical limit is obtained by replacing
the 15 generators $\tilde{T}^{ab}$ ($a,b={\sf 1},..,{\sf 4}$) by their expectation
values in a given state $|\psi\rangle$:
\begin{equation}
  \label{eq:15}
  \tilde{T}^{ab}\rightarrow\mathsf{T}_{ab}=\langle\psi|\tilde{T}^{ab}|\psi\rangle = 4(\psi^\dagger\psi)_{ab}-\frac{1}{2}\delta_{ab},\quad\mbox{i.e.}\quad
\mathsf{T}=4\psi^\dagger\psi-\frac{1}{2}{\rm Id}.
\end{equation}
We have in turn $\mathsf{T}^\dagger=\mathsf{T}$ and ${\rm Tr}\mathsf{T}=0$.

We now proceed to finding lower and upper
bounds on ${\rm Tr}\mathsf{T}^2$, as these will be important for the minimization of the energy. 
We will show that $0\leq{\rm Tr}\mathsf{T}^2\leq1$. 
To do so, we consider the
eigenvalues of $\mathsf{T}$. Since $\mathsf{T}$ is hermitian, its eigenvalues $t_n$ are real. 
Using the inequality $\sum_{n=1}^Nt_n^2\geq \frac{1}{N}\left(\sum_{n=1}^N t_n\right)^2$, we
find the lower bound
\begin{equation}
  \label{eq:28}
  {\rm Tr}\mathsf{T}^2\geq\frac{1}{4}({\rm Tr}\mathsf{T})^2=0,
\end{equation}
which is saturated for example for $\mathsf{T}_{\rm lower}=0_4$, and
corresponds to $\psi_{\rm
  lower}=\frac{1}{2\sqrt{2}}\sigma^0(i\tau^y)$. 
The specific form of
$\mathsf{T}$ in terms of the square of $\psi$ imposes a stringent upper bound. Indeed, the antisymmetry
of $\psi$ makes the latter diagonalizable, and that combined with its even dimension
imposes that its eigenvalues come in pairs $\pm y_{1,2}$. In turn the
eigenvalues of $\mathsf{T}$ are doubly degenerate and equal to
$4|y_{1,2}|^2-1/2$. Therefore,
\begin{equation}
  \label{eq:29}
  {\rm Tr}\mathsf{T}^2=2\sum_{i=1,2}\left(4|y_{i}|^2-1/2\right)^2=32(|y_1|^4+|y_2|^4)-1,
\end{equation}
since ${\rm Tr}\mathsf{T}=8(|y_1|^2+|y_2|^2)-2=0$. Given ${\rm
  Tr}\psi^\dagger\psi=2(|y_1|^2+|y_2|^2)=1/2$, the maximum of ${\rm
  Tr}\mathsf{T}^2$ is reached for $\{|y_1|=0,|y_2|=1/2\}$ or
$\{|y_1|=1/2,|y_2|=0\}$ and equal to $1$. This is achieved for example
for $\mathsf{T}_{\rm upper}=\frac{1}{2}\sigma^0\tau^z$,
which corresponds to $\psi_{\rm upper}=\frac{1}{4}(i\sigma^y)(\tau^0+\tau^z)$.

In summary, a classical SU(4) ``spin'' $\mathsf{T}$ satisfies:
\begin{equation}
  \label{eq:16}
\mathsf{T}^\dagger=\mathsf{T},\qquad
{\rm Tr}\mathsf{T}=0,\qquad0\leq{\rm Tr}\mathsf{T}^2\leq1.
\end{equation}

In this formulation, in the classical limit the Hamiltonian is given
by:
\begin{equation}
  \label{eq:18}
  \mathsf{H}_{\rm Heis}=J\sum_{\langle ij\rangle}\sum_{a,b={\sf
  1}}^{{\sf 4}}\mathsf{T}_{ab}^i\mathsf{T}_{ba}^j=J\sum_{\langle ij\rangle}{\rm Tr}[\mathsf{T}_i\mathsf{T}_j].
\end{equation}
On the triangular lattice we can rewrite:
\begin{equation}
  \label{eq:25}
  \mathsf{H}=
  \frac{J}{4} \sum_{t\;triangle} \left[ {\rm Tr}\, \left(\sum_{i \in
        t} \mathsf{T}_i\right)^2 - {\rm Tr} \left( \sum_{i\in t}
      \mathsf{T}_i^2\right)\right].
\end{equation}
For antiferromagnetic coupling, $J>0$, to minimize the energy, we would like the first term
to vanish, and the second to be as negative as possible. 
%This can be achieved in at least one way, by extending the SU(2) solution as follows. 
Let us denote by $\mathbf{z},\mathbf{x}$ two real, orthogonal, three-dimensional unit vectors, and define
\begin{eqnarray}
  \label{eq:app18}
  \mathbf{n}(\theta) & = & \cos\theta\mathbf{z}+\sin\theta\mathbf{x}, \\
  \mathsf{T}(\theta) & = & \frac{1}{2}\sigma^0 (\mathbf{n}(\theta)\cdot{\boldsymbol{\tau}}).
\end{eqnarray}
Then, the matrices $\mathsf{T}_{l=1,2,3} =\mathsf{T}\left(\frac{2\pi
    l}{3}\right)$ satisfy $\sum_{l=1}^3 \mathsf{T}_l=0$ and ${\rm
  Tr}\,\mathsf{T}_l^2=1$, thus minimizing the energy on a triangle. 
The corresponding state $\psi_l$ can be chosen to be
\begin{equation}
  \label{eq:26}
 \psi_l=\frac{1}{4}(i\sigma^y)\left(\tau^0+\mathbf{n}(\theta_l)\cdot\boldsymbol{\tau}\right),
\end{equation}
where $\theta_l=2\pi l/3$, so that
 \begin{equation}
   \label{eq:34}
   |\psi_l\rangle=\frac{1}{2}\left[(|\mathsf{13}\rangle+|\mathsf{24}\rangle)+(|\mathsf{13}\rangle-|\mathsf{24}\rangle)\cos\theta_l+(|\mathsf{14}\rangle+|\mathsf{23}\rangle)\sin\theta_l\right].
\end{equation}
Indeed, choosing $\mathbf{z}=(0,0,1)$ and $\mathbf{x}=(1,0,0)$, we
have $\psi_0=\frac{1}{2}(i\sigma^y)\otimes\begin{pmatrix} 1 & 0 \\ 0 & 0 \end{pmatrix}=|\mathsf{13}\rangle$, and
  $\psi_l$ is obtained from $\psi_0$ through the rotation $\psi_l=R_l^T\psi_0R_l^{\vphantom{T}}$, with $R_l=\sigma^0 r_l$, where
  \begin{equation}
    \label{eq:27}
    r_l=\exp[\frac{i}{2}\frac{2\pi
      l}{3}\mathbf{y}\cdot\boldsymbol{\tau}]=\cos\frac{\pi
      l}{3}+i\mathbf{y}\cdot\boldsymbol{\tau}\sin\frac{\pi l}{3},
  \end{equation}
  where $\mathbf{y}=(0,1,0)$.
  
\subsection{Mapping between the SO(6) and SU(4) formulations}
\label{sec:su4-so6-mapping}

Here we describe the mapping between the SO(6) and SU(4) formulations and show that the classical ground state obtained in the two formulations is indeed the same state.

The six basis states $|\hat{n}\rangle$ in Eq.~\eqref{eq:55} correspond to the following $4\times 4$ antisymmetric matrices $\psi$:
\begin{align}
  \label{eq:map_su4_so6}
|\hat{1}\rangle& \to \psi_1=\frac{i}{2\sqrt{2}} \sigma^y\otimes \tau^z,\qquad 
|\hat{2}\rangle \to \psi_2=\frac{1}{2\sqrt{2}} \sigma^y\otimes \tau^0, \nonumber\\
|\hat{3}\rangle& \to \psi_3=\frac{i}{2\sqrt{2}} \sigma^y\otimes \tau^x,\qquad
|\hat{4}\rangle \to \psi_4=\frac{i}{2\sqrt{2}} \sigma^y\otimes \tau^y, \nonumber\\
|\hat{5}\rangle& \to \psi_5=\frac{1}{2\sqrt{2}} \sigma^0\otimes \tau^x, \qquad
|\hat{6}\rangle \to \psi_6=\frac{1}{2\sqrt{2}} \sigma^z\otimes \tau^z, 
\end{align}
Using this mapping one can translate the classical states that optimize the energy on the triangular lattice corresponding to $\mathbf{v}(\theta)$ in Eq.~\eqref{eq:app18} to the corresponding $\psi(\theta)$.
More explicitly, for $\mathbf{x}=\mathbf{e}_1$, $\mathbf{y}=\mathbf{e}_2$, $\mathbf{z}=\mathbf{e}_3$ with $\mathbf{e}_n$ denoting the unit vector along the $n$th dimension in $\mathbb{R}^6$ we obtain a state
\begin{equation}\label{eq:psi_theta}
\psi(\theta) = \frac{1}{4} \sigma^y \left(\tau^0-\mathbf{n}(\theta)\cdot {\boldsymbol{\tau}} \right),
\end{equation}
where $\mathbf{n}(\theta)=(\sin\theta, 0, \cos\theta)$ and ${\boldsymbol{\tau}}=(\tau^x,\tau^y,\tau^z)$.
Thus, $\psi(\theta_l)$, with $\theta_l=2\pi l/3+\pi$ reproduce the states in Eq.~\eqref{eq:26} up to an overall phase.

\section{Large Hund's coupling limit}
\label{app:LargeJH}

In the large Hund's coupling limit, i.e.\ $J_H/J\gg1$, the term  $-J_H\sum_i\mathbf{S}_i^2$ in
Eq.~\eqref{eq:H_hund} requires the total spin at each site to be in the $S=1$ representation of SU(2). 
The associated vector space is spanned by
\begin{flalign}
  \label{eq:app32}
&\qquad |S=1,s^z=1\rangle &=& \qquad |2\rangle &=& \qquad\frac{1}{\sqrt{2}}(|\hat{1}\rangle+i|\hat{2}\rangle),\nonumber \\
&\qquad |S=1,s^z=0\rangle &=& \qquad \frac{1}{\sqrt{2}}(|3\rangle+|4\rangle) &=& \qquad |\hat{3}\rangle,\nonumber \\
&\qquad |S=1,s^z=-1\rangle &=& \qquad |5\rangle &=&\qquad \frac{1}{\sqrt{2}}(-|\hat{1}\rangle+i|\hat{2}\rangle),& 
\end{flalign}
and thus the operator $\hat{\mathcal{P}}_{i,S=1}=\sum_{n=1}^3|\hat{n}_i\rangle\langle\hat{n}_i|$ projects the state on site $i$ onto the $S=1$ subspace.
Note also that $|S=1,s^x=0\rangle=|\hat{1}\rangle$ and $|S=1,s^y=0\rangle=|\hat{2}\rangle$, and thus the $S=1$ spin operators can be written as
\begin{equation}
  \label{eq:app31}
  S^z=\sqrt{2}\hat{A}^{21},\qquad S^x=\sqrt{2}\hat{A}^{32},\qquad S^y=\sqrt{2}\hat{A}^{13}.
\end{equation}

Denoting by $\hat{\mathcal{P}}_{S=1}=\prod_i \hat{\mathcal{P}}_{i,S=1}$, where $i$ runs over all lattice sites, to lowest order in $J/J_H$ the SO(6) ``Heisenberg'' Hamiltonian becomes:
\begin{eqnarray}
  \label{eq:app30}
  \hat{H}_{S=1} = \hat{\mathcal{P}}_{S=1} \hat{H} \hat{\mathcal{P}}_{S=1}  = J\sum_{\langle ij \rangle}\sum_{m,n=1}^3\hat{A}^{mn}_i\hat{A}^{mn}_j = J\sum_{\langle ij\rangle}\mathbf{S}_i\cdot\mathbf{S}_j.
\end{eqnarray}

\section{Additional numerical results}

\subsection{Probing magnetic order}\label{app:DMRG}

\begin{figure}
\centering
\includegraphics[width=0.7\linewidth]{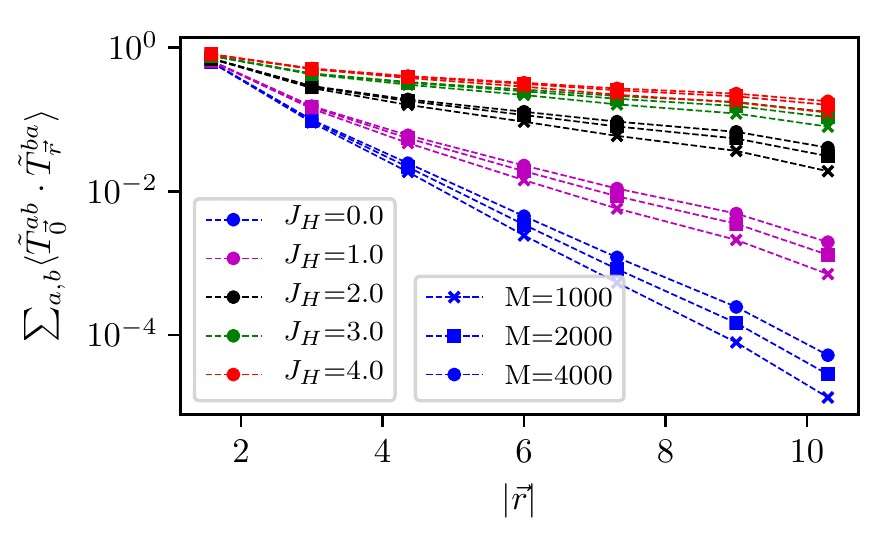}
\caption{Flavor-flavor correlations obtained using DMRG and shown on a logarithmic scale for a cylinder of circumference $N_y=3$ and length $N_x=12$. Different colors correspond to different values of $J_H$ (given in units of $J$) and different markers to different bond dimensions $M$. For $J_H=0$ the decay of the correlations is consistent with an exponential. }
\label{fig:dmrgSpinSpin}
\end{figure}

To complement the analysis presented in Sec.~\ref{sec:DMRG} of the main text, indicating that a finite Hund's coupling, $J_H$, is required to drive the system into a magnetically ordered state, we calculate flavor-flavor correlations in the absence of pinning fields at the ends of the cylinder. More specifically, the correlations calculated are $\sum_{a,b={\sf 1}}^{\sf 4} \langle \tilde{T}^{ab}_{\vec{0}}\tilde{T}^{ba}_{\vec{r}} \rangle$, where $\vec{0}$ denotes the origin which we choose to be at the left end of the cylinder, and we consider positions $\vec{r}$ on the lattice which correspond to the same sub-lattice as the site at $\vec{0}$ when $120^\circ$-order is present.  When more than one site on the lattice correspond to the same distance $|\vec{r}|$, a symmetrization is performed and an average value for the correlations is used.
Resulting correlations are shown in Fig.~\ref{fig:dmrgSpinSpin} for cylinders of circumference $N_y=3$ as $J_H$ is increased and the bond dimension is varied. For the maximal bond dimension used of $M=4000$, the truncation error was of order $10^{-4}$.

\subsection{Projection of the SU(4) spin model onto the subspace of singlet coverings using MPS}\label{app:ProjectionWithMPS}

As mentioned in the main text, to study the projection of the SU(4) Heisenberg model onto the subspace of nearest-neighbor singlet coverings, for system sizes of 12 sites and larger, we use MPS-based simulations.

We start by constructing the MPS representations of the nearest-neighbor singlet coverings. 
We note that the tensor product of two 6-dimensional vector representations of SO(6) is given by the sum of a symmetric traceless, antisymmetric and a one-dimensional representation (the singlet state). 
Therefore, the projection onto the singlet state, given by the operator $\hat{P}_{ij}$ (see Eq.~\eqref{eq:56b} in the main text), can be written as
\begin{align}
\hat{P}_{ij} \propto \left( \sum_{a,b={\sf 1}}^{\sf 4} \tilde{T}_i^{ab} \tilde{T}_j^{ba} + \hat{{\rm Id}}_{ij} \right) & \left( \sum_{a,b={\sf 1}}^{\sf 4} \tilde{T}_i^{ab} \tilde{T}_j^{ba} - \hat{{\rm Id}}_{ij} \right) = \nonumber \\
& \left( -\hat{Q}_{ij} + \hat{\Pi}_{ij} + \hat{{\rm Id}}_{ij} \right) \left( -\hat{Q}_{ij} + \hat{\Pi}_{ij} - \hat{{\rm Id}}_{ij} \right),
\end{align}
where the first (second) term in the product above projects out the anti-symmetric (symmetric) representation.
Given a nearest-neighbor covering $C=\{(i_k,j_k)\}_{k=1,..,N/2}$, as was defined in the main text,
we can obtain the corresponding singlet covering MPS by applying the matrix product operator (MPO) representation of the product of projectors $\prod_{k=1}^{N/2} \hat{P}_{i_kj_k}$  to a random initial MPS. Note that to allow for an $SU(4)$ singlet covering state on a system of width $N_y$ a bond dimension of $6^{N_y}$ is required for the MPS. Once the MPS representations of the singlet coverings are obtained, both the overlap matrix, required to solve the generalized eigenvalue problem, and the matrix elements of the projected Hamiltonian can be computed. For the latter, an MPO representation of the original spin Hamiltonian is used.
We then solve the generalized eigenvalue problem (since the dimension of the projected Hamiltonian is greatly reduced compared to the one of the original spin Hamiltonian, it can be easily diagonalized using standard sparse diagonalization), both to find the ground state of the projected Hamiltonian in terms of the singlet coverings, and to calculate the gap in the projected problem.

To calculate the overlap of the ground state of the projected Hamiltonian with the ground state of the original spin Hamiltonian, we obtain an MPS representation of the latter using DMRG.
For the results presented in Table~\ref{tab:GSOverlaps} in the main text, bond dimensions used for the calculation of the ground state were between $M=1000$ and $M=2000$ depending on system size, resulting in truncation errors $\epsilon$ smaller than $2\cdot10^{-3}$ in all cases. A finite truncation error gives rise to an error in the calculation of the overlap that we estimate to be of order $\sqrt{\epsilon}$.

\subsection{Exact diagonalization of the dimer model}

\subsubsection{Energies and excitation spectrum of the interpolated Hamiltonian}\label{app:DimerEDEn}

In Table~\ref{tab:edEnergies} we summarize the energies of the lowest energy states of the dimer Hamiltonians $H_{n=0,1,2}$ (where $n$ denotes the order of the expansion in $x$) obtained using ED. We list the energies of the lowest energy states in the topological sectors $\rm{TS}(0,0)$ and  $\rm{TS}(1,1)$, for three different system sizes with $N=12,16$ and $36$ sites. We note that the energies obtained for $H_0$ reproduce the ones presented in~\cite{Ralko2005} for $v/t=0$.

\begin{table}
\centering
\begin{tabular}{l | c | c | c} 
N=12 & $H_0$ & $H_1$ & $H_2$ \\
 \hline \hline 
(0,0) & -4.05317 & -3.25070 & -3.29407 \\
(1,1) & -4.37228 & -3.64575 & -3.76333 \\
\hline
$\Delta E$  & 0.31911 & 0.39505 & 0.46926
\end{tabular}
\\
\begin{tabular}{l | c | c | c} 
N=16 & $H_0$ & $H_1$ & $H_2$ \\
 \hline \hline 
(0,0) & -5.52971 & -4.43419 & -4.54785 \\
(1,1) & -5.42488 & -4.32630 & -4.42862 \\
\hline
$\Delta E$ & -0.10482 & -0.10789 & -0.11923
\end{tabular}
\qquad
\begin{tabular}{l | c | c | c} 
N=36 & $H_0$ & $H_1$ & $H_2$ \\
 \hline \hline 
(0,0) & -11.76017 & -9.40533 & -9.59950 \\
(1,1) & -12.03778 & -9.91507 & -10.08708 \\
\hline
$\Delta E$ & 0.27761 & 0.50974 & 0.48758
\end{tabular}
\caption{Energies of the lowest energy states in the two topological sectors, as well as the gap $\Delta E = E_{(0,0)}-E_{(1,1)}$, obtained using ED of the dimer models $H_n$ (where $n=0,1,2$ is the order of the expansion in the parameter $x=1/6$) for three different system sizes. }
\label{tab:edEnergies}
\end{table}
\begin{figure}
\begin{subfigure}[t]{.32\linewidth}
\includegraphics[width=\linewidth]{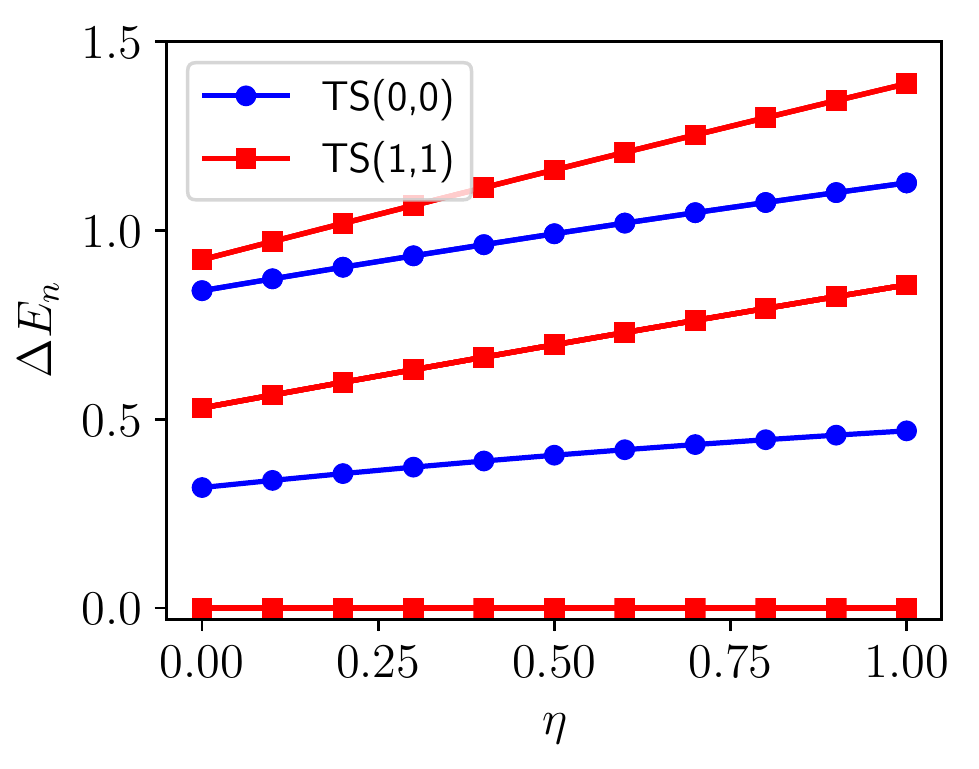}
\captionsetup{skip=-1.6in}
\caption{}
\end{subfigure}
\begin{subfigure}[t]{.32\linewidth}
\includegraphics[width=\linewidth]{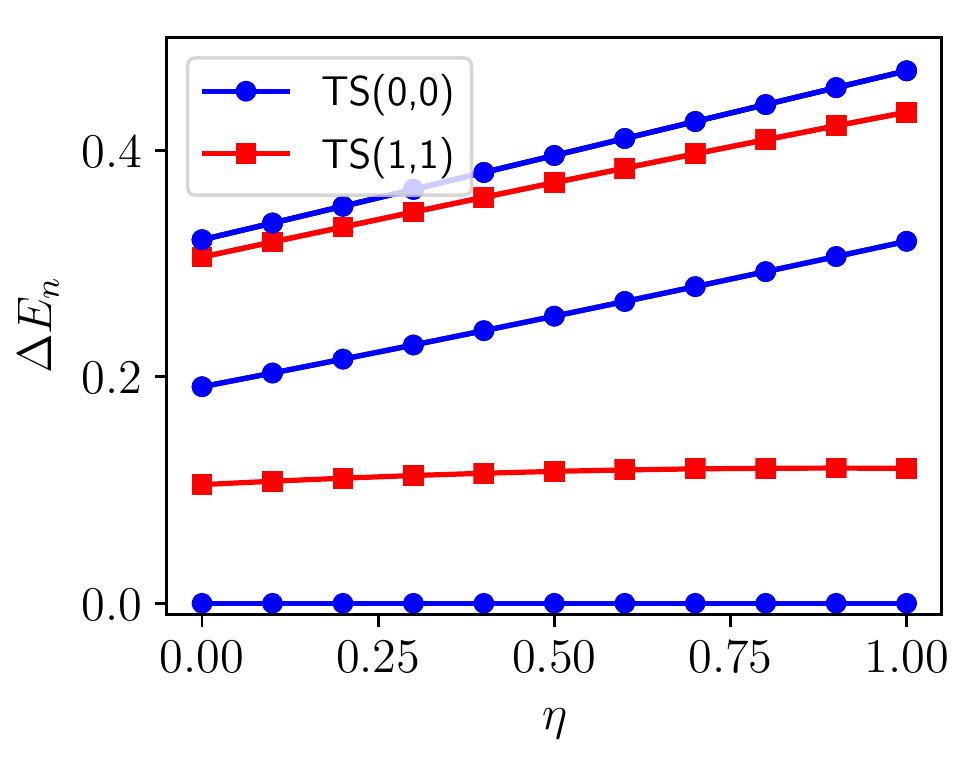}
\captionsetup{skip=-1.6in}
\caption{}
\end{subfigure}
\begin{subfigure}[t]{.32\linewidth}
\includegraphics[width=\linewidth]{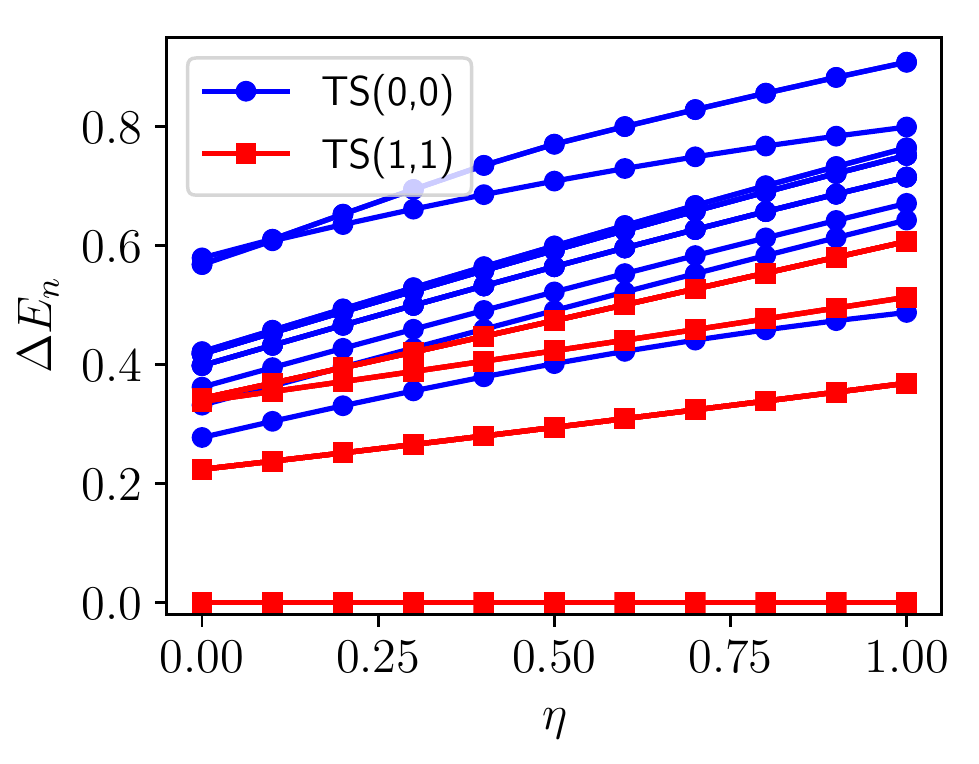}
\captionsetup{skip=-1.6in}
\caption{}
\end{subfigure}
\caption{Low energy excitation spectrum of the interpolated
  Hamiltonian $(1-\eta) H_0+\eta H_2$ for systems with $N=12,16,36$
  sites in (a,b,c) respectively. Energy states in the topological
  sector $\rm{TS}(0,0)$ are plotted in blue (circles), and the ones in
  $\rm{TS}(1,1)$ are plotted in red (squares).}
 \label{fig:edSpectrumInterpolation}
\end{figure}

\subsubsection{Dimer-dimer correlations}\label{app:DimerEDCorr}

\begin{figure}
\begin{subfigure}[t]{.5\linewidth}
\includegraphics[width=\linewidth]{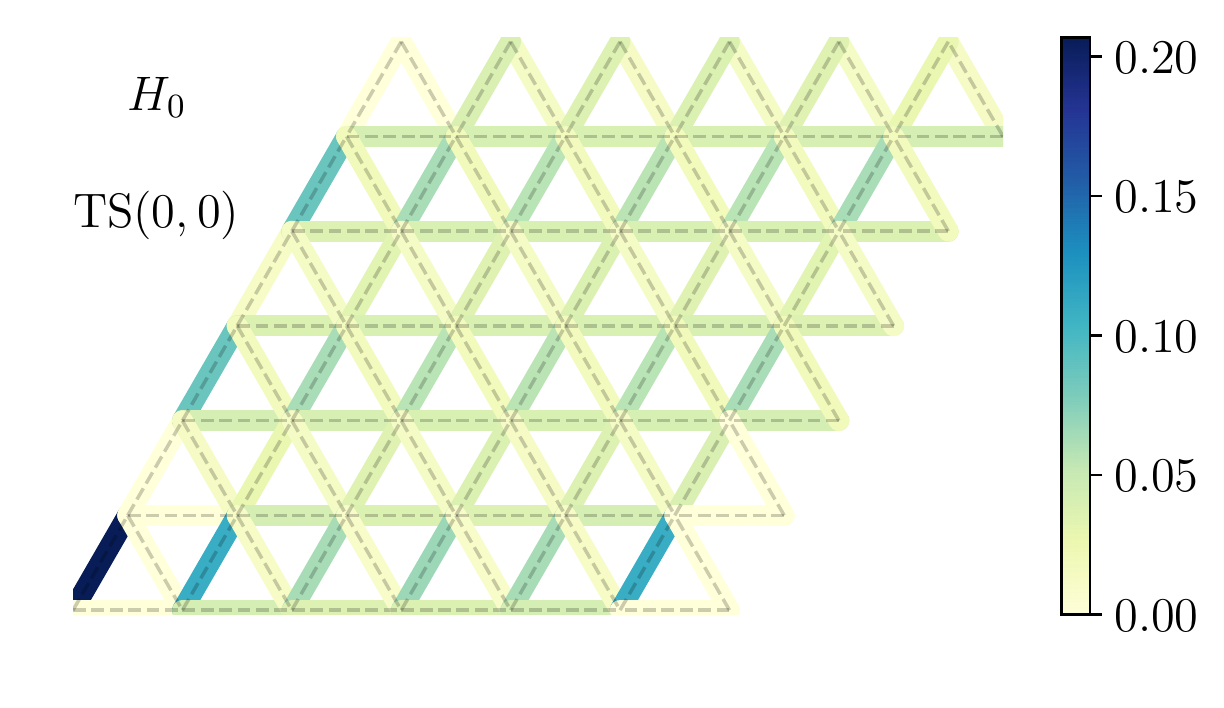}
\captionsetup{skip=-1.75in}
\caption{}
\end{subfigure}
\begin{subfigure}[t]{.5\linewidth}
\includegraphics[width=\linewidth]{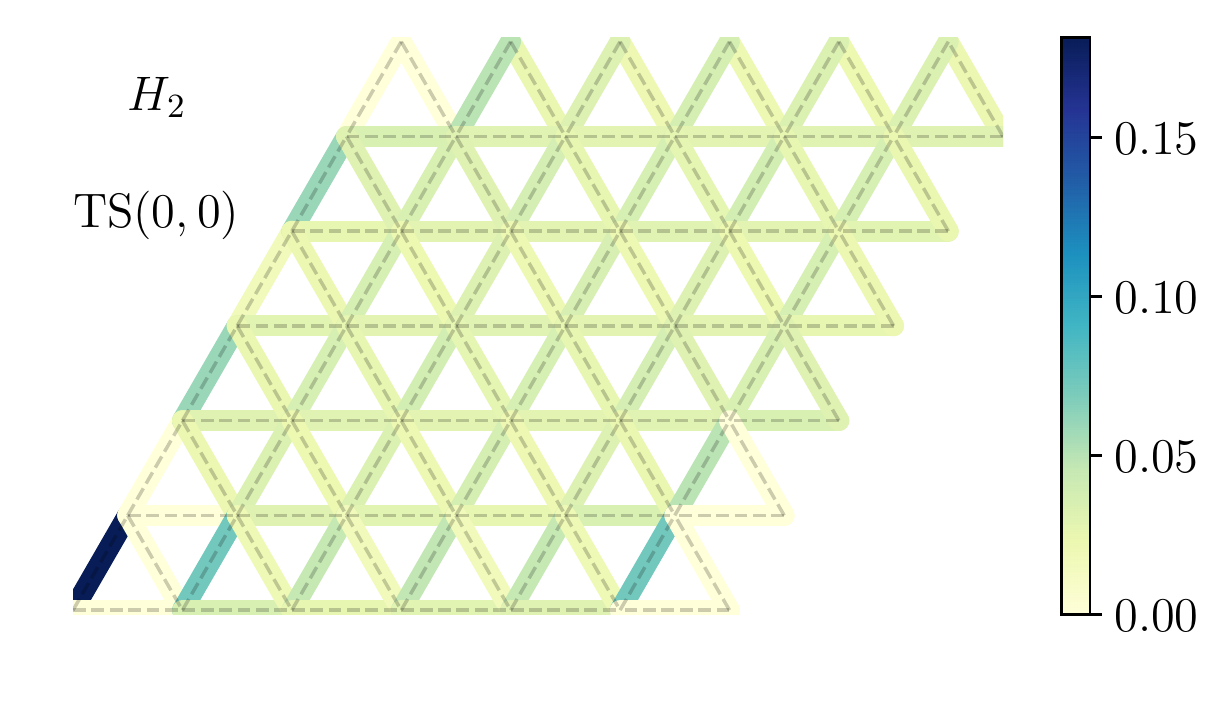}
\captionsetup{skip=-1.75in}
\caption{}
\end{subfigure} \\
\begin{subfigure}[t]{.5\linewidth}
\includegraphics[width=\linewidth]{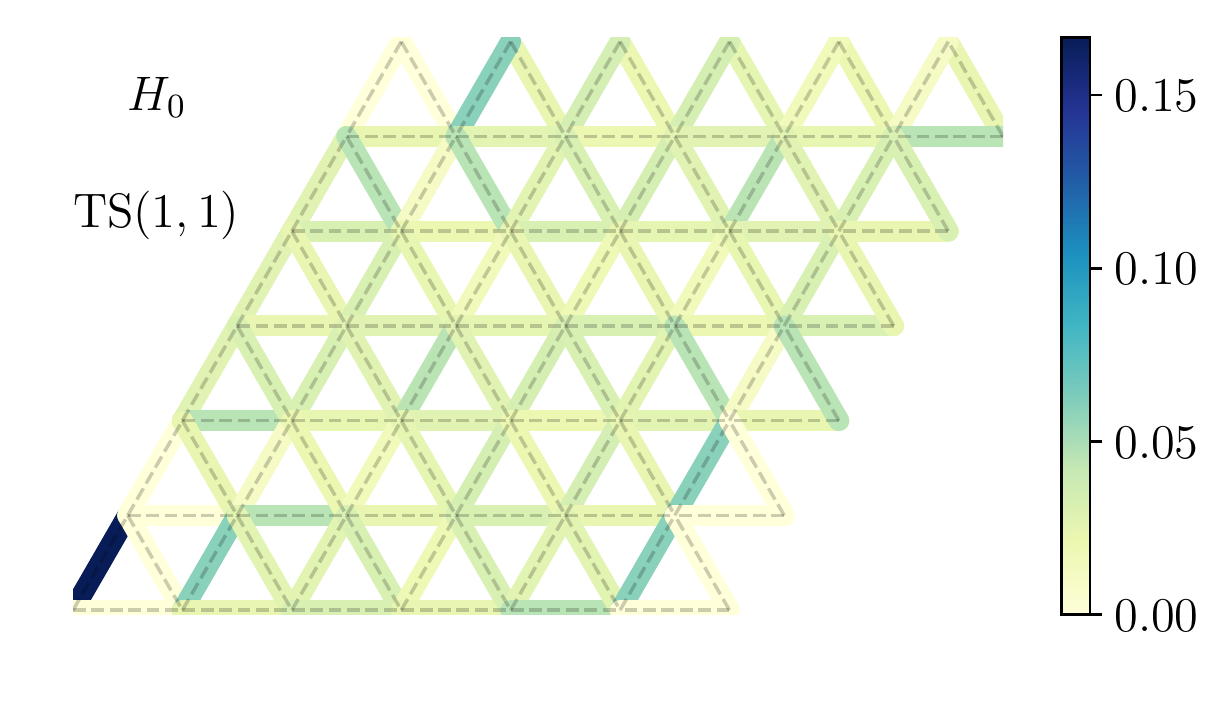}
\captionsetup{skip=-1.75in}
\caption{}
\end{subfigure}
\begin{subfigure}[t]{.5\linewidth}
\includegraphics[width=\linewidth]{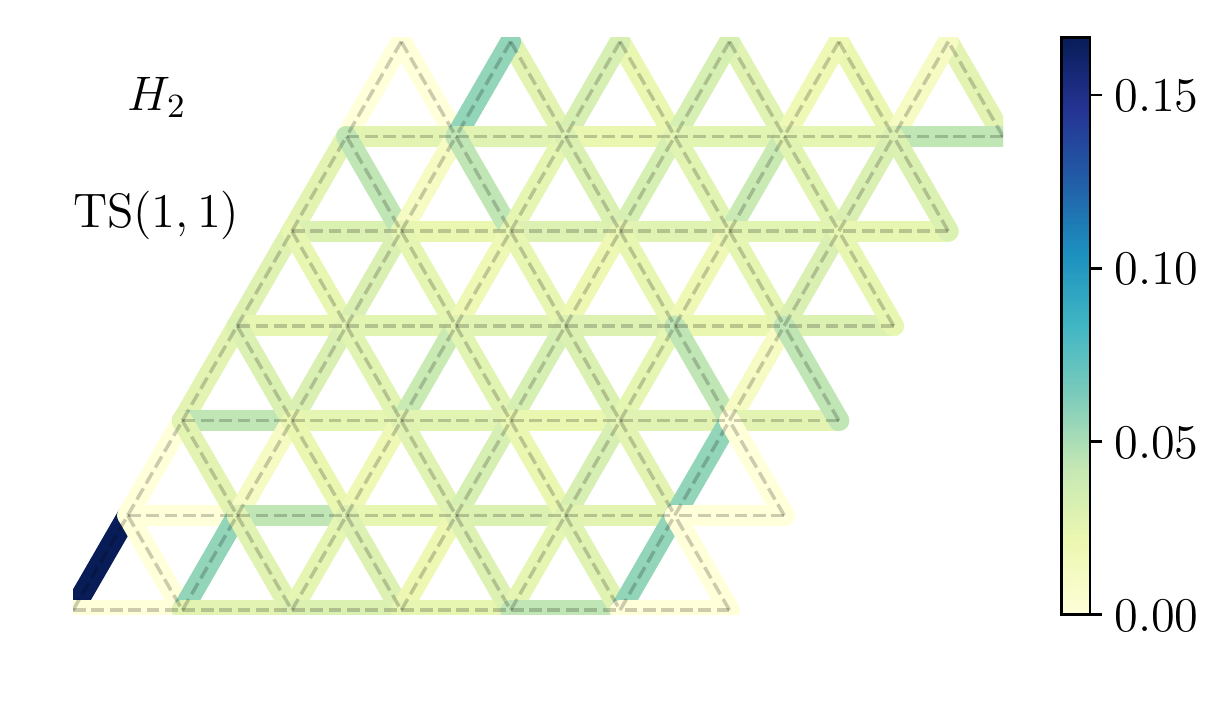}
\captionsetup{skip=-1.75in}
\caption{}
\end{subfigure}
\caption{Dimer-dimer correlations $\langle b_1 b_i \rangle$, in the
  lowest energy state in the topological sectors $\rm{TS}(0,0)$ and $\rm{TS}(1,1)$
  respectively for the standard dimer model with $v/t=0$, $H_0$, on
  the left, and of the extended dimer model $H_2$, on the
  right.}
 \label{fig:edBondBond_H0}
\end{figure}

In Fig.~\ref{fig:edBondBond_H0} we present side by side the real space dimer-dimer correlations for the lowest energy state of $H_0$ and $H_2$ respectively, in the two topological sectors $\rm{TS}(0,0)$ and $\rm{TS}(1,1)$. As was mentioned in the main text the correlations in $\rm{TS}(0,0)$ become more uniform for $H_2$, while for $\rm{TS}(1,1)$ the correlations remain practically unchanged.

\end{appendix}

\bibliography{su4bib.bib,su4numerics.bib}

\nolinenumbers

\end{document}